\documentclass{aa}
\usepackage{natbib}
\bibpunct{(}{)}{;}{a}{}{,}
\usepackage{graphicx}
\usepackage{subfig}
\usepackage{txfonts}
\usepackage{epsfig}
\usepackage{longtable}
\usepackage{lscape}
\usepackage{amssymb}
\usepackage{xspace}
\usepackage{placeins}

\usepackage[colorlinks=true,citecolor=blue]{hyperref}
\usepackage{refcount}
\usepackage{xcolor}

 \definecolor{arancio}{rgb}{1,0.5,0}
 \definecolor{viola}{rgb}{0.7,0,1}
 \definecolor{verde}{rgb}{0.2,0.7,0.7}
\definecolor{cobalt}{rgb}{0.0, 0.28, 0.67}
\definecolor{airforceblue}{rgb}{0.36, 0.54, 0.66}
\definecolor{ballblue}{rgb}{0.13, 0.67, 0.8}
\definecolor{battleshipgrey}{rgb}{0.52, 0.52, 0.51}
\definecolor{darkgreen}{rgb}{0.0, 0.2, 0.13}

\def\inte{{\em INTEGRAL}}
\def\xmm{\textit{XMM-Newton}\xspace}

\def\chan{{\em Chandra}}
\def\asca{{\em ASCA}}

\def\rxte{{\em RXTE}}
\def\swift{\textit{Swift}\xspace}

\def\suzaku{{\em Suzaku}}
\def\nustar{{\em NuSTAR}}
\def\uhuru{{\em Uhuru}}
\def\pn{\textrm{EPIC-pn}\xspace}
\def\moso{\textrm{MOS1}\xspace}
\def\most{\textrm{MOS2}\xspace}

\def\ferg{\mathrm{erg\,s^{-1}\,cm^{-2}}}

\def\u4{{4U\,1907+09}}
\def\J393{{IGR\,J16393$-$4643}}
\def\IGR19{{IGR\,J19140$+$0951}}
\def\2IGR{{IGR\,J17503$-$2636}}
\def\3IGR{{IGR\,J18410$-$0535}}
\def\4IGR{{IGR\,J11215$-$5952}}
\def\5IGR{{IGR\,J17315$-$3221}}
\def\xte{{XTE\,J1855$-$026}}
\def\6IGR{{IGR~J17329$-$2731}}

\begin{document}

   \title{\xmm\ and \swift\ observations of supergiant high mass X-ray binaries}
   \author{C. Ferrigno
          \inst{1}
          \and
          E. Bozzo\inst{1,2}
          \and
          P. Romano\inst{3}
          }
   \institute{Department of astronomy, University of Geneva, chemin d'\'Ecogia, 16, CH-1290, Versoix, Switzerland\\
              \email{carlo.ferrigno@unige.ch}
              \and
              INAF-OAR, Via Frascati, 33, 00078 Monte Porzio Catone RM, Italy
              \and
              INAF, Osservatorio Astronomico di Brera, Via E.\ Bianchi 46, I-23807, Merate, Italy
             }
   \date{\today}

\abstract{Wind-fed supergiant X-ray binaries are precious laboratories not only to study accretion under extreme gravity and magnetic field conditions, but also to probe still highly debated properties of massive star winds. These includes the so-called clumps, originated from the inherent instability of line driven winds, and larger structures. In this paper, we report on the results of the last (and not yet published) monitoring campaigns that our group has been carrying out since 2007 with both \xmm\ and the \swift\ Neil Gehrels observatory. Data collected with the EPIC cameras on-board \xmm\ allow us to carry out a detailed hardness ratio-resolved spectral analysis that can be used as an efficient way to detect spectral variations associated to the presence of clumps. Long-term observations with the XRT on-board \swift,\ evenly sampling the X-ray emission of supergiant X-ray binaries over many different orbital cycles, are exploited to look for the presence of large scale structures in the medium surrounding the compact objects. These can be associated either to corotating interaction regions or to accretion/photoionization wakes, as well as tidal streams. The results reported in this paper represent the outcomes of the concluded observational campaigns we carried out on the supergiant X-ray binaries \u4,\ \J393,\ \IGR19,\ and \xte,\ as well as the supergiant fast X-ray transients \2IGR,\ \3IGR,\ and \4IGR.\ All results are discussed in the context of wind-fed supergiant X-ray binaries and shall ideally serve to optimally shape the next observational campaigns aimed at sources in the same classes. We show in one of the paper appendices that \5IGR,\ preliminary classified in the literature as a possible supergiant X-ray binary discovered by \inte,\ is the product of a data analysis artifact and should thus be disregarded for future studies.}


   \maketitle

\section{Introduction}
\label{sec:intro}

Supergiant X-ray binaries (SgXBs) are a sub-class of high mass X-ray binaries (HMXBs) hosting most commonly a neutron star (NS) accreting from the wind of an OB supergiant. Apart from a few exceptions, the bulk of known systems in this class the compact object accretes from the in-flowing material that the massive companion looses through a fast and dense wind. The interest for SgXBs has been revived in the past years due to the recognition that these are key laboratories to investigate properties of the still highly debated massive star winds using the NS as a probe, especially in the domain of macro-clumping and large scale structures \citep[see, e.g.,][and discussions therein]{nunez17, bozzo16}.
\begin{table*}
\setlength{\tabcolsep}{2pt}
 \centering
  \caption{Properties of the sources studied in this paper.}\label{tab:windfed}
  \tiny
  \renewcommand{\arraystretch}{1.2}
\begin{tabular}{@{}lccccccccccc@{}}
\hline
ID & Companion & Distance & Orbital period &   Spin period &  Super-orbital period  &  $T_{\pi/2}$ & $e$ & $\omega$ & $a\sin{i}/c$ & $B_{\rm NS}$ & Type\\
& & kpc & (d) & (s) & (d) & (MJD) &  & (deg) &  (lt-s) & 10$^{12}$~G & \\
\hline
IGR\,J11215$-$5952 & B0.5Ia & $6.5^{+1.1}_{-1.5}$ & 164.6$\pm$0.1 & 186.78$\pm$0.3 & --- & 57925.5$\pm$0.5 & --- & --- & --- & --- & SFXT \\
IGR\,J16393$-$4643  & OB$^{a}$ & 12$^{a}$ & 4.2380$\pm$0.0005 & 904.0$\pm$0.1 & 14.9805$\pm$0.0022$^{a}$ & 53418.3$\pm$0.1 & --- & --- & --- & 2.5$\pm$0.1 & Class. \\
IGR\,J17503$-$2636  & OB$^{a}$ & 10$^{a}$ & --- & --- & --- & --- & --- & --- & --- & 2.0$^{a}$ & SFXT \\
IGR\,J18410$-$0535 & B1 Ib & 3.2$^{a}$ & --- & --- & --- & --- & --- & --- & --- & --- & SFXT \\
XTE\,J1855$-$026 & BN0.2 Ia & 10$^{a}$ & 6.0724$\pm$0.0009 & 361.1$\pm$0.4 & --- & 51495.25$\pm$0.002 & 0.04$\pm$0.02 & 226$\pm$15 & 80.5$\pm$1.4 & --- & Class. \\
4U\,1907$+$097 & O8/O9 Ia & 5$^{a}$ & $8.3753^{+0.0003}_{-0.0002}$ & 440.341$^{0.012}_{-0.0017} $ & --- & $50134.76^{+0.16}_{-0.20}$ & $0.28^{+0.10}_{-0.14}$ & $330\pm20$ & $83\pm4$ & 2.1 & Class. \\
IGR\,J19140$+$0951 & B0.5Ia/d & 2--5$^{a}$ & 13.5527$\pm$0.0001 & 5937$\pm$219\tablefootmark{a} & --- & 52061.42\tablefootmark{b} & --- & --- & --- & --- & Class. \\
\hline
\label{tab:sources}
\end{tabular}
\tablefoot{ We indicated with orbital epoch ($T_{\pi/2}$) either the mid-eclipse time or the reference orbital epoch used for the ephemerides if the source is not eclipsing. In the case of IGR~J11215$-$5952, we indicated as $T_{\pi/2}$ the mid-time of the 2017 outburst as reported by \citet[][but see also \citealt{romano09d}]{sidoli17}.
    In the case of IGR\,J16393$-$4641, $T_{\pi/2}$ corresponds to the epoch for which the minimum of the source lightcurve folded at the best known orbital period is at phase zero. For IGR\,J19140$+$0951, the reported orbital period is the one derived in Appendix~\ref{app:period}.
    We also reported, for the source for which it was possible, the orbit eccentricity $e$, the longitude of periastron $\omega$, the projected semi-major axis length $a\sin{i}/c$, and the estimated NS magnetic field strength ($B_{\rm NS}$, if derived from a confirmed CRSF). References are given in the text.
    \tablefoottext{a}{Tentative to be confirmed.}
    \tablefoottext{b}{This is the apparent time of the minimum of the orbital profile in the \swift/BAT band, chosen conventionally.}
}
\end{table*}

In the framework of this renewed interest, our group has started a number of monitoring program of several SgXBs with both \xmm\ and \swift\ to look for spectral variability in the X-ray emission of these sources that could be ascribed to the presence of massive structures in the stellar winds approaching the compact object and causing episodes of enhanced X-ray emission and/or obscuration of the high energy source \citep[either the so-called ``clumps'' or even larger structures; see, e.g.,][and later in this section]{puls08}. Our monitoring program covers both sources within the sub-class of the ``classical'' SgXBs, showing a moderate X-ray variability, up to a factor of $\sim$10$^3$ between quiescent and more active periods, and the supergiant fast X-ray transients (SFXTs), showing a much more pronounced X-ray variability up to a factor of 10$^6$ between quiescence and the brightest outbursts \citep[see, e.g.,][and references therein]{walter2015, romano15sb}.

As discussed by \citet{bozzo17b}, hunting for spectral variations during SgXBs flares and outbursts to study the smaller stellar wind clumps requires sufficiently long and uninterrupted observations of these sources with X-ray instruments endowed with a large effective area in the soft X-ray domain ($\sim$3~keV). This is because the time intervals for the spectral extraction have a typical duration of few hundreds to thousands of seconds and enough X-ray counts need to be collected to perform meaningful spectral fits and disentangle both continuum and absorption column density variations. These integration times are set by the usual duration of flares and outbursts, which is limited to a few hours at the most, and the need of having as many probed as possible time intervals along the event rise and decay to study the dynamics of the accretion process \citep[see, e.g.,][]{bozzo11,bozzo13b}. So far, the EPIC cameras on-board \xmm\ \citep{jansen01aa} have proven to be the most effective instruments to pursue this goal \citep{bozzo17b}. The techniques we deployed to look for the spectral variability in the \xmm\ data of classical SgXBs and SFXTs comprise an adaptively rebinned hardness ratio (HR) of the source energy-resolved lightcurves and a Bayesian block automatized selection of the time intervals corresponding to the most significant changes in the HR for the spectral extraction. These techniques are exhaustively described in a number of previous papers of ours, where we also illustrate the results obtained from several classical SgXBs and SFXTs \citep[see, e.g.,][and references therein]{bozzo10, bozzo11, bozzo13b, bozzo15, bozzo17b, ferrigno20}. The HR-resolved spectral analysis that we have carried out so far revealed that during sufficiently bright flares and outbursts from the SgXBs we recorded an increase of the absorption column density preceding the brightening event and a decrease of the absorption close to the peak of the event. In several cases, we also observed a new increase of the absorption column density toward the end of the flare/outburst, as well as a change in the centroid energy of the iron line feature that is produced due to the fluorescence of the accretion X-rays onto the surrounding stellar wind material. In a few observations, episodes of enhanced absorption of the X-rays from the NS have been observed for as long as few hundreds of seconds also without being in coincidence with either a flare or an outburst.
The physical picture that emerged from these results is that accretion in all analyzed SgXBs is compatible with occurring from a clumpy wind, where dense structures approach the compact object before being accreted and cause the local absorption column density to rise before the flare/outburst. The decrease of the absorption column density toward the peak of the event is ascribed to the photoionization effect of the enhanced X-rays onto the clump material, while the recombination following the beginning of the X-ray flux decay after the peak of the flare/outburst can explain the subsequent re-increase of the local absorption column density back to pre-rebrightening values. This scenario is more quantitatively confirmed for those cases where also a change in the centroid energy of the iron line is measured, as this provides a clearer identification of the ionization status of the stellar wind around the compact object \citep[see, e.g.,][and references therein]{bozzo11}. Those episodes in which only a transient absorption of X-rays is visible without being associated with a rebrightening event are commonly ascribed to clumps passing in front of the compact object along the line of sight of the observer without intercepting its orbit and being (at least) partly accreted.

The results obtained from several observations of classical SgXBs and SFXTs suggest that the features measured through the HR-resolved spectral analysis are relatively similar for both classes of objects. Although  clumps are thus a key ingredient for the accretion in both sub-classes of sources, it seems necessary to assume that additional mechanisms are at work in SFXTs to explain their much more prominent X-ray variability. So far, considered mechanisms include gatings due to the NS rotation and magnetic field \citep{bozzo08, bozzo17} and the onset of a long-standing settling accretion regime \citep{shakura12}.

It should be mentioned that a number of flares, especially from the SFXTs, did not show evidence of substantial absorption column density enhancements \citep[see, e.g.,][]{bozzo15}. These events have been either interpreted as triggered by the above ``additional mechanisms'' without the (significant) intervention of clumps or as events observed through a direction such that our line of sight did not pass through (or intercept) the clump. As of today, the geometrical shape of the clumps (as well as other relevant parameters like the spatial extends in general, the masses and densities) are poorly constrained and thus geometrical effects have to be folded in the study of the accretion process as additional uncertainties.

Beside clumps, other massive structures are known to populate the accretion environments of NSs in SgXBs and can thus contribute to enhance the X-ray variability of these systems. Among such structures, there are the so-called ``corotating interaction regions" (CIRs) which are extended structures originating from the surface of the supergiant star and extending up to several stellar radii. The CIRs are characterized by mild over-density ratios compared to the rest of the stellar wind and since they do not perfectly co-rotate with the supergiant star, the interception of the NS with one of these structures could result in periodic enhancements of the X-ray luminosity and absorption column density at specific orbital phases \citep[thus giving rise also to super-orbital modulations; see][and references therein]{bozzo16}.

Similar enhancements at fixed orbital phases can also be produced by the presence of accretion and photoionization wakes, as well as tidal streams. Accretion wakes are dense structures partly surrounding the compact object and produced by the focusing of the stellar wind medium by the NS gravitational influence, while photoionization wakes are regions where the over-density (compared to the surrounding accretion medium) is due to the X-ray photoionization of the stellar wind and its partial stagnation. Finally, tidal streams are possible only in those systems where the supergiant companion nearly fills its Roche lobe \citep[see, e.g.,][and references therein]{manousakis11,grinberg17, kre19}.

The study of orbital-phase dependent structures requires long-term observations covering as many orbital periods as possible. The reason is that the eventual spectral variability recorded from the data on a single specific orbital phase is likely to be dominated by the effect of short-term variations of the accretion environment associated to clumps (thousands of seconds to hours). Averaging data at the same orbital phase but collected over many different orbits ensures that the short-term variability of the clumps is ``washed away'' and spectral changes can be most likely ascribed to the presence of large scale stable structures. For these reasons, relatively short snapshots with the narrow field instrument  X-ray Telescope \citep[XRT, ][]{burrows05} on-board the Neil Gehrels  \swift\ Observatory \citep[\swift, ][]{Gehrels2004} carried out over several months (considering that the typical orbital period of an SgXB is of few tens of days at the most)  provide us the most effective strategy to collect the required data for the analysis of large scale wind structures in classical SgXBs and SFXTs.

In this paper, we report on several \xmm\ observations of classical SgXBs and SFXTs that were either not
published nor yet analyzed with our techniques (see above) to reveal spectral variations that can be
associated to the presence of clumps. These results complement those reported previously in our papers on
this topic. Furthermore, we report for the first time on the analysis of monitoring observational campaigns
performed with \swift\,/XRT on several classical SgXBs in order to investigate possible spectral
variability as a function of the orbital phase. Although some of these data were already reported
elsewhere, we systematically apply a technique that can most effectively be used to reveal the presence of
large scale structures around the compact objects in these systems. We discuss then our results in the
framework of wind accretion in neutron star SgXBs.

The sources investigated in this paper are listed in Table~\ref{tab:sources}.
    We note that only in the case of \u4\ both kinds of observations to study
the short and long-term spectral variability associated to clumps and larger wind scale structures are
available. In all other cases, either focused \xmm\ observations or longer-term \swift/XRT data have been
collected. For \xte,\ XRT data have been exploited to search for the short-term spectral variability
associated to clumps because the observations were performed during a few rare bright outbursts of the
source. In the case of \2IGR,\ the data from the XRT observational campaign could not be folded on the
source orbital period as this is still unknown.

In Sect.~\ref{sec:redux}, we describe the general analysis methods for
    \swift\ and \xmm.
    In Sect.~\ref{sec:sample} we briefly describe each sources in our sample with an overview of the
    data-sets available for each of them, we detail specific analysis methods,
    the results we obtained, and their discussion.
    In Sect.~\ref{sec:conclusions}, we draw the conclusions of our work.

\section{General data analysis methods}
\label{sec:redux}

\subsection{\swift}
All \swift\ data were uniformly processed and analyzed using the standard software
({\sc FTOOLS}\footnote{\href{https://heasarc.gsfc.nasa.gov/ftools/ftools_menu.html}{https://heasarc.gsfc.nasa.gov/ftools/ftools\_menu.html.}} v6.29b),
calibration (CALDB\footnote{\href{https://heasarc.gsfc.nasa.gov/docs/heasarc/caldb/caldb_intro.html}{https://heasarc.gsfc.nasa.gov/docs/heasarc/caldb/caldb\_intro.html.}}  20210915), and methods. The \swift/XRT data were
processed and filtered with the task {\sc xrtpipeline} (v0.13.6).

For all sources with available XRT data, we first extracted the average spectrum in
each XRT observation to measure the 0.3--10\,keV flux.
Then, we computed, for each observation, the HR value using the source count-rate in the 0.3--4\,keV
and 4--10\,keV energy bands.
If an orbital period was available  (\u4, \J393, \IGR19), we  calculated the orbital phase corresponding to
each XRT observation and
chose eight phase bins that would yield a comparable number of source counts (\u4: $\sim 2700$ counts;
\J393: $\sim 1300$ counts;  \IGR19: $\sim 900$ counts).
Using the above energy bands,
we calculated the hardness ratio for each phase bin and plotted the hardness ratio as a function of the
orbital phase.

We note that the XRT observations  are generally composed of up to three snapshots with typical exposures
of 300--1000~s. Therefore, none of the XRT observations is suitable to look for the known pulsations of
some of the target sources (see Sect.~\ref{sec:sample}); however, computing the HR in each of them rather
than in the single snapshots ensures us that the effect of the pulse period energy dependence on the HR is
averaged out. This is further strengthened by the fact that several different observations are averaged
together in order to compute the HR in the defined orbital phase bins.

In order to investigate more in depth possible spectral variability in different orbital phase bins, as
suggested by the HR variations, we extracted different source spectra for each of these bins, grouped them
so as to have at least one count per bin, and fit them by using a simple absorbed power-law model in

\textsc{xspec} version 12.12.0. For the absorption column density, we adopted the {\sc Tbabs} component
with {\em wilm} abundances \citep{wilms00} and {\em vern} cross sections \citep{vern96}.

\subsection{\xmm}
All \xmm\ observation data files (ODFs) were processed by using the \xmm\ Science Analysis System (SAS 20.0.0), following standard procedures\footnote{
\href{http://www.cosmos.esa.int/web/xmm-newton/sas-threads}{http://www.cosmos.esa.int/web/xmm-newton/sas-threads}.}. We filtered out background flaring time intervals due to
cosmic protons by extracting the 10--12\,keV \pn lightcurve binned at 100\,s and identifying a count rate with less than 0.1\% probability to belong to a
Gaussian distribution. We retained, however, any time interval with a count rate below 0.6 counts per second.

The regions adopted for the extraction of all source and background scientific products (lightcurves and spectra) were chosen by checking where the instrument point spread function includes
an optimal fraction of the photons specific to each source, or by picking the maximum allowed radius when the observation was carried out in small-window mode (details for each source are specified in the following sub-sections).
We extracted the spectra and lightcurves of the three EPIC camerasm together with the associated ancyliary and response matrices, using standard procedures.

We grouped all EPIC spectra using the algorithms described in \citet{Kaastra} and adopted
as baseline model a power-law affected at the lower energies by photoelectric absorption.
In addition to what used for \swift/XRT data, we generally add an absorber partly covering the source (\texttt{pcfabs} in \textsc{xspec}).
To find the best-fit parameters, we minimized Cash statistics with background correction (\texttt{Cstat} in \textsc{Xspec}),
using a standard Levenberg-Marquardt algorithm based on the \texttt{CURFIT} routine from Bevington and then computed uncertainties
using a Monte Carlo Markov chain exploiting the Goodman-Weare algorithm.
We used 60 walkers, a burning phase of 6000, and a chain length of 36\,000. Uncertainties were found by computing the appropriate
percentiles in the posterior distributions.
As customary, we used a logarithmic prior for normalization and
column density, and a linear prior for the slope and covering fraction.

For each source observed by \xmm,\ we extracted the \pn, \moso, and \most light curves
in two energy bands by computing the band limits as the median photon energy in the \pn camera to optimize the computation of the hardness ratio (HR).
We summed together all EPIC lightcurves of each source to improve the statistical uncertainties and compute the adaptively
rebinned hardness ratio \citep[HR; see][for a description of the adaptive rebinning method]{bozzo13}.
A typical minimum signal to noise ratio (S/N) of 25 was achieved for the different sources in each of the time bins of the summed lightcurves.
We determined periods in
which the HR changes significantly by a Bayesian block analysis optimized to identify the most relevant HR variations
with a negligible number of false positives \citep[we used the fitness for point measurements and \texttt{ncp\_prior}=3.98, see][Sect 3.3]{Scargle2013}.
In each of the identified intervals, we extracted the corresponding spectra using appropriate extraction regions as for the averaged spectra and constructed also the corresponding ancillary and response matrices.

All uncertainties in the measured parameters from each source in the following sub-sections are indicated at 68\,\% confidence level, unless stated otherwise.

\section{Description of targets, analysis results, and discussion}
\label{sec:sample}

\subsection{ \u4}
\label{sec:4u}

\u4\ is a classical supergiant X-ray binary discovered in the '70s by the \uhuru\ satellite \citep{forman78apj, 1972ApJ...178..281G} and hosting
a slowly rotating NS \citep[the spin period is $\sim$437~s;][]{makishima84} orbiting a O8/O9 Ia supergiant \citep{kerk89, cox05}.
The system orbital period is measured at $\sim$8.38~d \citep{marshall80, zand98}.
The NS in this system is known to undergo episodes of spin torque reversal \citep{fritz06, inam09},
to display occasional quasi-periodic oscillations at a frequency of about 65~mHz
\citep{zand98, muke01}, and to be endowed with a magnetic field strength of
$\sim$2$\times$10$^{12}$~G \citep[the estimate is provided by the presence of a well known cyclotron scattering feature in the source X-ray spectrum with a centroid energy of $\sim$19~keV; see, e.g.,][and references therein]{hemphill13, varun19}.
As most highly magnetized NSs in HMXBs, the compact object in \u4\ features a complex pulse profile,
which is both energy and luminosity dependent.
This has been the central subjects of several literature works on the source \citep[see, e.g.,][]{muke01, rivers10, tr12, fuerst12}.
In the X-ray band, this source displays a dipping behavior which has been known for decades but is yet not clearly understood \citep{zand97,doro12}.

\subsubsection{Data analysis and results}
\begin{figure}
\vspace{-1.4truecm}
\hspace{-0.2truecm}
   \includegraphics[width=1.15\columnwidth]{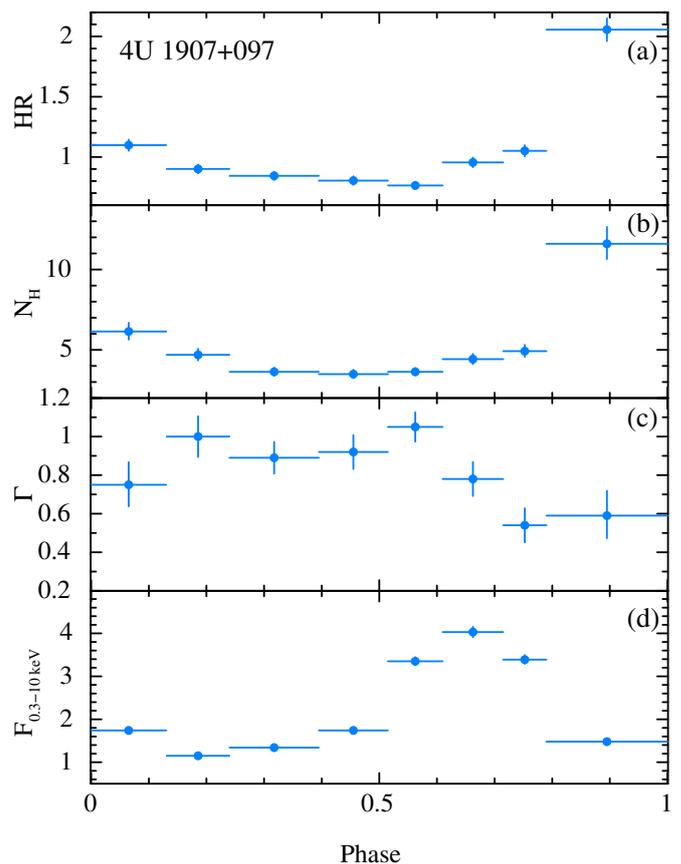}
   \caption{\swift/XRT hardness ratio of  \u4,  and best-fit parameters as a function of orbital phase
(absorption column density $N_{\rm H}$  in units of $10^{22}$\,cm$^{-2}$, power-law photon index $\Gamma$,
and 0.3--10\,keV flux not corrected for absorption in units of $10^{-10}$ erg\,cm$^{-2}$\,s$^{-1}$).
}
   \label{fig:u1907_spec_by_phase}
\end{figure}

Our yet unpublished monitoring campaign on \u4{} with \swift\,/XRT was performed
with a pace of two observations per week, each 1\,ks long, spanning from February to September 2015 (ObsID 33483).
The full log of the XRT observations is provided in Table~\ref{4u1907:tab:swift_xrt_log}.
The results of fitting the average spectrum of the source in each XRT observation with an absorbed power law
and the corresponding 0.3--10\,keV flux are reported in Table~\ref{4u1907:tab:swift_xrt_log}.
We calculated the orbital phase corresponding to each XRT observation by using the source ephemerids
published by \citet[][see Table~\ref{tab:sources}]{zand98} and grouped these observations in eight phase bins yielding a
comparable number of source counts in each bin ($\sim 2700$ counts).
Fig.~\ref{fig:u1907_spec_by_phase}a shows the hardness ratio (4--10\,keV / 0.3--4\,keV) calculated for each of the eight time bins
as a function of the orbital phase.
\begin{figure}
        \includegraphics[width=\columnwidth]{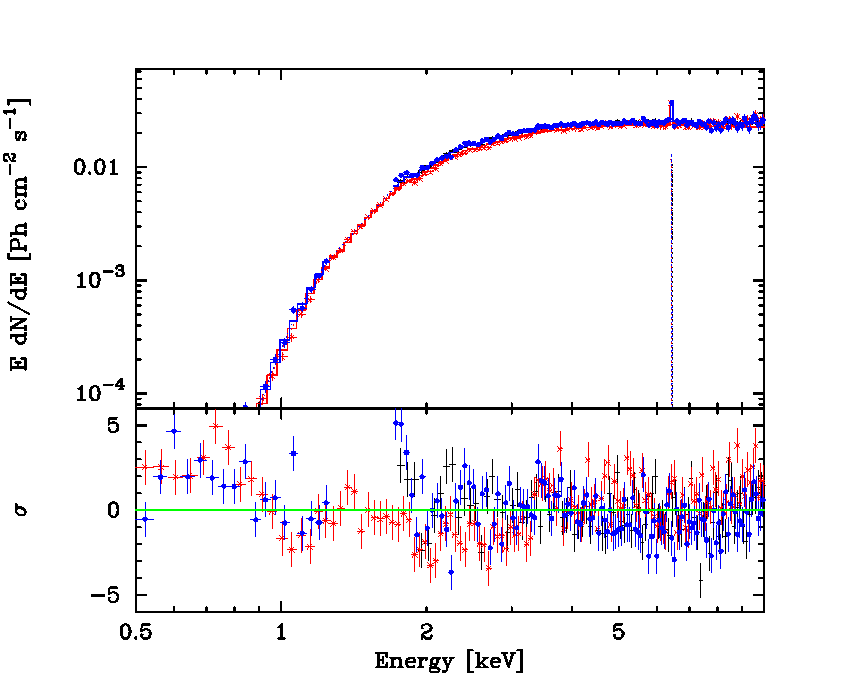}
    \caption{\u4\ unfolded spectra obtained by using the entire exposure time available in the \xmm\ OBSID.~0555410101 for the \pn (black points), \moso (red crosses), and \most (blue circles) cameras. The residuals from the best fit model are reported in the bottom panel.}
    \label{fig:av-spec}
\end{figure}

We also provide the first detailed analysis of an \xmm\ observation that was previously
reported  by \citet{gimenez15} extracting only the average spectrum for a study aimed primarily at the iron line emission.
\xmm\ observed \u4\ close to the epoch of periastron passage from 2009-04-18 at 12:16:25 to 2009-04-18 at 18:08:50 UT (OBSID 0555410101)
for a total exposure time of about 20~ks. The \pn and \moso cameras were operated in timing mode, while the \most in small window.
The observation was not affected
by any flaring background time interval and thus we retained the full observation exposure for the scientific analysis.

We extracted the \pn events in a region encompassing 80\,\% of the source net signal (from RAWX 27 to 45 included) and a background region
with a width of 15 RAWX units (RAWX 7--22). We extracted the \moso source events in a circular region with a radius of 800 pixels,
centered at the source best position, while we chose a background region with a radius of 1200 pixels in an external CCD.
We extracted the \most source events encompassing 80\% of the source net signal (RAWX 290--318 included)
and a background box 2400$\times$9600 pixels big located on
an external CCD, unaffected by the source signal.

We found a satisfactory fit to the source averaged \pn, \moso, and \most spectra using an absorbed power-law model (we adopted the \textsc{TBabs} component
as for \swift/XRT) and a partial covering (\textsc{pcfabs} in {\sc Xspec}).
We also found a clear evidence of a prominent iron line at $\sim$6.4~keV that was modeled in the fit using a Gaussian line with zero width.
Due to the known calibration limitations for the different operating modes of the EPIC cameras, we restricted the fit to the energy range
1.1--10\,keV for the \pn, 0.5--10\,keV for the \moso, and 0.5--1.1\,keV plus 1.8--10\,keV for the \most.
Even though these choices exclude the obvious residuals linked to calibration uncertainties, scattered points remained
visible especially below $\sim$2\,keV (see Fig.~\ref{fig:av-spec}). We added a 2\% systematic error on the spectrum to obtain a fit acceptable at the
4\,$\sigma$ level, as the scattered residuals did not suggest the presence of an additional spectral component ($\chi^2=466$ for 352 degrees of freedom,
hereafter d.o.f.)\footnote{The $\chi^2$ test statistics is appropriate here as in each spectral bin, there are at least 50 counts.}.
The best-fit parameters are reported in Table~\ref{tab:av-par}.
Here and in the following, $N_\mathrm{H}$ is the absorption column density along the direction to the source (including the Galactic absorption),
$N_\mathrm{H, pc}$ is the column density of the partial absorber (representing the absorption column density local to the source),
$f$ is the covering fraction of the partial absorber, $\Gamma$ is the power-law photon index,
$E_\mathrm{Fe}$ and $\mathrm{norm}_\mathrm{Fe}$ are the centroid energy and normalization of the Gaussian line representing the iron emission,
and F$_\mathrm{2-10~keV}$ is the measured 2-10~keV power-law flux in units of $10^{-12}\,\mathrm{erg\,s^{-1}\,cm^{-2}}$.
Based on the current knowledge of the EPIC cameras calibrations,
we consider the 7\% difference in the \moso instrument normalization
within the expected systematic calibration uncertainties of the different modes.
\begin{table}
    \renewcommand*{\arraystretch}{1.2}
    \caption{Best-fit parameters obtained from the \xmm\ data of \u4\ collected during the observation 0555410101.}
    \label{tab:av-par}
    \centering
\begin{tabular}{lr@{}ll}
   \hline
    \hline
    Parameter & \multicolumn{2}{c}{value} & units \\
    \hline
    $E_\mathrm{Fe}$ & 6.412 &$_{-0.005}^{+0.009}$ & keV \\
    $\mathrm{norm}_\mathrm{Fe}$ & 1.73&$\pm$0.10 & $\mathrm{10^{-4}\,Ph\,s^{-1}\,cm^{-2}}$\\
    $N_\mathrm{H}$ & 2.26 &$\pm$0.11 & cm$^{-2}$\\
    $N_\mathrm{H, p.c.}$ & 3.23 &$_{-0.16}^{+0.20}$ &  cm$^{-2}$\\
    $f$ & 0.63 &$\pm$0.04 & \\
    $\Gamma$ & 1.139 &$\pm$0.009 & \\
    F$_\mathrm{2-10\,keV, EPIC-pn}$\tablefootmark{a} & 351.5 &$_{-1.0}^{+1.1}$ & $10^{-12}\,\mathrm{erg\,s^{-1}\,cm^{-2}}$ \\
    F$_\mathrm{2-10\,keV, MOS1}$ & 325 &$\pm$1 & $10^{-12}\,\mathrm{erg\,s^{-1}\,cm^{-2}}$\\
    F$_\mathrm{2-10\,keV, MOS2}$  & 352 &$\pm$1 & $10^{-12}\,\mathrm{erg\,s^{-1}\,cm^{-2}}$\\
    $\chi^2$/d.o.f. &  466&/353 & \\
    \hline
\end{tabular}
\tablefoot{
\tablefoottext{a}{Fluxes are not corrected for absorption.}}
\end{table}

\begin{figure*}
    \begin{minipage}{0.5\textwidth}
    \includegraphics[width=1.0\textwidth]{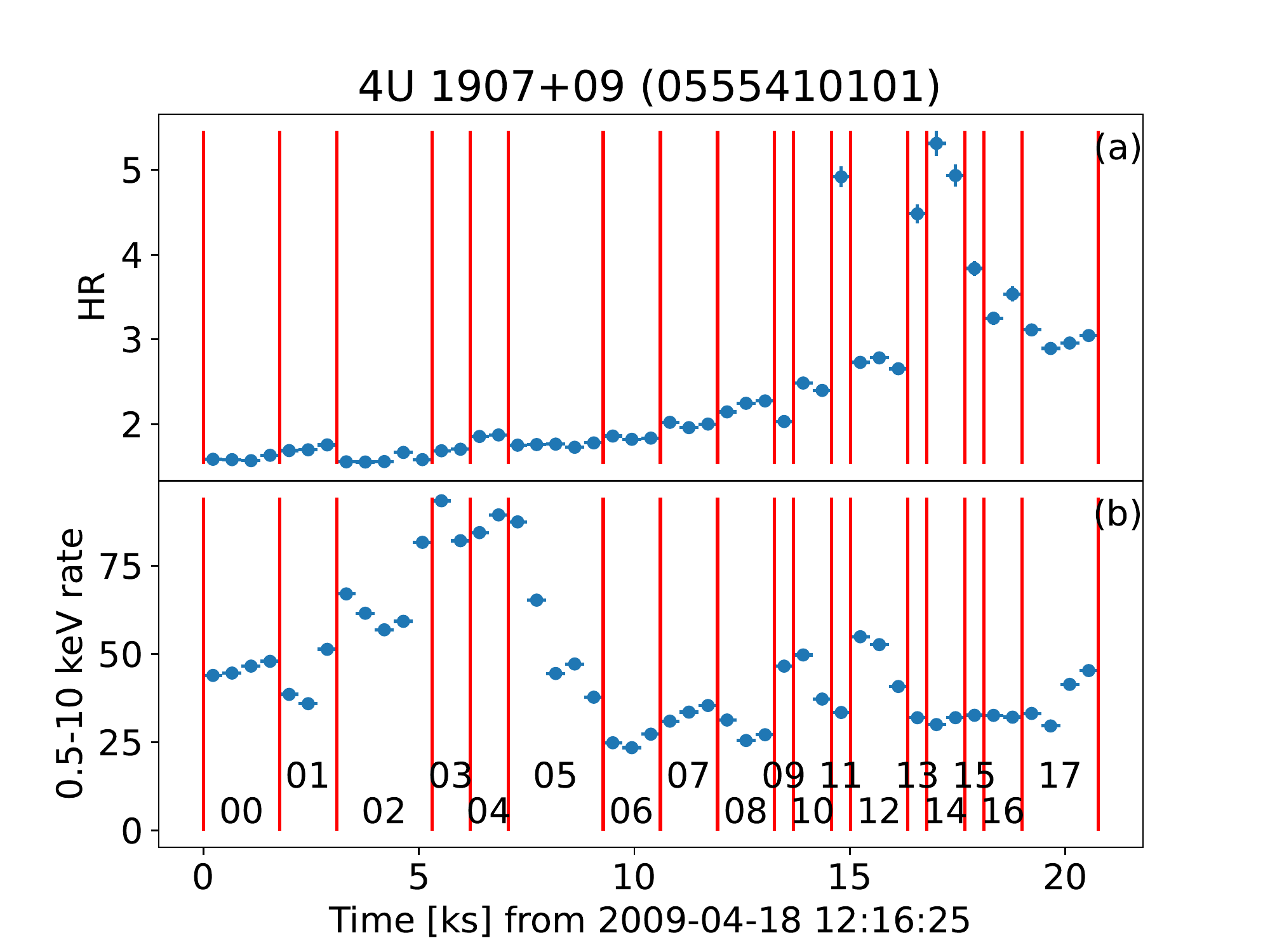}
     \includegraphics[width=1.0\textwidth]{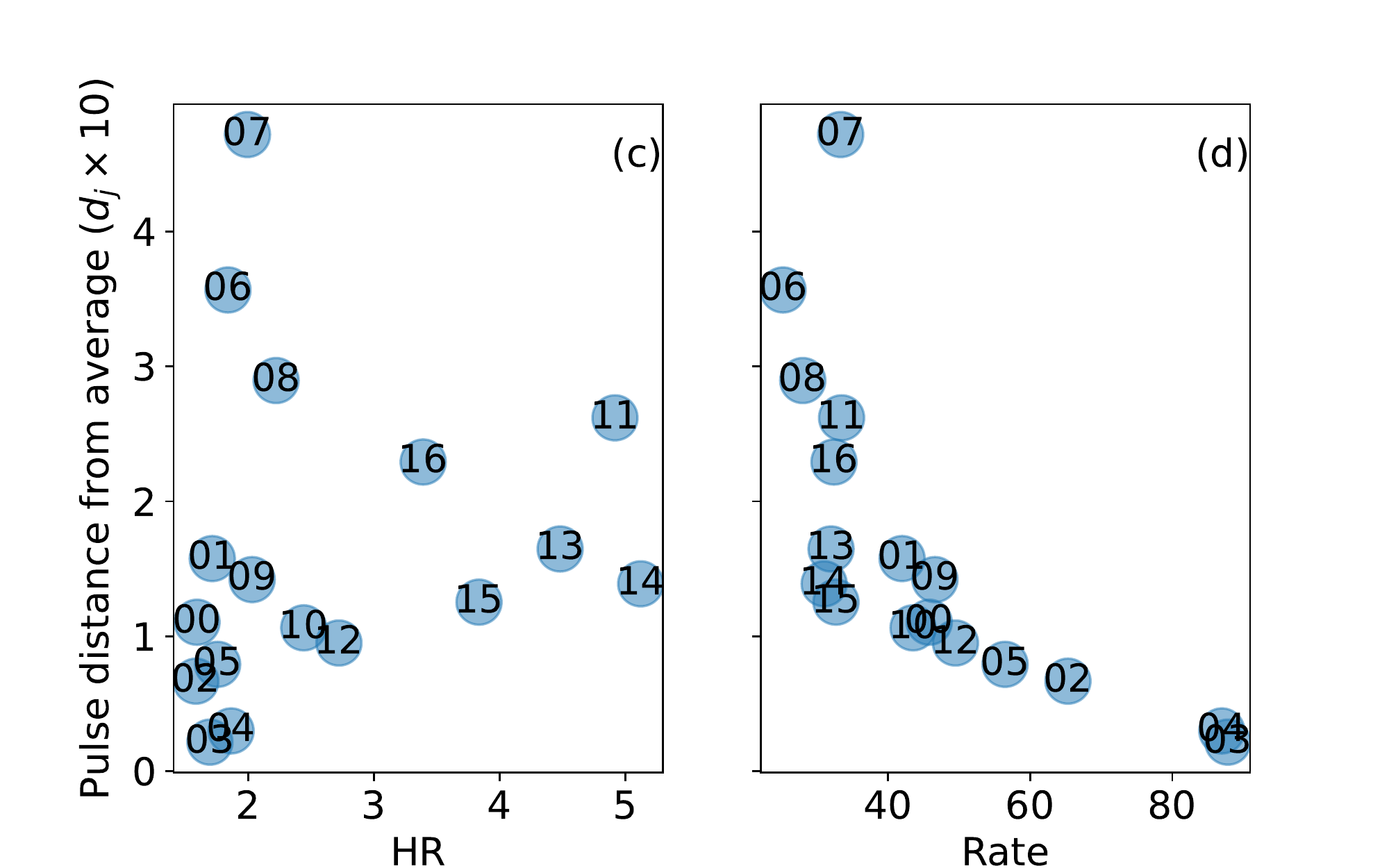}
    \end{minipage}
    \hspace{-0.5cm}
    \begin{minipage}{0.5\textwidth}
    \vspace{1.4cm}
    \includegraphics[width=1.07\textwidth]{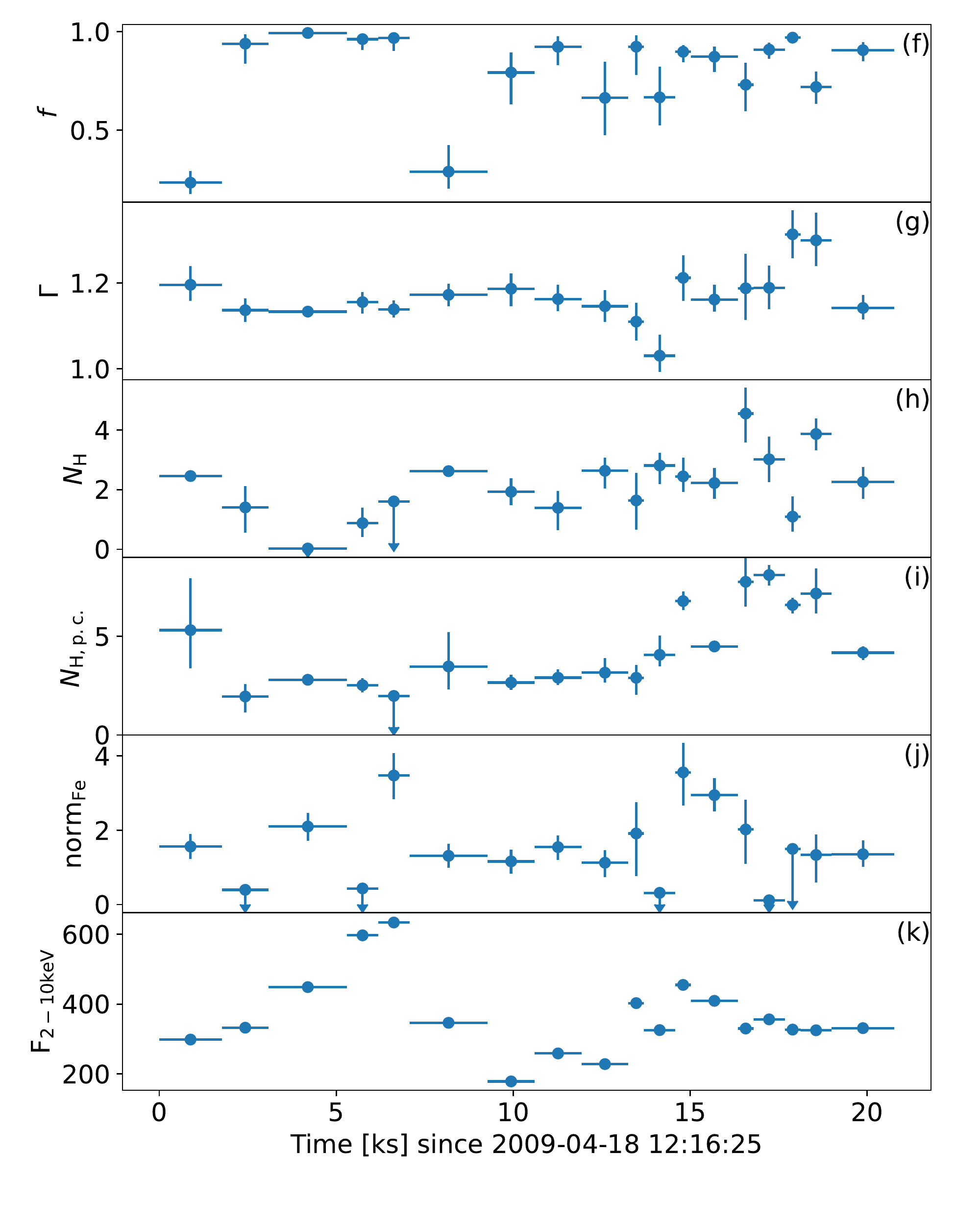}
    \end{minipage}\hfill
    \caption{\textit{Top left, panel (a)}: hardness ratio of the combined \pn, \moso, \most light curves (3--10\,keV/0.5--3\,keV).
    \textit{Top left, panel (b)}: total rate in the 0.5--10\,keV energy range.  \textit{Bottom left, panel (c)}: distance $d_j$ from the
    average for each pulse profile as function of HR. \textit{Bottom left, panel (d)}: distance $d_j$ from the
    average for each pulse profile as function of the total rate. The numbers indicate the
    different intervals identified in panel (b) (see eq.~\ref{eq:distance} for the definition of $d_j$). \textit{Right}: plot of the best-fit parameters as a function of
    time obtained from the HR-resolved spectral analysis of the \xmm\ observation of \u4.\ The parameters reported in the different panels are those introduced in Table~\ref{tab:av-par} and described in the text.}
    \label{fig:4u-av-par}
\end{figure*}

From the cleaned \pn source event file list, we determined the best source spin period using the epoch-folding technique \citep[see, e.g.,][]{dai11} at 2.2631(7)\,mHz and then extracted the background-subtracted energy-resolved lightcurves of the source in the 0.5--3\,keV and 3--10\,keV for all EPIC cameras binned at the above period, such that the variability eventually observed can be ascribed to the accretion environment and not the energy dependence of the source pulse period.
The hardness ratio obtained after combining \pn, \moso, and \most data  is shown in panel (a) of Fig.~\ref{fig:4u-av-par}, while the corresponding count-rate is given in panel (b). The
source displays a remarkable variability, with the largest changes in the HR visible toward the end of the observation, when the source enters a lower X-ray emission state.

We highlight as red vertical lines in panels (a) and (b) of  Fig.~\ref{fig:4u-av-par} the time intervals with
significant variations of the HR in which we extracted spectra to investigate the possible origin of this variability.
We show the best-fit spectral parameters as function of time in Fig.~\ref{fig:4u-av-par}. The centroid energy of the iron line is not shown because it remained stable at the value measured from the average spectrum to within the associated uncertainties.
For each time interval, the \pn, \moso, and \most spectra were extracted
and fit together with the same model used for the averaged spectrum (see earlier in this section).
We note that fixing for all spectra the value of the {\sc Tbabs} absorption column did not result in acceptable fits,
so we left this parameter free to vary also for the HR-resolved spectral analysis.
It is evident from this figure that the time intervals during which the higher HR has been recorded are characterized
by an overall enhanced local absorption column density ($N_{\rm H, p.c.}$),
while the covering fraction remains relatively stable around the average value and the powerlaw photon index displays only a modest softening compared
to the initial part of the observation. Interestingly, the lowest values of the absorption column density are measured
close to the peaks of the brightest emission episodes. We verified that the hardness is driven mainly by the variation of the
column density in the partial-covering component by checking the linear correlation between $HR$ and $N_\mathrm{H, pc}$ that is significant at
99\% confidence level using the $r^2$ statistics on a sample of 1000 bootstrapped data-sets, distributed according to the actual measurements.
Other parameters are not significantly correlated to the $HR$. We investigated also the correlations between spectral parameters, but could not find any significant linear trend.
\begin{figure}
    \includegraphics[width=\columnwidth]{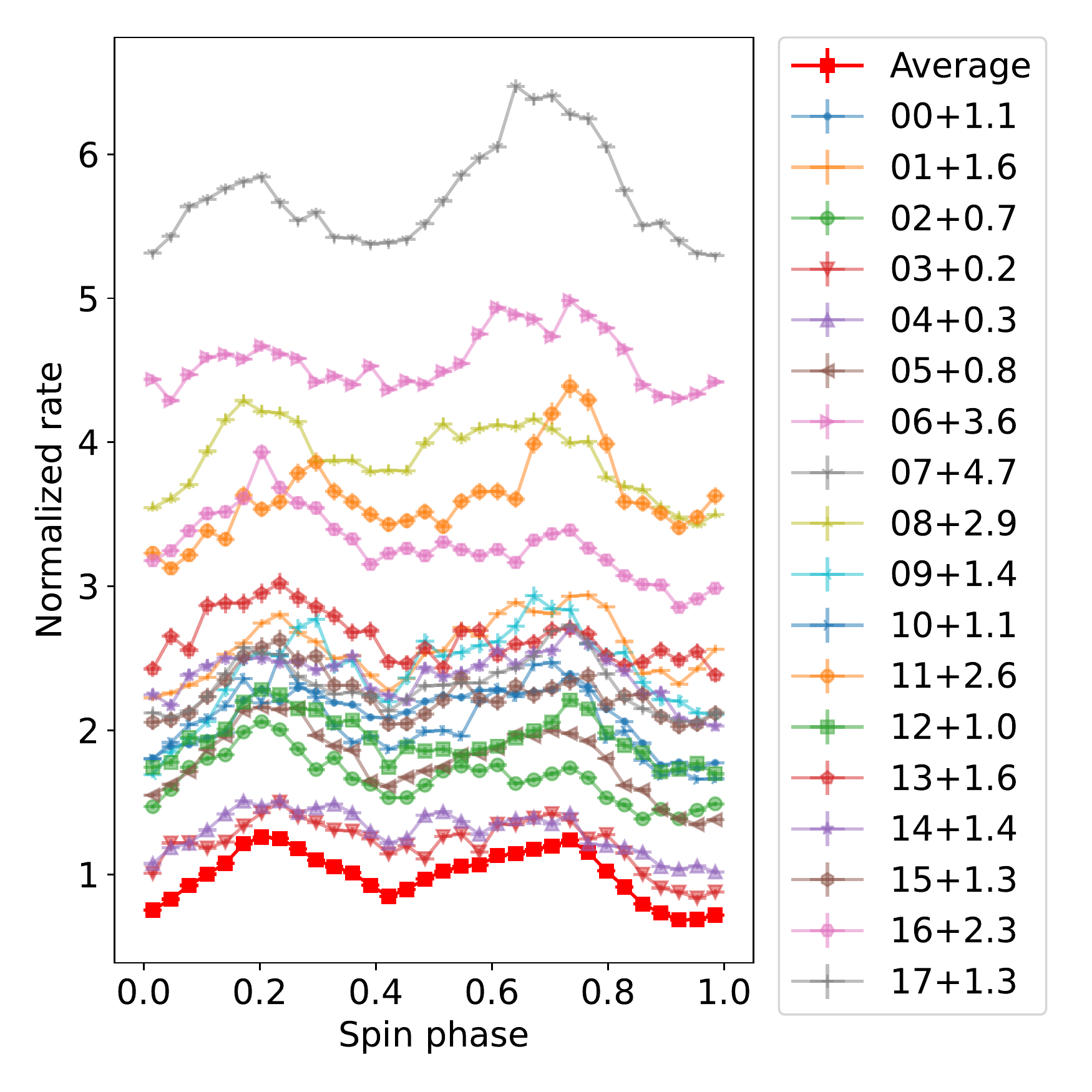}
    \caption{Average and HR-resolved pulse profiles of \u4\ divided by their mean and vertically displaced by ten times their distance from the average,
    as computed in eq.~(\ref{eq:distance}). Labels represent the order of pulse profile as labelled in panel (a) of Fig.~\ref{fig:4u-av-par}. }
    \label{fig:4u-pulses}
\end{figure}

For each time interval of different hardness ratio in panels (a) and (b) of  Fig.~\ref{fig:4u-av-par},
we folded the total light curve at the NS spin period to extract the pulse profiles.
In Fig.~\ref{fig:4u-pulses}, we show the pulse profiles after dividing them for their average and displacing them vertically by ten times their distance from the average profile, computed as
\begin{equation}
\label{eq:distance}
d_j = \sum_{i=1}^{32} \frac{\left| p_{i,j} - \bar p_i\right|}{\sigma_{i,j}}\,,
\end{equation}
\noindent
where $j$ represent the pulse-profile number, $p_{i,j}$ is the $i$-th bin of the $j$-th pulse profile, $\sigma_{i,j}$ the uncertainty,
and $\bar p_i$ is the average $i$-th bin of the average pulse profile. The label of each line represents the number of interval $j$ from zero to 17, as
indicated in panel (b) of Fig.~\ref{fig:4u-av-par}. In this figure, we also show for each HR interval the pulse variability as function of the HR (panel c) and of the total rate (panel d).

The time intervals in which the source is brighter are characterized by pulse profiles that are more similar to the average profile.
However, we note a pronounced variation in the shape of the pulse profile, compared to the average shape, during the lower luminosity HR intervals.
The most peculiarly shaped pulse profiles are recorded during the HR intervals 6--8, immediately following the brightest source emission phase toward the middle of the \xmm\ observation.
Intervals 13--15 are instead characterized by the highest HR values (and average fluxes) but the corresponding pulse profiles are relatively similar to the average one.
Interval 11 has one of the highest recorded HR values, a relatively low flux, and is characterized by a shape of the pulse profile
with intermediate properties between the average profile and those of the most peculiar intervals 6--8.
\begin{table*}
    \caption{Results of the orbital phase-resolved spectral analysis conducted on the \swift/XRT observations of \IGR19.}
   \label{19140:tab:swift_xrt_spe}
    \centering
    \tiny
    \renewcommand{\arraystretch}{1.3}
\begin{tabular}{l llllllll}
    \hline
    \hline
   \noalign{\smallskip}
    Parameter & \multicolumn{8}{c}{Orbital phase}  \\
   \noalign{\smallskip}
 \hline
 \noalign{\smallskip}
                                              &0--0.205 &0.205--0.270 &0.270--0.375 &0.375--0.500 & 0.500--0.625&0.625--0.750 &0.750--0.835 &0.835--1.000 \\
   \noalign{\smallskip}
 \hline
 \noalign{\smallskip}
   $n_\mathrm{H}$                   & $0.8^{+0.1}_{-0.1}$ & $1.6^{+0.2}_{-0.2}$  & $2.7^{+0.3}_{-0.3}$    & $4.3^{+0.5}_{-0.4}$ & $1.6^{+0.2}_{-0.2}$ &$1.0^{+0.1}_{-0.1}$ & $1.2^{+0.2}_{-0.1}$ & $1.1^{+0.2}_{-0.1}$      \\
    $\Gamma$                       & $0.8^{+0.2}_{-0.2}$ & $1.0^{+0.2}_{-0.2}$  & $1.2^{+0.2}_{-0.2}$   & $1.3^{+0.3}_{-0.3}$ & $1.1^{+0.2}_{-0.2}$ &$1.0^{+0.2}_{-0.2}$  & $1.1^{+0.2}_{-0.2}$ & $1.1^{+0.2}_{-0.2}$  \\
   F$_\mathrm{0.3-10\,keV}$    & $1.8^{+0.1}_{-0.1}$ & $10.3^{+0.5}_{-0.4}$ & $10.4^{+0.5}_{-0.5}$ & $4.9^{+0.3}_{-0.2}$ & $4.7^{+0.2}_{-0.2}$ & $4.1^{+0.2}_{-0.2}$ & $2.6^{+0.2}_{-0.1}$ & $3.0^{+0.2}_{-0.2}$ \\
    Cstat/d.o.f.                 & 479.7/488           & 440.6/521             & 423.5/511             & 387.6/464           & 417.3/520    & 383.4/478          & 352.7/445           & 405.2/457            \\
   \noalign{\smallskip}
 \hline
 \noalign{\smallskip}
\end{tabular}
\tablefoot{We report the measured value of
    the absorption column density ($N_{\rm H}$, in units of $10^{23}$\,cm$^{-2}$), the power-law photon index ($\Gamma$), and the
    0.3--10\,keV flux (not corrected for absorption, in units of $10^{-11}$ erg\,cm$^{-2}$\,s$^{-1}$).}
\end{table*}

\subsubsection{Discussion of the results}

The \swift/XRT orbital monitoring of the source provided some interesting results. As shown in Fig.~\ref{fig:u1907_spec_by_phase}, the flux is remarkably peaked around phase 0.5--0.8,
remaining virtually constant during the rest of the orbit.
The source also displays a strongly variable absorption column density and photon index along the orbital phase.
In particular, the evident spectral hardening around phase 0.8--1.0 is driven by both an increase by a factor of $\sim$2
in the local absorption column density and a flattening of the spectral index.
As the phase 0 in our calculation was assumed to be the same as that of \citet{zand98},
this point correspond to the epoch of the mean longitude 90$^{\circ}$, that is when the NS is beyond the companion right along the line of sight to the observer.
This occurs slightly after the periastron
passage that corresponds to phase $\sim$0.7.
Compatible results were obtained in the past by using \asca\ and \rxte\ data \citep{roberts01,sahiner12}, although the monitoring presented here with XRT provides
an extension of the coverage at the softer X-rays (down to 0.3\,keV) and the orbital phase-resolved spectral
analysis is carried out by averaging multiple observations over several orbital cycles rather than
making use of single short pointings (few ks) carried out at specific orbital phases within the same orbital revolution.

The flux increase close to periastron passage is to be expected in a moderately eccentric wind-fed SgXB, especially
if the photoionization of the stellar wind by the accreting NS is relatively low.
This is due to the higher density of a line-driven wind closer to the supergiant star,
as well as the decrease in the relative velocity between the NS and the companion's wind, both effects leading
to an enhanced mass accretion rate onto the compact object \citep[see, e.g.,][and references therein]{bozzo21}.
The interpretation of the spectral hardening around phase 0.8--1.0 is less straightforward, and was largely
debated also in the past \citep[see][and references therein]{roberts01}. The most widely accepted explanation
is the presence of a trailing gas stream connecting the supergiant with the NS, similarly to what was considered
to explain the case of the eccentric HMXB GX\,301$-$2 \citep{leahy91, leahy08, kostka10}.
Although the statistics of the XRT data does not allow us to establish a firm conclusion,
our findings are fully consistent with those reported in the past.
Future observations carried out with grating instruments (as those on-board \xmm\ and \chan) might be exploited
to further corroborate this scenario looking for changing photoionization lines as a function
of the orbital phase as done previously in the case of the SgXB Vela\,X-1 \citep{watanabe06}.

The pointed \xmm\ observation of \u4\ allowed us to study the variability of the source emission over short timescales,
typical of the wind accretion process (see Sect.~\ref{sec:intro}). Once the \xmm\ lightcurve is binned at multiples of the best determined spin period,
the observed variability can be ascribed to changes in the accretion environment surrounding the compact object and we noticed
in Fig.~\ref{fig:4u-av-par} that the highest recorded values of the local absorption column density occur during periods where the source intensity is low.
A similar behavior has been observed in other SgXBs and commonly ascribed to the presence of massive clumps passing occasionally in front of the NS
along the line of sight to the observer without being accreted and thus without producing an enhanced X-ray emission.
The HR-resolved spectral analysis reported in Fig.~\ref{fig:4u-av-par}  supports this conclusion because it shows that $N_{\rm H, p.c.}$
is the main driver of spectral variability during the second half of the observation. From the same figure, we also observe that
there is a decrease of the local absorption column density around the peaks of the flares taking place during the first $\sim$10~ks
of the observation and that there are two sharp decreases on the covering fraction at the beginning of the observation
(before the bright flares go off) and right after the end of the flaring period. Drops of the local absorption column density during the brightest emission periods are usually
interpreted as due to the photoionization of the clumpy stellar wind by the X-rays from the compact object,
while the initial low value of the covering fraction could indicate that there was a progressive fragmentation
of a dense clump before the flaring period began. The recorded drop of the covering fraction after the flare
could be explained in this context as the residual (non-accreted) portion of the fragmented dense clump moving away from the compact object \citep[see, e.g.,][and discussions therein]{bozzo11}.

The average source pulse profile, mediated using the entire exposure time available of the \xmm\ observation shows
a spin-phase variability virtually identical to what has been reported previously \citep[see, e.g.,][]{zand98}.
A study of the variability of the pulse profile with the source intensity, at the best of our knowledge,
was attempted in the past by \citet{roberts01} using four different \asca\ observations \citep[see also][]{in01,fritz06,rivers10}.
These data spanned a total range in the source X-ray luminosity of a factor of $\sim$60 and highlighted some possible change
in the pulse profile shape as a function of the intensity, but the statistics was rather low to perform any meaningful comparison among the different
profiles. As shown in Fig.~\ref{fig:4u-pulses} and \ref{fig:4u-av-par}, the relatively high statistics and uninterrupted exposure of the \xmm\ observations
allowed us to report here for the first time an analysis of the pulse profile resolved in intensity and HR. We found that the shape of the pulse profile
displays a remarkable variability compared to the average profile, with the largest deviations being recorded during the intervals 6, 7, and 8 (see panel
(d) of Fig.~\ref{fig:4u-av-par}). These intervals correspond to a period of relatively faint emission from the source (combined count-rate from the EPIC
cameras of $\lesssim$30\,cts~s$^{-1}$) and immediately precedes the going off of two faint flares from the source during (and after) which the highest HR
values are measured (see panels (a) and (b) of Fig.~\ref{fig:4u-av-par}). As our HR-resolved spectral analysis demonstrated that the HR variations are
driven by the increase of the local column density and since changes in the shape of the pulse profile in HMXBs are known to be associated to switching
between different accretion geometries \citep[see][for an early discussion]{Parmar1989}, we infer that the encounter of a stellar wind clump with a NS can
sometimes alter the way in which the material is accreted and not only the mass accretion rate or the absorption.
However, the two phenomena are not necessarily connected.
Although this is a somewhat standard assumption in theoretical models proposed to interpret wind-fed systems \citep[see, e.g.,][and references
therein]{bozzo08,shakura12}, we are not aware of similar other evidence in SgXBs. This result could only be obtained by combining the HR-resolved spectral
analysis with a HR-resolved study of the source pulse profile, a technique that our group plans now to exploit for additional bright sources within the
SgXB class.

It is worth mentioning here that additional evidence in favor of the scenario proposed above could be obtained by performing a pulse-phase resolved
spectral analysis of the source emission within each identified HR time interval in Fig.~\ref{fig:4u-av-par} (panels (a) and (b)). Unfortunately, the
statistics of the \xmm\ data is not sufficient to carry out such analysis (only one single EPIC spectrum can be extracted for each HR time bin in order to achieve a meaningful spectral fit). This limitation cannot be overcome by using deeper observations with any currently available facility as the required statistics is limited by the X-ray photons collection capability of the existing instruments. This is directly related to the available effective area of the instrument, and under this respect the \xmm\ EPIC cameras are already providing the best performances. Future instrumentation endowed with a much larger area in the soft X-ray domain, as the XIFU and WFI on-board {\it Athena} \citep{xifu,wfi} or the LAD on-board the {\it eXTP} \citep{extp} or the {\it STROBE-X} \citep{strobex} missions could provide the necessary advancements to unveil the nature of the pulse profile changes in \u4\ and other wind-fed systems.

\subsection{ \IGR19}
\label{sec:19140}

\IGR19\ was discovered by \inte\ in 2003 \citep{hanni04} and later classified as a classical SgXB thanks to the identification of the optical counterpart as a B0.5 Ia star \citep{torrejon10} and the detection of an orbital period of 13.5~d \citep[see, e.g.,][and references therein]{wen06}. The distance to the B0.5 Ia star has been recently determined using {\it Gaia} data from the originally estimated 3.6~kpc \citep{torrejon10} to 2.8$_{-2.3}^{+1.3}$ \citep{arnason21}. \citet{sidoli16} reported about the discovery of a possible pulsation period from a \chan\ observation of the source at $\sim$5937~s and a quasi periodic oscillation (QPO) from an \xmm\ observation of the source at a frequency of $\sim$1.46~mHz. The QPO has been interpreted as due to the onset of a quasi-spherical accretion regime.

\subsubsection{Data analysis and results}

Our \swift\,/XRT  monitoring campaign on \IGR19\ was carried out in 2015
with a pace of two observations per week, each 1 ks long, spanning from February to September 2015 (ObsID 30393).
These observations (see Table~\ref{i19140:tab:swift_xrt_log}), summing up to a total exposure of $\sim$60~ks,
cover a bit less than 17 revolutions of the system.
The log of the XRT observations, Table~\ref{i19140:tab:swift_xrt_log}, also shows for each XRT observation the orbital phase
according to the ephemerides reported in Table~\ref{tab:sources}.
We note that as the previous ephemerides of the source were published by \citet{corbet04atel} using \rxte/ASM data
up to 2004 and since then additional 8 years of ASM data plus about 15 years of \swift/BAT monitoring data on the source
have been made available, so we could refine the determination of the orbital period to the value reported in
Table~\ref{tab:sources} (all analysis details are provided in Appendix~\ref{app:period}).
We also note that in the case of \IGR19\ the spin period is not know and the tentatively reported one by \citet{sidoli16} is far too
long to allow any meaningful search in the short XRT observations.

\begin{figure}
\vspace{-1.4truecm}
\hspace{-0.2truecm}
   \includegraphics[width=1.15\columnwidth]{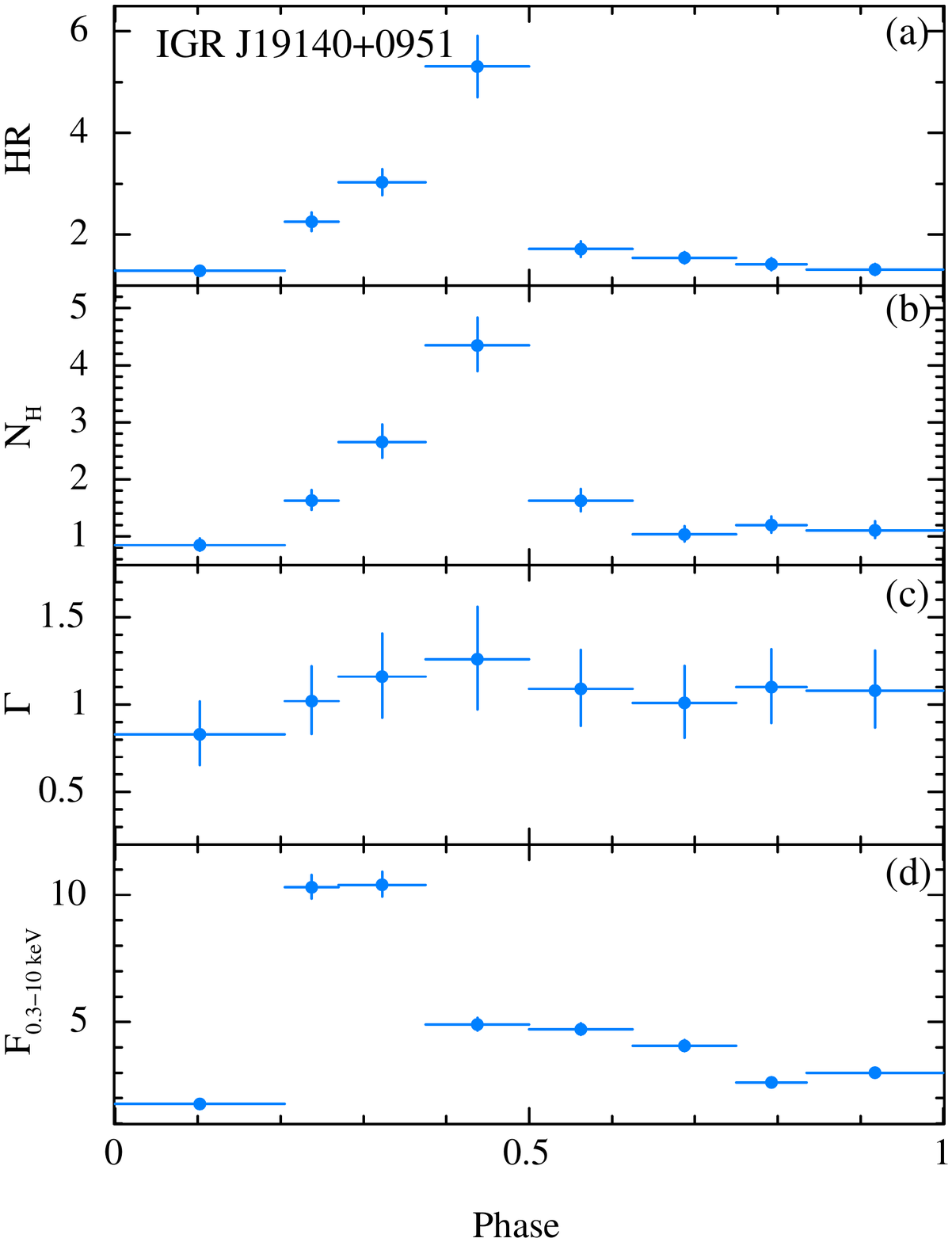}
   \caption{Same as Fig.~\ref{fig:u1907_spec_by_phase} but realized for \IGR19.\ Here the flux is given in units of $10^{-11}$ erg\,cm$^{-2}$\,s$^{-1}$.}
   \label{fig:19140_spec_by_phase}
\end{figure}

The plot of the source HR as a function of the orbital phase is shown in Fig.~\ref{fig:19140_spec_by_phase}a.
We fit the eight spectra extracted from the 8 phase bins with an absorbed power-law and report the results in Table~\ref{19140:tab:swift_xrt_spe}
and Fig.~\ref{fig:19140_spec_by_phase} (panels b--d).

For completeness, we mention here that the sole \xmm\ observation of the source published by \citet{sidoli16} does not allow any HR-resolved spectral investigation, as the source was relatively faint during the time spanned by the \xmm\ data and no flare was apparent from the source lightcurve \citep[see Fig.~3 in][]{sidoli16}.

\subsubsection{Discussion of the results}

The plot of the spectral parameters and HR as a function of the orbital phase (Fig.~\ref{fig:19140_spec_by_phase}) obtained from the XRT monitoring of the source displays a rather intriguing variability with the peak of the X-ray flux measured at phase 0.2--0.4, which is consistent with the sharply peaked average orbital profile obtained in the 15--50\,keV band using 15 years of continuous monitoring by \swift/BAT (Fig.~\ref{fig:orbit_bat}). We notice a steep increase in the absorption column density immediately following and reaching the maximum at phase 0.4--0.5.  A similar increase of the source local absorption column density at specific orbital phases was already reported in the past by \citet{prat08}. However, their measurements as a function of the orbital phase were affected by large uncertainties mainly due to the limited coverage below 3\,keV. Our results show that the orbital variability of the flux and absorption column density in \IGR19\ is much more extreme than what was observed before and it is remarkably similar in what is observed from the much better studied source \u4.\ In both systems, there is a shift in phase of about $\sim$0.2 between the maximum of the flux and the maximum of the absorption column density, and the profile of the flux variability is also peaked at a restricted interval in the orbital phase ($\lesssim$0.2). In both sources, there is also a relatively modest to no change in the power-law slope. We thus conclude that, most likely, the same scenario that has been proposed for \u4\ is also applicable to the case of \IGR19:\ this source shall be characterized by a non-negligible eccentricity and there should be a large structure located close to the NS and moving with the compact object, possibly a gas stream as indirectly proved in the case of \u4.\ Although we cannot firmly exclude alternative scenarios, the data so far are compatible with this hypothesis and the similarity with the case of \u4\ provides support in the right direction.

A measurement of the eccentricity in the case of \IGR19\ would be best obtained by following the evolution of the source spin period at different orbital phases. However, the possibly long spin period of the source (over 5~ks, to be confirmed, see Table~\ref{tab:sources}) would make such measurements extremely time consuming for any sufficiently sensitive facility capable of performing uninterrupted observations of several tens of ks (e.g., the EPIC cameras on-board \xmm). A validation of our proposed scenario for \IGR19\ through the direct comparison with \u4\ could be obtained by exploiting also deep X-ray observations of the source at specific orbital phases, especially around the peak of the absorption column density where emission/absorption lines can provide a measurement of the ionization status of the absorbing material and its position compared to both the NS and the supergiant companion.

\subsection{\4IGR}
\label{sec:11215}

\4IGR\ is the only SFXT source displaying regular outburst at the periastron passage and having a well measured spin and orbital periods. The systems hosts a $\sim$187~s spinning NS orbiting every $\sim$165~days around the B0.5 Ia companion, located at a distance of 6.5$_{-1.5}^{+1.1}$~kpc \citep[see][and references therein]{sidoli20, arnason21}. The source has been observed many times with virtually all available X-ray facilities and the regularity of its luminosity variations allowed the scheduling of targeted X-ray observations right at the peak of the brightest emission periods. These observing campaigns were aimed at obtaining high S/N data and look for CRSFs, but so far no firm detection was reported \citep{sidoli17, sidoli20}. The role of clumps in the accretion process undergoing in this system is of wide interest because the supergiant star in \4IGR\ showed evidence of a magnetized stellar wind and, when this magnetic field is carried by the clumps, it can lead to reconnections and subsequent bright episodes of X-ray emission. Such mechanism is one of the proposed scenarios to interpret the peculiar behavior of the SFXTs in X-rays \citep{hubrig18}.

\subsubsection{Data analysis and results}

We report on a new \xmm\ observation of the source performed from 2021-01-25 at 23:32:04 to 2021-01-26 at 05:01:25 (OBSID 0862410301).
Out of the 19.8\,ks of \pn exposure, we retained 15.6\,ks after removing periods of background flaring.
For the \pn, we use a source extraction region with radius on 0.53 arcmin, based on the radius at which the
surface brillance of the source equals the surrounding background and an adjacent background region with
radius of 1 arcmin. For \moso, we use a source extraction region fixed to lie within the small windows of 0.47 arcmin and
a background region with radius 1 arcmin in an external CCD. For \most, we used RAWX from 282 and 322 for the source, which encompasses 60\% of the PSF, to
avoid the contribution from a field source, and a background region with the shape of a box 8$\times$2 arcmin.

\begin{figure}
    \includegraphics[width=\columnwidth]{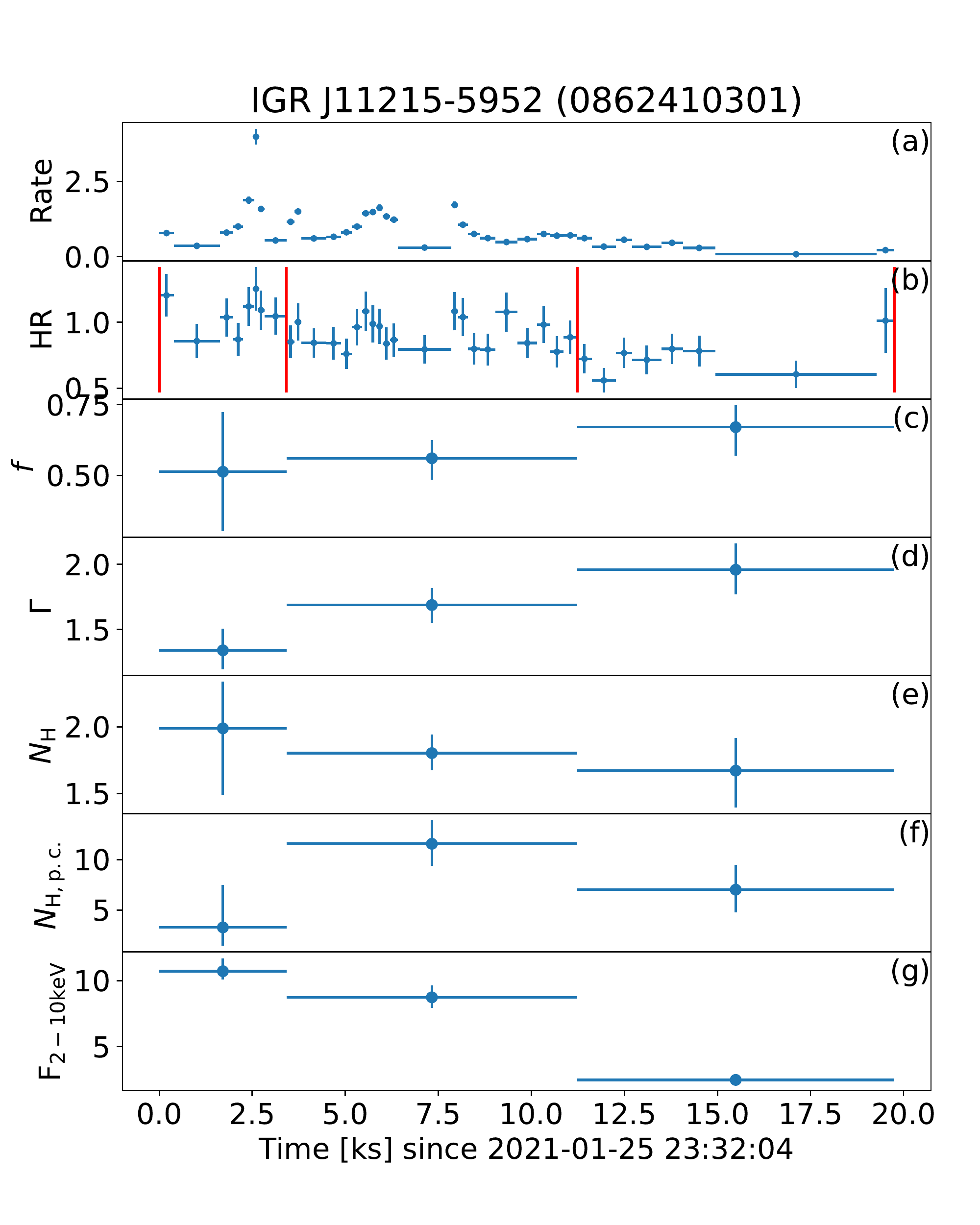}
     \caption{Plot of light curve, HR, and best-fit parameters as a function of time obtained from the HR-resolved spectral analysis of the \xmm\ observation of \4IGR.\
\textit{Panel (a)}: the source count rate in the 0.5--10\,keV bands, after adaptive rebinning with minimum S/N of 15 in the light curve extracted for 0.5--3\,keV;
\textit{Panel (b)}: HR between the 3--10\,kEv and 0.5--3\,keV energy-resolved lightcurves with the identified time intervals for the spectral extraction (marked with red lines);
\textit{Panel (c)}: $f$, the covering fraction of the partial absorber;
\textit{panel (d)}: $\Gamma$, the power-law photon index;
\textit{panel (e)}: $N_\mathrm{H}$, the absorption column density along the direction to the source (including the Galactic absorption);
\textit{panel (f)}: $N_\mathrm{H, pc}$, the absorption column density of the partial absorber (representing the absorption column density local to the source; {\sc pcfabs} in {\sc Xspec});
\textit{panel (g)}: F$_\mathrm{2-10~keV}$, the measured power-law flux in the 2--10~keV energy range not corrected for absorption for the \pn
in units of $10^{-12}\,\mathrm{erg\,s^{-1}\,cm^{-2}}$.    \label{fig:4igr_hr}
    }
\end{figure}

We extracted light curves in the 0.5--3.5 and 3.5--10\,keV bands to compute the HR.
We caught the source in an episode of decreasing flux characterized by a softening spectrum as evidenced in Fig.~\ref{fig:4igr_hr}.
We managed to describe both the average and HR-resolved spectra using a power-law modified by full neutral absorption
(\textsc{TBabs}) and a partial covering component (\textsc{pcfabs}). The best C-stat is 238 as compared to 295 obtained using
an absorbed power-law. To test the significance of the improvement, we simulated, for each best-fit model, 100 spectra with equivalent
exposure and background using parameters extracted from the chain used to compute uncertainties. For each simulated spectrum,
we performed a fit with the same model. For the absorbed power-law, 20\% of the simulated spectra have
higher C-stat than the real spectrum, while adding a partial covering component this fraction raises to 93\%, indicating a significant
improvement.
As it can be appreciated in Fig.~\ref{fig:4igr_hr}, the parameter showing the strongest variability is the power-law photon index, suggesting
that the softening is mostly linked to intrinsic change in the emission rather than to intervening absorption. We investigated the
parameter linear correlations, but the small number of available points prevents any significant detection.

\subsubsection{Discussion of the results}

During the new \xmm\ observation, \4IGR\ displayed a progressively fading lightcurve with a single weak flare occurring at $t\sim2$~ks and lasting only a few hundreds of seconds (see Fig.~\ref{fig:4igr_hr}). The flare was far too faint to detect any significant HR variation.  There is an intriguing ``dip'' in the X-ray emission visible 7.5~ks after the beginning of the observation, but the count-rate was too low to detect any HR variability within the dip itself. As summarized in Fig.~\ref{fig:4igr_hr}, our Bayesian analysis technique identified three different time intervals with a decreasing trend in the overall HR. This trend seems to be mostly driven by a softening of the power-law photon index, although the error bars associated to the best fit parameters in all intervals are relatively large due to the limited statistics of the data. The significant detection of a partial covering component in the spectrum of the source is consistent with the clumpy wind accretion scenario (see Sect.~\ref{sec:intro}), although confirmation through the detection of spectral variability along the rise and decay of brighter flares/outbursts would help us strengthen this  conclusion.

\subsection{ \3IGR}
\label{sec:18410}

\3IGR\ was discovered by \asca\ in 1994 \citep[AX\,J1841.0-0536; see][]{bamba01} and associated to the class of the SFXT thanks to the discovery of repeated bright sporadic outbursts by \inte\ \citep{rodriguez04,sguera06, walter2007b}. The optical counterpart was identified as a B1 Ib supergiant at roughly 3.2~kpc \citep{nespoli08}. Although the system is thought to host a NS, neither the compact-object pulse period nor the binary orbital period has been firmly measured. During a bright flare caught by \xmm\ in 2011, the source displayed one of the clearest observational evidence to date in favor of clumpy winds playing a major role in the X-ray variability of SgXBs \citep{bozzo11}. In 2019, an \xmm\ observation caught the source in the faintest observed state
with a upper limit of $2\times10^{-14}\,\mathrm{erg\,s^{-1}\,cm^{-2}}$ at 90\% confidence level in the 1--10\,keV energy band \citep{ferrigno20}.

\subsubsection{Data analysis and results}

\begin{figure}
    \includegraphics[width=1.1\columnwidth]{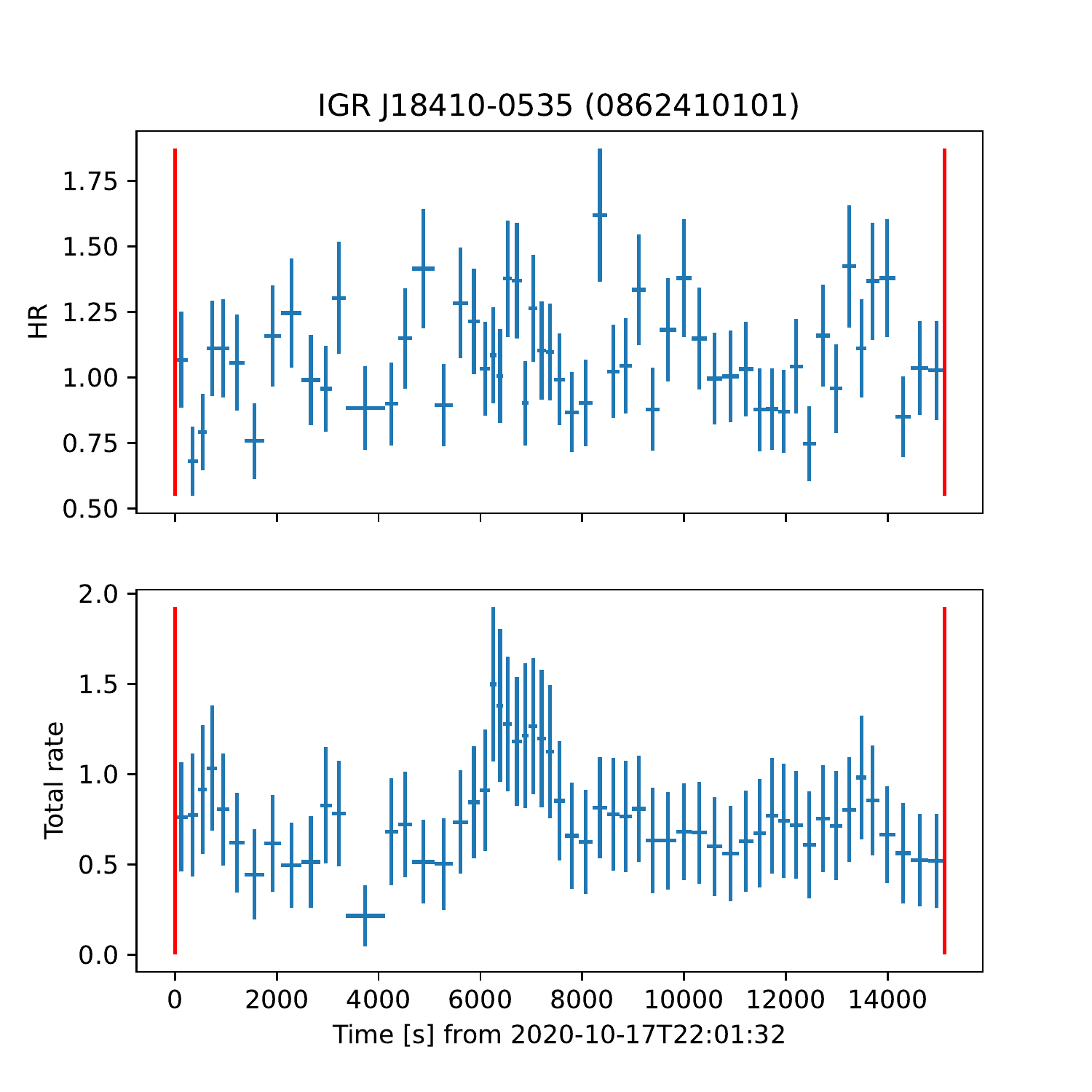}
    \caption{\textit{Lower panel}: combined \pn+\moso+\most light curve of the 2020 observation of \3IGR\ (OBSID 0862410101) in the 0.5--10\,keV energy range. \textit{Upper panel}: hardness ratio between the light curves in the bands 3.5--10\,keV and 0.5--3.5\,keV rebinned at a minimum S/N of 8. There are no significant variations of the hardness ratio and thus the Bayesian algorithm identified a single time interval for the spectral extraction (marked with red lines and corresponding to the whole observation).}
    \label{fig:0862410101}
\end{figure}

Here, we report on a snapshot that \xmm\ performed on \3IGR\ from 2020-10-17 at 20:00:11
to 2020-10-18 at 02:23:31 UT (OBSID 0862410101). The \pn was operated in full frame, the \moso in small window, and the \most in timing mode.
The observation was marginally affected by flaring background, so we retained a good time of 14.6\,ks out of an elapsed time of 15\,ks. The source had an average count rate of 0.4\,cts/s in the \pn 0.5--10\,keV band
and did not show any appreciable variation of hardness ratio (Fig.~\ref{fig:0862410101}). We modelled the average spectrum extracted from the three cameras with an absorbed power-law with best fit parameters reported in Table~\ref{tab:3IGR}.

\begin{table}
    \caption{Best-fit parameters of the \xmm\ observation of \3IGR\ carried out in 2020 (OBSID 0862410101).}
    \label{tab:3IGR}
    \centering
    \tiny
    \begin{tabular}{lr@{}ll}
        \hline
        \hline
        Parameter & \multicolumn{2}{c}{Value} & unit\\
        \hline
        $N_\mathrm{H}$ & 4.06 &$_{-0.13}^{+0.17}$ & 10$^{22}$~cm$^{-2}$ \\
        $\Gamma$ & 1.62 &$_{-0.04}^{+0.05}$ &  \\
        Flux (2-10 keV) & 6.17 &$\pm$0.14 & $10^{-12} \, \mathrm{erg\,s^{-1}\,cm^{-2}}$\\
        \hline
    \end{tabular}
\end{table}

\subsubsection{Discussion of the results}

Despite the fact that \3IGR\ is one of the best known SFXTs to have shown the clearest evidence of clumpy wind accretion due to a bright outburst in 2011 \citep{bozzo11}, our targeted \xmm\ observations to the source were not able to catch additional bright events. Following the deep upper limit on the source flux that we obtained with an observation in 2019, we could only detect during the additional pointing in 2020 a moderately faint flare in the EPIC lightcurves (see Fig.~\ref{fig:0862410101}). During the rise and decay of this flare, as well as during the remaining part of the \xmm\ observation, we could not record any significant variation of the HR that could have indicated a change in the spectral parameters and thus provide evidence in favor of clumpy wind accretion during the low emission state of the source. Given the peculiarity of the 2011 bright outburst, the source clearly deserves further attention and more flares at intermediate luminosity between outburst and quiescence are needed to complete our investigation of the clumpy wind accretion in this source.

\subsection{ \J393}
\label{sec:393}

\J393\ is a classical SgXB discovered by \inte\ in 2004 \citep{bird04} and associated with the previously known X-ray source AX\,J1639.0-4642 \citep{sugizaki01}. The system hosts a $\sim$910~s spinning NS \citep{bodaghee06} orbiting around a still poorly known OB companion, and the measured orbital period is of $\sim$4.24~d \citep[see][and references therein]{corbet13}. The usually heavy extinction measured from X-ray observations in the direction of the source led to the inclusion of this system within the so-called highly obscured X-ray pulsars, a class of objects that has been largely unveiled thanks to \inte\ observations \citep[see, e.g.,][and references therein]{walter2015}. \J393\ has also shown evidence for a super-orbital modulation, although no firm conclusion has been reached yet \citep{corbet21}. The discovery of a CRSF at $\sim$29.3~keV led to the determination of the NS magnetic field strength at 2.5$\times$10$^{12}$~G \citep{bodaghee16}.

\subsubsection{Data analysis and results}

Our monitoring campaign of \J393\ was carried out in 2016 with \swift\,/XRT from January to June 2016,
with a pace of one 1\,ks observation per day (ObsID 34135, for a total of 130\,ks)
thus covering slightly more than 31 revolutions of the system.
A selection of these data ($\sim70$\,ks) have been previously reported by \citet{kab20}.
These authors focused on the properties of the suspected source X-ray eclipse and suggested that
 this might not be a true eclipse, although it regularly occur at the same orbital phase.
Here, we reanalyze the whole data set to carry out an orbital-phase dependent HR-resolved spectral analysis
with the main goal of identifying spectral changes that could point toward the existence of large scale
structures in the stellar wind around the compact object.

\begin{figure}
\vspace{-1.4truecm}
\hspace{-0.2truecm}
   \includegraphics[width=1.15\columnwidth]{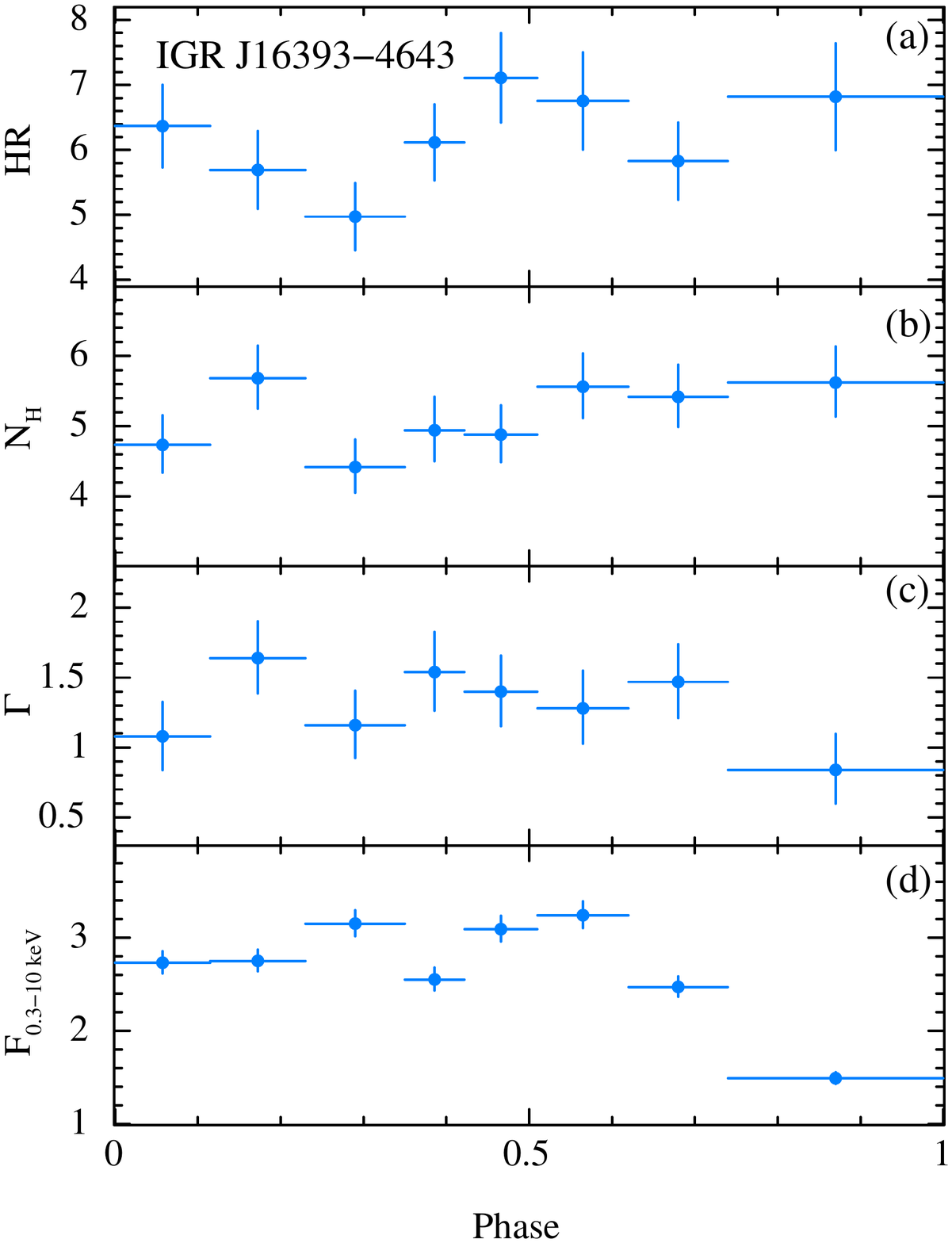}
   \caption{Same as Fig.~\ref{fig:u1907_spec_by_phase} but realized for \J393.\ Here the flux is given in units of $10^{-11}$ erg\,cm$^{-2}$\,s$^{-1}$.}
   \label{fig:i393_spec_by_phase}
\end{figure}

The \swift\ observing logs for \J393\ are reported in Table~\ref{i16393:tab:swift_xrt_log},  where we also
indicated the orbital phase estimated for each observation
by adopting the same ephemerides as in \citet[][see Table~\ref{tab:sources}]{kab20}.
The equivalent of Fig.~\ref{fig:u1907_spec_by_phase} but realized in the case of \J393\ is shown in Fig.~\ref{fig:i393_spec_by_phase}a.
In order to investigate possible spectral variability in different orbital phase bins,
we extracted different source spectra for each of these bins and fit them by using a simple absorbed power-law model .
The results are shown in Table~\ref{i393:tab:swift_xrt_spe} and Fig.~\ref{fig:i393_spec_by_phase}  (panels b--d).
\begin{table*}
    \caption{Results of the orbital phase-resolved spectral analysis conducted on the \swift/XRT data of \J393.}
    \label{i393:tab:swift_xrt_spe}
    \centering
    \tiny
    \renewcommand{\arraystretch}{1.3}
\begin{tabular}{l llllllll}
    \hline
    \hline
   \noalign{\smallskip}
    Parameter & \multicolumn{8}{c}{Orbital phase}  \\
   \noalign{\smallskip}
 \hline
 \noalign{\smallskip}
                                              & 0.0--0.115         & 0.115--0.23        & 0.23--0.35        & 0.35--0.422      & 0.422--0.51        & 0.51---0.62      & 0.62--0.74         & 0.74--1.0             \\
   \noalign{\smallskip}
 \hline
 \noalign{\smallskip}
   $n_\mathrm{H}$                   &$4.7^{+0.4}_{-0.4}$  &$5.7^{+0.5}_{-0.4}$ &$4.4^{+0.4}_{-0.4}$ &$4.9^{+0.5}_{-0.4}$ &$4.9^{+0.4}_{-0.4}$  &$5.6^{+0.5}_{-0.4}$ &$5.4^{+0.5}_{-0.4}$ &$5.6^{+0.5}_{-0.5}$   \\
    $\Gamma$                         &$1.1^{+0.2}_{-0.2}$  &$1.6^{+0.3}_{-0.3}$ &$1.2^{+0.2}_{-0.2}$ &$1.5^{+0.3}_{-0.3}$ &$1.4^{+0.3}_{-0.2}$  &$1.3^{+0.3}_{-0.3}$ &$1.5^{+0.3}_{-0.3}$ &$0.8^{+0.3}_{-0.2}$   \\
   F$_\mathrm{0.3-10 keV}$    &$2.7^{+0.1}_{-0.1}$  &$2.7^{+0.1}_{-0.1}$ &$3.2^{+0.1}_{-0.1}$ &$2.6^{+0.1}_{-0.1}$ &$3.1^{+0.1}_{-0.1}$  &$3.2^{+0.2}_{-0.1}$ &$2.5^{+0.1}_{-0.1}$ &$1.5^{+0.1}_{-0.1}$   \\
    Cstat/d.o.f.                  & 452.9/515           &521.0/506           &492.3/515           &434.7/477           &409.3/495            &469.6/507           &415.3/526           & 442.1/546            \\
   \noalign{\smallskip}
 \hline
 \noalign{\smallskip}
\end{tabular}
\tablefoot{We report the measured value of
    the absorption column density ($N_{\rm H}$, in units of $10^{23}$\,cm$^{-2}$), the power-law photon index ($\Gamma$), and the
    0.3--10\,keV flux (not corrected for absorption) in units of $10^{-11}$ erg\,cm$^{-2}$\,s$^{-1}$.}
\end{table*}

We also checked in the \xmm\ archive suitable observations of \J393\ to carry out a HR-resolved spectral analysis aimed at discovering possible spectral variability. The source was observed twice by \xmm\ in 2004 and in 2010. However, both observations are not suitable to our goal. The observation in 2004 is relatively short ($\sim$8~ks) and it shows 8 peaks and valleys corresponding to the source pulse period \citep{bodaghee06}. There is no flaring behavior that could be analyzed to look for HR variations. During the slightly longer observation carried out in 2010 ($\sim$10~ks), the source was a factor of $\gtrsim$10 fainter and the lightcurve does not show any flaring behavior that is bright enough to carry out a meaningful HR-resolved spectral analysis \citep[see also][]{pragati18}.

\subsubsection{Discussion of the results}

Our detailed study of the source orbital phase-resolved HR and spectral properties extend and complete
the previous work from \citet{kab20}. We analyzed all XRT data of our monitoring campaign with a uniform technique exploited
for all the SgXBs in this paper (to facilitate a direct comparison) and computed the HR value of the folded XRT data over the
source orbital period in 8 phase bins containing virtually the same number of photons. The spectra extracted in each of these
bins could be well fit with a simple absorbed powerlaw model, and the plot of these parameters (as well as the HR) as a
function of the orbital phase in Fig.~\ref{fig:i393_spec_by_phase} does not show any prominent variability. There is a
potentially interesting V-shaped feature in the profile of the source flux between phases 0.3--0.5, but such feature does not
seems to be connected with either a significant change of the powerlaw slope or a variation of the local column density. The
source is heavily obscured along the entire orbit ($N_{\rm H}\gg$10$^{23}$~cm$^{-2}$) and the limited band-pass of the XRT
energy coverage did not reveal significant variations of the HR (given also the relatively large associated error bars).

The relatively sharp drop of the source X-ray flux around phase 0.75--1.0 corresponds to the suspected eclipse studied also by
\citet{kab20} and \citet{islam15}. In agreement with their results, our analysis also evidences that, despite the decrease by
a factor of $\gtrsim$2 in the source flux, the other spectral parameters did not show at this particular orbital phase a
dramatic variation compared to other phases. We recorded a flattening of the powerlaw photon index that, in principle, is
expected in case of an X-ray eclipse. However, this is not accompanied by a drop in the local absorption column density. Such
drop is expected because during the eclipse the source of X-rays is hidden from the direct view of the observer who is looking
rather to the remaining diffuse fluorescence emission of the X-rays from the occulted NS onto the surrounding wind material
spread all around the binary system \citep[see, e.g.,][and references therein]{bozzo09}.

The XRT data are thus raising concerns against the interpretation of the drop in flux at phases 0.75--1.0 as an X-ray eclipse,
but the statistics is far to low to allow a deeper analysis. \citet{kab20} proposed that this drop in flux could be caused by
a grazing eclipse and absorption of the X-rays in the stellar corona. At present, however, alternative possibilities cannot be
excluded. In the context of the corotating interaction regions, there could be the intriguing possibility that the obscuration
is due to one of these large structures that is tilted away from the plane of the NS orbit and
attenuate the X-ray emission along the line of sight to the observer mainly through scattering with a non-measurable
enhancement of photoelectric absorption as for the grazing eclipse
\citep[this idea was put forward already in][]{bozzo16}. Interestingly, the source has also displayed evidence for an at least
transitory super-orbital period, which origin could  be also related to corotating interaction regions \citep{corbet21}. The
peculiar orbital phase 0.75--1.0 of \J393\ definitively deserves a dedicated observation with larger effective area
instruments (as the EPIC cameras on-board \xmm) able to eventually detect emission/absorption lines, providing details about
the physical conditions of the material causing obscuration, as well as revealing modest variations of the spectral parameters
that could go undetected due to the limited statistics of the \swift/XRT data.

\subsection{ \xte}
\label{sec:1855}

\xte\ was discovered by the \rxte\ satellite in 1998 \citep{corbet99} and it is known to host a $\sim$361~s spinning NS orbiting every 6.1\,d a BN0.2 Ia supergiant star located at roughly 10~kpc \citep[see][and references therein]{galan15}. The optical companion was identified thanks to the refined \swift\ position obtained with an arcsec level accuracy \citep{romano08atel}. \xte\ is known to be eclipsing \citep[see][for the most updated source ephemerides]{falanga15, coley15} and it has long been classified as a classical SgXB, although its behavior is partly anomalous compared to other objects of this class due to the emission of sporadic bright X-ray outbursts. These have been observed a few times with \inte\ and \swift\ \citep[see, e.g.,][and references therein]{watanabe10, krimm12}.

\subsubsection{Data analysis and results}

We report on a detailed analysis of the yet unpublished \swift\ data (see Table~\ref{x1855:tab:swift_xrt_log})
collected during the outburst observed in 2011 \citep{krimm12}.
This is the only outburst for which data in the soft X-ray domain ($\lesssim$1-2~keV) are available.
\begin{figure}
\vspace{-0.6truecm}
\includegraphics[width=1.1\columnwidth]{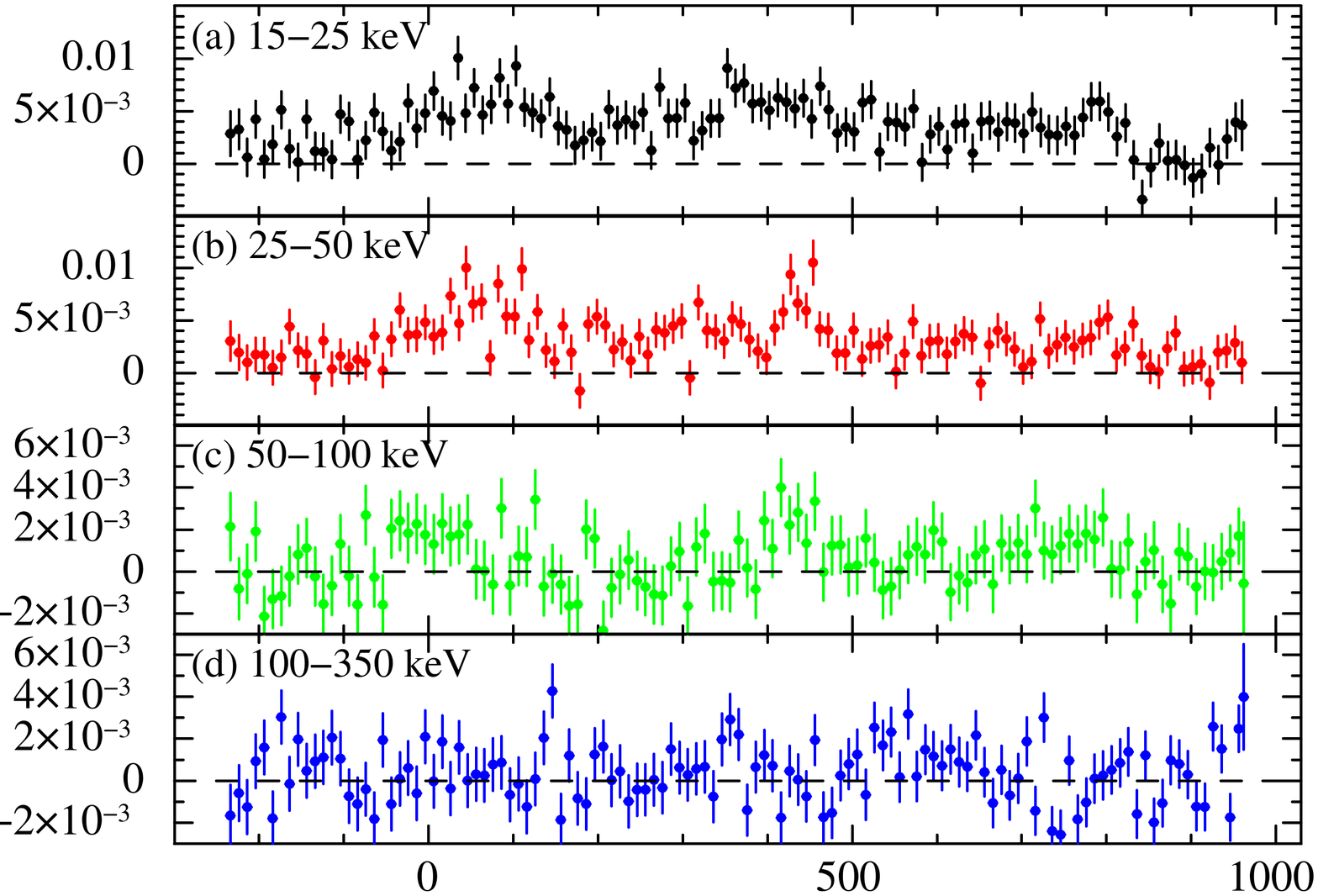}
\includegraphics[width=1.1\columnwidth]{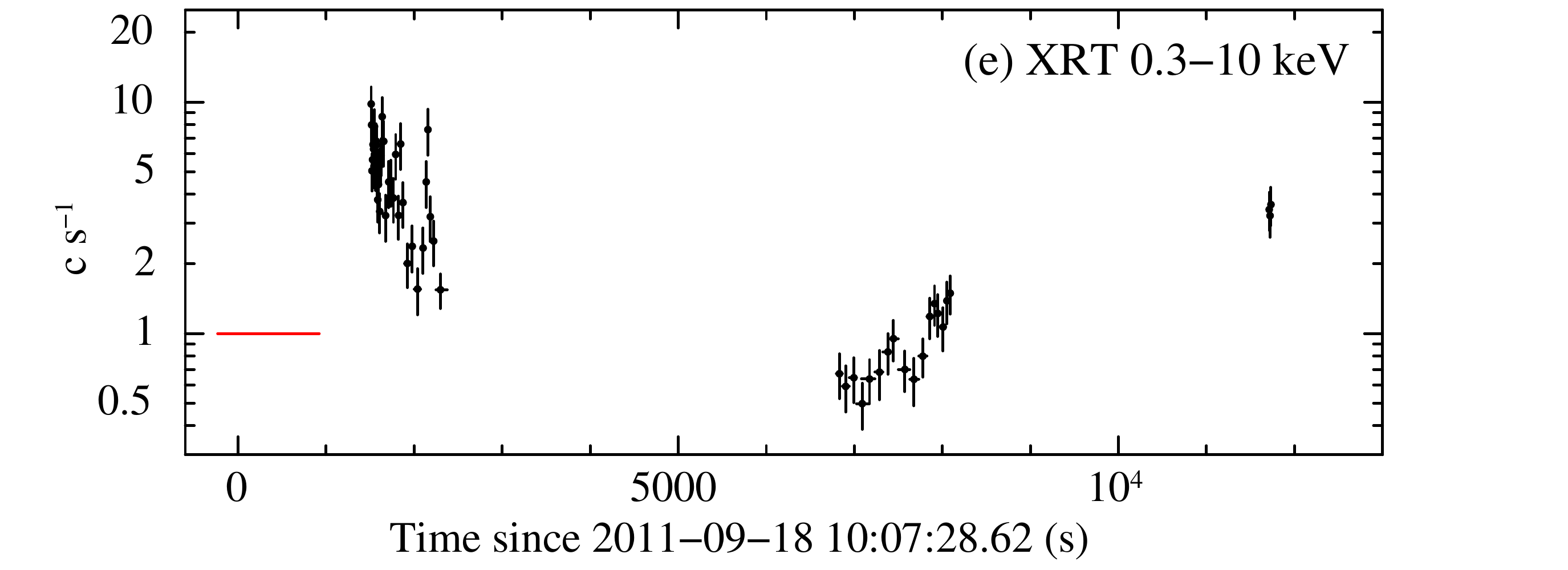}
    \caption{Light curves of the 2011 September 18 outburst of \xte\ caught by BAT and XRT.
    \textit{Panels (a--d)}: BAT light curves (c s$^{-1}$, at 10\,s binning);
   \textit{panel (e)}: XRT light curve (0.3--10\,keV c s$^{-1}$); the horizontal line marks the time where BAT event data were collected.
   Note the different scale in the x-axis.  }
    \label{fig:lcv_1855_tot}
\end{figure}
\begin{table*}
    \tabcolsep 4pt
    \caption{Spectral analysis conducted on the \swift\ observations of \xte.}
    \label{x1855:tab:swift_bat_xrt_spec}
    \centering
    \tiny
    \renewcommand{\arraystretch}{1.3}
\begin{tabular}  {cr cc cccc}
  \hline
 \hline
 \noalign{\smallskip}
 Spectra (model)    & Time range (s)        &  $N_\mathrm{H}$            &   $\Gamma$   &   E$_{\rm cut}$ & F$_\mathrm{15-50\,keV}$      &       F$_\mathrm{0.3-10\,keV}$  & $\chi^2$/d.o.f. \\
                &   since trigger        &  ($10^{23}$ cm$^{-2}$ )  &                      &   (keV)           &  ($\times10^{-9}$)                       &      ($\times10^{-10}$)             &  Cstat/d.o.f. \\
  \noalign{\smallskip}
 \hline
 \noalign{\smallskip}
 BAT evt (PL)     &   $-$239--960        &     -                      &  $2.48\pm0.09$        &  -                       & $2.27\pm0.08$      & -                          &  30.90/24 \\ 
 BAT evt (CPL)   &   $-$239--960        &     -                     &  $0.48^{+0.58}_{-0.64}$  &  $16_{-3}^{+9}$   & $2.31\pm0.08$     &  -                         &  16.63/23 \\  
 BAT DPH1  (PL) &      1504--1804      &     -                      &   $2.45^{+0.15}_{-0.14}$ &  -                      & $1.88\pm0.15$     & -                          &  26.08/4   \\   
 BAT DPH2  (PL) &      1804--2104      &     -                     &   $2.44^{+0.24}_{-0.21}$ &  -                      & $1.32\pm0.14$     & -                          &  6.81/4    \\ 
 BAT DPH3 (PL)  &      2104--2382      &     -                     &   $2.16^{+0.18}_{-0.17}$ &  -                      & $1.49\pm0.14$     & -                          &  9.98/4    \\  

\noalign{\smallskip}
   WT1 (PL)        &  1510--1627           &     $1.8^{+0.5}_{-0.4}$ & $0.4^{+ 0.4}_{-0.4}$   & -                         & -                            & $9.9^{+0.8}_{-0.7}$   &   $279/380$ \\
   PC1  (PL)        &   1629--2376          &     $1.2^{+0.3}_{-0.2}$ & $-0.2^{+ 0.3}_{-0.3}$ & -                         & -                            & $7.2^{+0.6}_{-0.5}$   &    $260/307$\\
   WT2 (PL)        &   6788--8160          &     $4.0^{+0.9}_{-0.7}$ & $1.3^{+ 0.6}_{-0.5}$   & -                         & -                            & $1.8^{+0.2}_{-0.2}$   &    $240/298$\\
   PC2  (PL)        &   6795--8154          &     $3.0^{+1.2}_{-0.8}$ & $0.6^{+ 0.7}_{-0.6}$   & -                         & -                            &  $1.3^{+0.2}_{-0.1}$  &    $162/144$\\
   WT3  (PL)       &   11708--11740     &     $0.8^{+0.9}_{-0.3}$ & $0.7^{+ 0.9}_{-0.6}$   & -                         & -                            & $4.4^{+1.1}_{-0.9}$   &   $62/85$\\
\noalign{\smallskip}
\hline
 \noalign{\smallskip}

DPH1 $+$ WT1  (CPL)  &  -            &      $0.9\pm0.1$             &  $-0.46_{-0.10}^{+0.09}$ &  $10\pm1$     &  $2.2_{-0.2}^{+0.1}$    &  $10.1_{-0.8}^{+1.3}$    &    355.49/389     \\ 
DPH2,3 $+$ PC1 (CPL) &  -            &      $0.6\pm0.2$             &  $-0.96_{-0.35}^{+0.38}$ &   $9_{-1}^{+2}$ &  $1.5_{-1.2}^{+0.1}$    &   $7.1_{-7.0}^{+0.1}$   &      336.19/322     \\ 

DPH1 $+$ WT1  (CPL+Fe)  &  -      & $0.8\pm0.1$                   &  $-0.45_{-0.10}^{+0.09}$ &  $10\pm1$     &  $2.2_{-0.2}^{+0.1}$    &  $9.9_{-0.6}^{+0.5}$    &    347.62/389     \\ 
DPH2,3 $+$ PC1 (CPL+Fe) &  -      & $0.6\pm0.2$ &  $-0.11_{-0.09}^{+0.10}$ &  $13_{-1}^{+3}$ &  $1.44_{-1.44}^{+0.04}$    & $\la$5.8    &      276.45/322     \\ 
\noalign{\smallskip}
 \hline
 \noalign{\smallskip}
 \end{tabular}
\tablefoot{ The power-law model is indicated with PL, the
    exponential cut-off with CPL. We report the measured value of
    the absorption column density ($N_{\rm H}$, in units of $10^{23}$\,cm$^{-2}$), the power-law photon index ($\Gamma$),
    the cut-off energy  (in keV) and the fluxes
    in the 15--50\,keV energy band from BAT (in units of $10^{-9}$ erg\,cm$^{-2}$\,s$^{-1}$),
    and in the  0.3--10\,keV energy band from XRT (not corrected for absorption) in units of $10^{-10}$ erg\,cm$^{-2}$\,s$^{-1}$.}
\end{table*}

\xte\ triggered the BAT (image trigger 503434) on 2011 September 18 at 10:07:28.6 UT
(T$_0$, MJD 55822.42186, orbital phase 0.61); \swift\ slewed to the target so that  the narrow-field
instruments started observing at $T_0+$1510\,s (see Table~\ref{x1855:tab:swift_xrt_log}).
The BAT event data were analysed using the standard BAT analysis software within FTOOLS.
Mask-tagged BAT light curves, covering the time range  $T_0-239$ to $T_0+963$\,s,
were created in the standard energy bands
(15--25, 25--50, 50--100, 100--350\,keV), and rebinned to fulfill at least one
of the following conditions: reaching a signal-to-noise ratio (S/N) of 5 or bin length of 10\,s.
The light curves of the first orbit of data are shown in Fig.~\ref{fig:lcv_1855_tot}.
A mask-weighted spectrum was also extracted from the events collected during the first orbit;
we applied an energy-dependent systematic error vector to the data and created response matrices
with {\sc batdrmgen} using the latest spectral redistribution matrices.
This spectrum, when fit in the energy range 15--70\,keV with a simple power law,
yields a photon index of $2.48\pm0.15$ ($\chi^2/$d.o.f.$=30.90/24$) and a
15--50\,keV flux of $(2.27\pm0.08)\times 10^{-9}$\,erg\,cm$^{-2}$\,s$^{-1}$
(see Table~\ref{x1855:tab:swift_bat_xrt_spec}).
This fit, however, shows residuals that indicate a curvature, so a fit with a
cut-off power-law model was also made, yielding a
photon index of $0.48^{+0.58}_{-0.64}$ and a cut-off energy E$_{\rm cut}=16_{-3}^{+9}$\,keV ($\chi^2/$d.o.f.$=16/23$).
We note that there are no XRT data simultaneous with this BAT spectrum.
\begin{figure}
\vspace{-0.8truecm}
\hspace{-0.2truecm}
        \includegraphics[width=1.15\columnwidth]{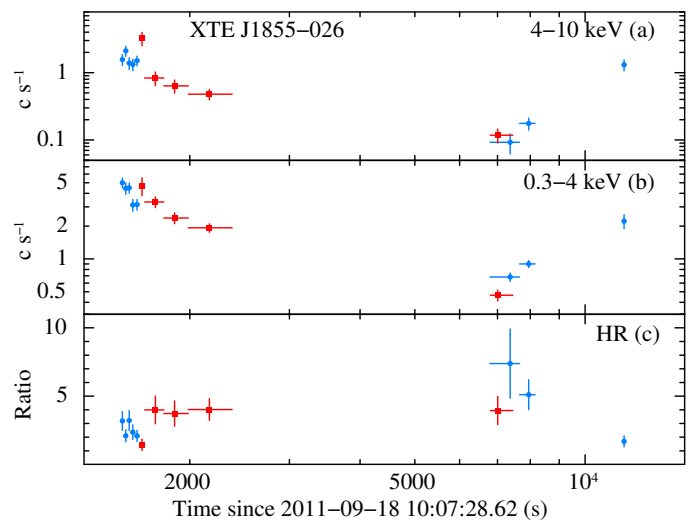}
   \caption{\textit{Panel (a)}: XRT hard band light curve (4--10\,keV c s$^{-1}$);
   \textit{panel (b)}: soft band light curve (0.3--4\,keV c s$^{-1}$);
   \textit{panel (c)}: hardness ratio (4--10\,keV/0.3--4\,keV). WT data are shown in blue, PC data in red.
   \label{fig:lcv_1855_hr}}
\end{figure}
The XRT light curve, split in 3 snapshots, starts at $T_0+$1510\,s and shows
a dynamic range of $\sim 20$ in $\sim6$\,ks
(0.3--10\,keV, see Fig.~\ref{fig:lcv_1855_tot}).
The hardness ratio, calculated in the 0.3--4 and 4--10\,keV bands is shown in Fig.~\ref{fig:lcv_1855_hr}.
XRT spectra were extracted in each observing mode and each of the snapshots that comprise the data set,
thus obtaining three window timing (WT) and
two photon counting (PC) spectra, and were fit with the a simple absorbed power-law model and Cash statistics.
A summary of the spectral results is shown in Table~\ref{x1855:tab:swift_bat_xrt_spec}.

BAT survey data products, in the form of detector plane histograms (DPH), are also available,
and were analysed with the standard FTOOLS software. Since BAT survey data are accumulated on-board
with typical integration times of 300\,s, pairing simultaneous BAT survey data with the XRT ones is not straightforward.
Therefore we extracted spectra (6 energy bins)
that most closely matched the XRT data:
{\it i)} DPH1 from $T_0+1504$\,s to $T_0+1804$\,s;
{\it ii)} DPH2 in the time range  $T_0+1804$--2104\,s;
{\it iii)} DPH3 in the time range   $T_0+2104$--2382\,s.
Each BAT spectrum was fit with a simple power-law model (results in Table~\ref{x1855:tab:swift_bat_xrt_spec}).
In order to ensure the closest to simultaneity, we chose to fit the following BAT and XRT groups: {\it a)} DPH1$+$WT1,
{\it b)} DPH2$+$DPH3$+$PC2.
When fitting these BAT and XRT spectra together, a constant needs to be used to model both the difference
of exposure and the non strict simultaneity.
Furthermore, the XRT spectra are fit by minimizing Cash statistics,
while the BAT ones with $\chi^2$ statistics.

\begin{figure}
\vspace{-0.8truecm}
\hspace{-0.2truecm}
        \includegraphics[width=1.15\columnwidth]{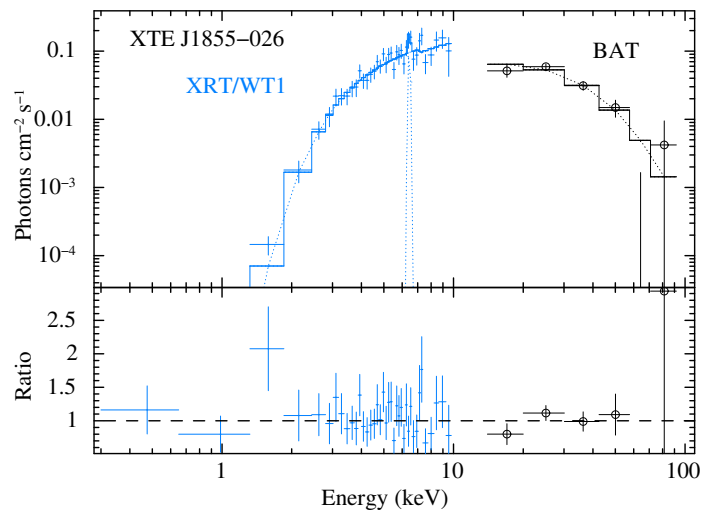}
   \caption{Spectroscopy of the 2011 September 18 outburst of \xte.
   Top panel: unfolded spectra of the nearly-simultaneous XRT/WT1 data (blue crosses) and BAT DPH1 data (empty black circles)
fit with an absorbed cut-off power-law  model.
   Bottom panel: data/model ratio of the fit.
    \label{fig:spec_1855_1b}}
\end{figure}

A fit to the DPH1$+$WT1 pair with an absorbed power-law model resulted in residuals suggesting a spectral curvature,
so we also performed a fit using an absorbed cut-off power law.
This yielded a $N_{\rm H}=0.9\pm0.1 \times10^{23}$ cm$^{-2}$, a
photon index of $-0.46_{-0.10}^{+0.09}$
and a cut-off energy E$_{\rm cut}=10\pm1$\,keV (inter-calibration constants C$_{\rm DPH1}=1$ fixed and C$_{\rm WT1}=2.6_{-0.3}^{+0.4}$).
This is reported in Table~\ref{x1855:tab:swift_bat_xrt_spec}.
The addition of an Iron line, represented by a Gaussian model with energy $\sim 6.4$\,keV, yields a continuum fit
with  $N_{\rm H}=0.8\pm0.1 \times10^{23}$ cm$^{-2}$,
a photon index of $-0.45_{-0.10}^{+0.09}$
and a cut-off energy E$_{\rm cut}=10\pm1$\,keV (inter-calibration constants C$_{\rm DPH1}=1$ (fixed) and C$_{\rm WT1}=2.4_{-0.3}^{+0.4}$),
while the line is characterised by a centroid energy $E_\mathrm{Fe}=6.4_{-0.7}^{+0.8}$\,keV and
a width consistent with zero ($<1.52$\,keV)
and an equivalent width  EW=$0.18_{-0.17}^{+0.24}$\,keV (Fig.~\ref{fig:spec_1855_1b}).

\begin{figure}
\vspace{-0.8truecm}
\hspace{-0.2truecm}
        \includegraphics[width=1.15\columnwidth]{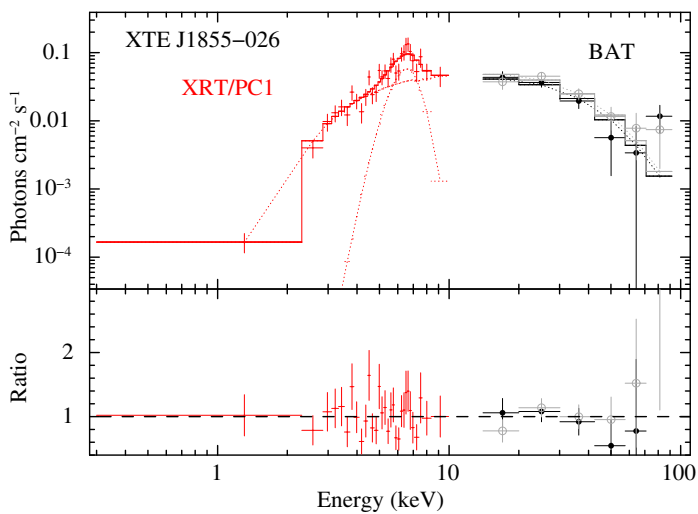}
   \caption{Spectroscopy of the 2011 September 18 outburst of \xte.
   Top panel: unfolded spectra of the simultaneous XRT/PC1 data (red crosses), the BAT DPH2 data in black (filled circles)
and the BAT DPH3 data in grey (empty circles) fit with an absorbed cut-off power-law  model.
   Bottom panel: data/model ratio of the fit.
   \label{fig:spec_1855_2b}}
\end{figure}

Similarly, a fit to the DPH2$+$DPH3$+$PC2 group with an absorbed power-law model indicated the presence of a possible spectral curvature in the residuals. An absorbed cut-off power law fit yielded a $N_{\rm H}= 0.6\pm0.2\times10^{23}$ cm$^{-2}$, a
photon index of $-0.96_{-0.35}^{+0.38}$,
and a cut-off energy E$_{\rm cut}=9_{-1}^{+2}$\,keV
(inter-calibration constants C$_{\rm DPH2}=1$ (fixed),
C$_{\rm DPH3}=1.2_{-0.1}^{+0.2}$, and
C$_{\rm PC1}= 3.8_{-0.9}^{+1.4}$).
This is reported in Table~\ref{x1855:tab:swift_bat_xrt_spec}.
The addition of an Iron line, represented by a Gaussian model with energy $\sim 6.4$\,keV, yields a continuum fit
with  $N_{\rm H}=0.6\pm0.2_{-0.1}^{+0.2}\times10^{23}$ cm$^{-2}$,
a photon index of $-0.11_{-0.09}^{+0.10}$
and a cut-off energy E$_{\rm cut}=13_{-1}^{+3}$\,keV
(inter-calibration constants C$_{\rm DPH1}=1$ (fixed)
C$_{\rm DPH3}=1.2\pm0.1$, and
and C$_{\rm PC1}= 1.3_{-0.3}^{+0.5}$), while the line is characterised by a
centroid energy $E_\mathrm{Fe}=6.4\pm0.2$\,keV,
a width of $0.85_{-0.11}^{+0.14}$\,keV
and  EW$\gg1$\,keV (Fig.~\ref{fig:spec_1855_2b}).

\subsubsection{Discussion of the results}

 The outburst of the source in 2011 was the only one where the broad-band emission from \xte\ could be studied combining data from a large field of view instrument (\swift/BAT) with those collected in the soft X-rays by a focused telescope (\swift/XRT). Although the XRT lightcurve during the outburst is fragmented due to the observational strategy of the satellite, the coverage of the outburst allows us to study possible changes in the spectral properties of the source X-ray emission from the onset of the event down to the return to the usual emission level.

We see from Fig.~\ref{fig:lcv_1855_hr} that XRT pointed the source about 1.5~ks after the onset of the outburst. The instrument recorded a progressive rise of the absorption column density, starting at $\lesssim$2$\times10^{22}$~cm$^{-2}$ (WT1 and PC1 data in Table~\ref{x1855:tab:swift_bat_xrt_spec})  and reaching up to $4\times10^{22}$~cm$^{-2}$ a few ks after the beginning of the monitoring (WT2 and PC2 data in Table~\ref{x1855:tab:swift_bat_xrt_spec}). Interestingly, XRT also recorded a new drop of the absorption column density down to $0.8\times10^{22}$~cm$^{-2}$ about 11~ks after the onset of the event when the flux decay was interrupted by a new brightening of the source (WT3 data in Table~\ref{x1855:tab:swift_bat_xrt_spec}). This behavior resembles what is typically observed in clumpy wind accreting systems, but in the case of the 2011 outburst of \xte,\ we have likely missed the initial increase in the local absorption column density as the XRT only began observing about 1.5~ks after the onset of the event. Nevertheless, we clearly detect the progressive increase of the local absorption column density during the decay of the outburst associated to the fading of the source and the diminishing effect of the photoionization onto the clumpy wind. The drop of the local absorption column density about 11~ks after the onset of the flare further strengthens this conclusion as it can be ascribed to the renewed effect of the photoionization when the source underwent a second (fainter) brightening.

The onset of the outburst up to 1.5~ks was observed by \swift\ only with the BAT. The hard X-ray spectrum above 15~keV could be described reasonably well with a simple power-law and the measured photon index is compatible with that recorded during the outbursts observed with the hard X-ray imager IBIS/ISGRI on-board \inte\ \citep{watanabe10}. However, the BAT data showed evidence for a possible curvature in the hard energy spectrum, with a cut-off energy at about 16~keV. When the XRT data are combined with the (quasi-)simultaneous BAT survey data during the later stages of the outburst development (data DPH1+WT1 and DPH2,3+PC1 in Table~\ref{x1855:tab:swift_bat_xrt_spec}), the value of the measured curvature slightly decreases toward $\sim$10~keV, although the associated uncertainties remained quite large due to the limited statistics of the data. Such curvature is rather ubiquitous in accreting SgXBs \citep[see, e.g.][and references therein]{walter2015} and was also reported previously in the case of \xte\ \citep{corbet99}. In the broadband fit of the XRT+BAT data, we find evidence for the presence of a iron line at a centroid energy of 6.4~keV. Although the statistics of the XRT data in the energy range of the iron line is limited, such feature is expected in the case of an SgXB. The iron line at 6.4~keV is commonly observed in wind-fed systems, due to the fluorescence of the X-rays from the accreting compact object onto the surrounding stellar winds \citep[see, e.g.][for recent reviews]{torrejon10,gimenez16}. Iron lines with compatible parameters as those measured by XRT were already reported in the case of \xte\ by \citet{jincy18} using \suzaku\ data.

Given its classification as a classical SgXBs and the presence of peculiar bright short outbursts similar to those of the SFXTs, \xte\ remains today an intriguing object that has likely not drawn sufficient attention from the community. It remains relatively poorly studied, as no detailed orbital monitoring in the soft X-rays has been carried out yet and we are missing sufficiently long exposure observations with the large area X-ray instruments (as the EPIC cameras) to probe possible emission/absorption lines in the high energy domain to investigate the properties of the stellar wind material within the binary. The evidence reported in this paper about the clumpy wind accretion in \xte\ provides good perspectives to renew the interest in this system.

\subsection{ \2IGR}
\label{sec:17503}

\2IGR\ is an SFXT discovered by \inte\ in 2018 \citep{chenevez18}.
The source underwent a bright X-ray flare lasting several hours before rapidly going back to a quiescent emission level \citep[the measured dynamical range in the X-ray domain is of $\sim$300; see][]{cha18, cha18b, ferrigno19}. The optical counterpart was tentatively identified as a heavily obscured supergiant located beyond the Galactic center \citep{masetti18}, although these findings require further consolidation \citep{collum18}. The \nustar\ data collected shortly after the discovery showed evidence for the presence of a cyclotron scattering feature in the broad-band spectrum of the source, suggesting that the compact object accreting in this system is a NS endowed with a magnetic field strength of $\sim$2$\times$10$^{12}$~G \citep{ferrigno19}.

\subsubsection{Data analysis and results}

We report on our \swift/XRT follow-up campaign until up to about one year after the discovery of the source,
spanning from April to June 2019 (with a pace of one 5\,ks observation per week, ObsID 10980), and not yet published elsewhere.
As the orbital period of the source is not known, we could only consider here the long term evolution of the source X-ray emission.

\begin{figure}
 \vspace{-0.4truecm}
 \includegraphics[width=1.1\columnwidth]{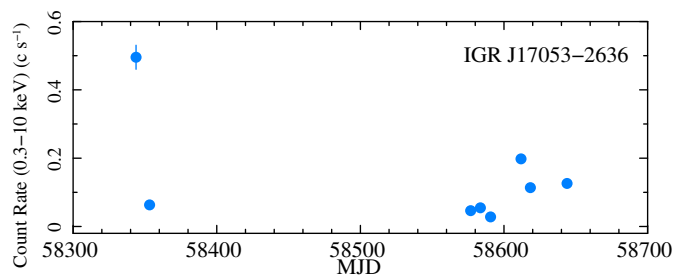}
       \vspace{-3.8truecm}
    \caption{Long-term light curve of \2IGR\ as measured by \swift\,/XRT.
    For completeness, we included the XRT observations already published in our previous paper on the source \citep[MJD 58343 and 58333,][]{ferrigno19}.
    }
    \label{fig:17503_xrt_lc}
\end{figure}

The summary of the available XRT observations of \2IGR\ is provided in Table~\ref{i17503:tab:swift_xrt_log}.
In the table we also report the measured flux of each XRT observation, together with the best fit parameters;
all spectra of \2IGR\ could be fit well with a simple absorbed power-law model.
We also included for completeness in the table one $\sim2$\,ks serendipitous observation in the
\swift\ archive (MJD 56224) preceding the discovery where the source and not yet published elsewhere.
During this pointing, the source was found in a low state (the measured count rate is $5.2\pm{0.7}\times10^{-1}$ c\,s$^{-1}$
corresponding to a 0.3--10\,keV flux of $0.5^{+0.3}_{-0.2}\times10^{-11}$ erg\,cm$^{-2}$\,s$^{-1}$).
The long-term X-ray light curve of \2IGR\ as measured by XRT is shown in Fig.~\ref{fig:17503_xrt_lc}.

\subsubsection{Discussion of the results}

Our XRT monitoring campaign of \2IGR\ demonstrates that the source behaves as most SFXTs, remaining at a fairly faint quiescent level for most of their lifetime outside rare and sporadic outbursts (so far only one outburst has been recorded). The source shows a rather stable spectral energy distribution over time during the low level emission state, as observed in several SFXTs \citep[see, e.g.,][and references therein]{walter2015}.
As the source was regularly detected during each of the performed XRT pointing at a flux of $\sim$10$^{-11}$~$\ferg$ or higher, our campaign seems to invalidate the tentative conclusion presented by \citet{ferrigno19} according to which the source is a faint SFXT due to its particularly high distance beyond the Galactic center (this was proposed to explain why no previous outbursts were ever detected from the source by the large field of view instruments on-board \inte\ and/or \swift).

The measured XRT fluxes correspond to a luminosity of roughly $\sim$10$^{35}$~erg~s$^{-1}$ at 10~kpc and the quiescent luminosity of several SFXTs has been measured to be as low as 10$^{32}$~erg~s$^{-1}$. The archival XRT observation that we reported here for the first time also confirmed that the source was already active at a similar flux level in 2012, about 6 years before the discovery of the first outburst by \inte.\ The reason why no previous outbursts have ever been detected before 2018 remains to be explained.
The intrinsic fragmentation of the XRT light curves (due to the pointing strategy of the satellite) did not allow us to study the SFXT typical behavior of flaring also at the lowest emission states, during which evidence of the clumpy wind accretion are commonly found (see Sect.~\ref{sec:intro}). For all these reasons, the source certainly deserves further observations to find clearer evidence of clumpy wind accretion and measure possible changes in the cyclotron line properties that could be associated with variations in the NS magnetic field configuration.

\section{Conclusions}
\label{sec:conclusions}

We reported in this paper on the outcomes from our concluded observational campaigns of several classical SgXBs and SFXTs with both \swift\ and \xmm.\ The goals of the data we collected were mainly to study structures in the stellar winds surrounding the compact objects in these systems, revealing details of the still highly debated macro-clumps, as well as other larger structures. In the field of macro-clumping, the results presented here complement those reported in our previous papers, while the longer-term XRT observational campaigns were reported here for the first time.

As a general conclusion, we found that the observational strategies set to study structures in the massive star winds with both \xmm\ and \swift\ have been so far successful, with the former focusing on the shorter term variability associated with the accretion of clumps (few thousands seconds to hours) and the latter on the longer time-scale variability (few to several days) associated with different orbital phases of these systems and driven by larger stellar wind structures (as accretion and photoionization wakes, CIRs, and accretion streams).

Short \xmm\ observations had been already exploited in a number of previous papers from our group and although we had been able to report in the past on significantly brighter flares than those described here, we could still advance the census of these kind of events from both classical SgXBs and SFXTs. In particular, the observation of the classical SgXB \u4\ allowed us to provide interesting evidence about how the encounter between the NS and a clump can affect the accretion process beyond the variations of the continuum emission and absorption column density studied before. The lack of spectral variability in the flares recorded from \3IGR\ and \4IGR\ increases the statistics of events that seem to occur without the interventions of clumps, likely triggered by additional mechanisms as the centrifugal and magnetic gating or the settling accretion regime. The lack of spectral variability is particularly puzzling in the case of \4IGR\ where the observed flare by \xmm\ reaches the previously reported threshold of a few cts~s$^{-1}$ above which variations are expected \citep{bozzo17}.

We also exploited in this paper XRT and BAT observations of \xte,\ caught during a rare bright outburst in 2011 (and yet unpublished). We could study for the first time the source broad band emission with high sensitivity and looked for possible spectral features and variability during the event. Although the source displayed a remarkable dynamic range in its X-ray luminosity, our analysis could not reveal any significant spectral variability. This source remains so far poorly studied and understood, being characterized by an intermediate behavior between classical SgXBs and SFXTs.

The study of orbital-phase spectral variability reported here for the first time with \swift/XRT has undoubtedly helped us understanding which among the observed sources are more likely to display measurable periodic spectral changes along their revolutions. In the case of \u4,\ our XRT campaign confirmed previous findings about the peculiar changes in the source flux, continuum slope, and absorption column density at different orbital phases, but we were able to extend literature studies exploiting the lower energy band-pass of XRT. This allowed us to reveal even more extreme variations especially of the local absorption column density to the source. Thanks to the XRT coverage at these low energies, we were able to reveal a completely analogous behavior in \IGR19,\ and proposed for this object a similar scenario as that studied already in much better details for \u4, involving the presence of a non-negligible eccentricity and a massive structure moving with the NS, likely a gas stream. No particularly striking variability has instead been observed from the orbital monitoring of \J393. This source is relatively faint even for XRT and although it is possible that some variability has gone undetected within the measured uncertainties of the different spectral parameters, we can definitively rule out prominent spectral changes as those measured from either \u4\ or \IGR19.\  In the case of the SFXT \2IGR, we could not study spectral variability along the different orbital phases as the system orbital period is not known yet. Our XRT observations revealed, however, that the source remained relatively stable in terms of X-ray flux and spectral energy distribution over time. Having followed the source up to nine months after its initial discovery in 2018 and having reported its detection in an archival XRT observation dating back in 2012, we consider that it is reasonable to assume that the source is displaying a low persistent luminosity with rare outbursts (only one detected so far) as it is common for the SFXTs. This further strengthens the previous conclusions about the nature of this source.

\begin{acknowledgements}

The data underlying this article are publicly available from the \xmm, \rxte, \swift, and \nustar\ archives. The full chain of the analysis of the \xmm\ and \inte\ data is available in the form of a python Notebook that can be run in a dockerized environment for \href{https://gitlab.astro.unige.ch/ferrigno/4u-1907-xmm/-/tree/3b2e7b905c5d5bc8049437b4494a4b46279f2531}{4U J1907$+$097}, \href{https://gitlab.astro.unige.ch/xmm-newton-workflows/0862410101/-/tree/fad1a1c7cbaacdb972a1cd0c84045cc5f9744946}{IGR J18410$-$0535}, and  \href{https://gitlab.astro.unige.ch/xmm-newton-workflows/0862410301/-/tree/2f6de5cacc6e08420f9349686a784ad203281262}{IGR J11215$-$5952}.

We thank the anonymous referee for detailed comments that helped us improve the paper.
The \swift\ data of our monitoring campaigns were obtained through contract ASI-INAF I/004/11/5 (PI P. Romano).
PR and EB acknowledge financial contribution from contract ASI-INAF I/037/12/0.
This work made use of data supplied by the UK Swift Science Data Centre at the
University of Leicester \citep[see][]{2007A&A...469..379E,2009MNRAS.397.1177E}
and by the multi-messenger online data analysis (\href{https://www.astro.unige.ch/mmoda/}{MMODA})
platform at the University of Geneva \citep[see][]{Neronov2021}.
CF is grateful to Dr. Volodymyr Savchenko for providing
assistance with MMODA and for sharing the tuning of \texttt{sextractor} to IBIS imaging.

\end{acknowledgements}

\bibliographystyle{aa}
\bibliography{bib}

\begin{appendix}
\section{Refinement of the \IGR19\ orbital period.}
\label{app:period}

To refine the early determination of the source ephemerides by \citet{corbet04atel}, we first downloaded the full ASM light curve from the online archive (from MJD 50088 to 55846) at the maximum available timing resolution. We computed the Lomb-Scargle periodogram \citep{Scargle2013} with
an oversampling factor of four and limiting ourselves to 0.4 times the average Nymquist frequency. Following \cite{Corbet2007},
we weighted each bin of the light curve with $(f\sigma+V_S)^{-1}$ and $f=1.2$.
By fitting the peak with a Sync curve we determined a period of 13.5550$\pm$0.0005\,d.

The source is also monitored by \swift/BAT and the up-to-date light curve in the 15--50 keV energy band (time resolution of 1 satellite orbit) can be downloaded from the Swift/BAT Hard X-ray Transient Monitor on-line page  \citep{Krimm2013}\footnote{\url{https://swift.gsfc.nasa.gov/results/transients/}}. The arrival time of the collected photons was converted to the solar system barycenter frame using the \texttt{earth2sun} program. We excluded data from 2019, as
we found that the orbital periodicity is not significant in the corresponding data. We then computed the
Lomb-Scargle periodogram with a weight of $\sigma^{-1}$ as in \citet{dai11}; the fit to the peak yields an orbital period of 13.55273$\pm$0.00005\,d in the time range MJD 53416--59597. The periods from ASM and \swift/BAT differ
by almost five equivalent Gaussian sigma, while the improvement of the uncertainty is linked to the higher signal of
the source in \swift/BAT combined with the slightly longer baseline.
To look for a possible decrease of the orbital period, we sliced the \swift/BAT lightcurve in several intervals with duration ranging from four months to three years, but no significant trend was detectable. Therefore, we conclude that the different orbital periods from ASM and \swift/BAT is most probably spurious and linked to the different data quality.

\begin{figure}[hb!]
    \hspace{-0.2truecm}
    \includegraphics[width=1.12\columnwidth]{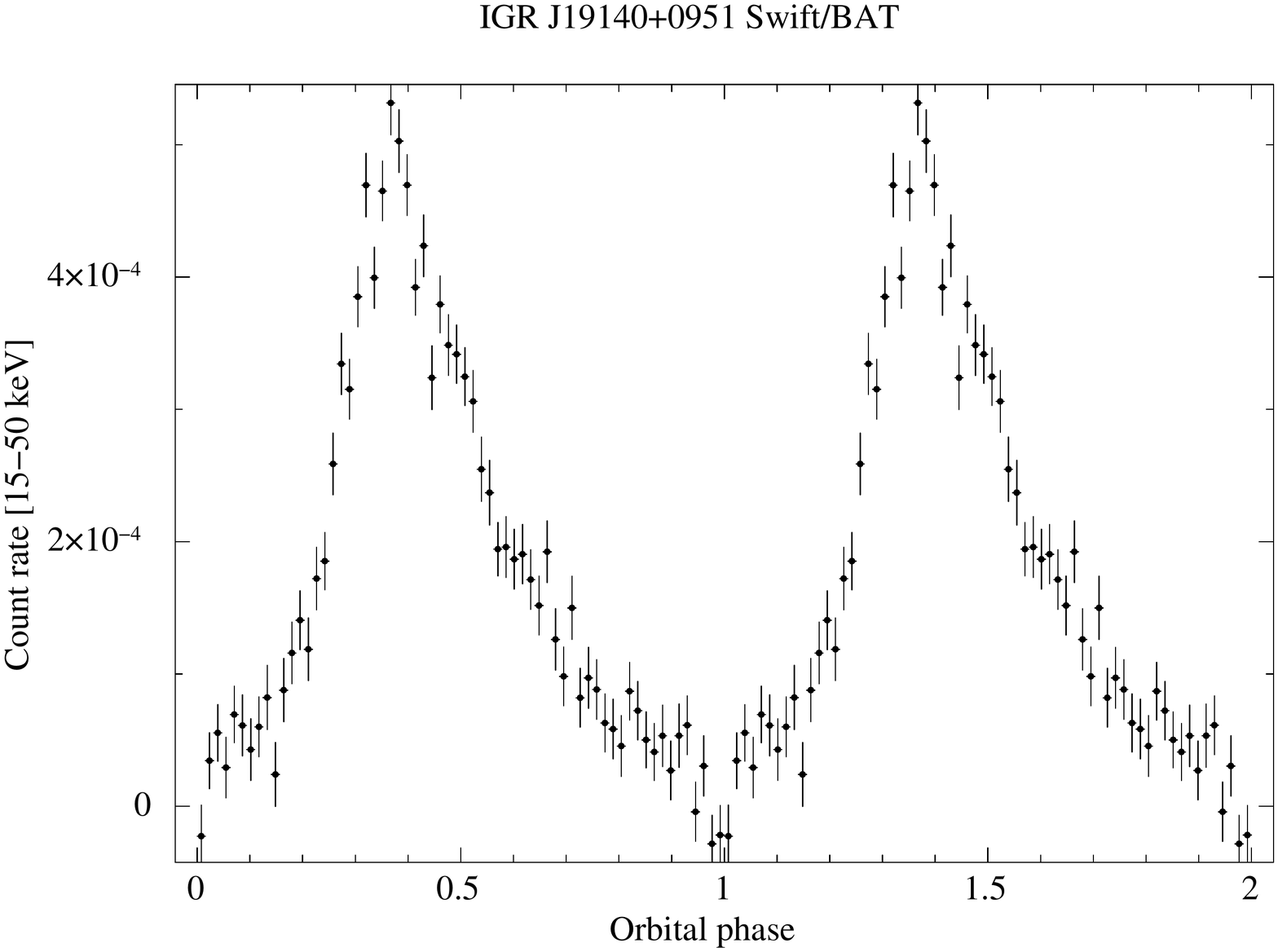}
    \caption{\IGR19\ orbital profile obtained from the \swift/BAT light curve (15--50\,keV energy range) folded at the newly refined period of 13.55273\,d over 64 bins. The assumed reference time is MJD\,52061.42 and data span from 15 Feb 2005 to 18 Jan 2022. The profile is displayed twice for clarity.
        \label{fig:orbit_bat}}
\end{figure}

In Fig.~\ref{fig:orbit_bat}, we show the orbital profile as measured from the BAT data in the 15--50\,keV band using our refined orbital period estimate.

\section{ \5IGR}
\label{sec:17315}

\5IGR\ was reported for the first time in the catalogue of the \inte\ sources published by \citet{krivonos12} and suspected to be another HMXB discovered by \inte,\ possibly an SgXB \citep[see also][]{clavel19}. The source is reported in the catalogue at a best determined position of RA=262.818, DEC=-32.306, with an associated positional accuracy of 2~arcmin. The estimated average flux in the 17-60~keV energy band was  4$\times$10$^{-12}$~erg~cm$^2$~s$^{-1}$. We show here that the reported discovery of this source is most likely to result from an artifact of the \inte\ data analysis and a re-analysis of all available \inte\ data show no evidence of a significant source detection at a position consistent with that reported previously for \5IGR.\
 \begin{table}
\tabcolsep 4pt
 \begin{center}
 \caption{{\it Swift}/XRT observation log for IGR~J17315$+$3221.}
 \label{i17315:tab:swift_xrt_log}
\scriptsize
\begin{tabular}{ lllll }
 \hline
 \hline
 \noalign{\smallskip}
 Sequence        & MJD                      & Start time  (UT)                     & End time   (UT)                      & Exp.   \\
                        &                             &     &      &(s)              \\
  \noalign{\smallskip}
 \hline
 \noalign{\smallskip}
00034659002	&	57608.31510	&	2016-08-08 07:33:36	&	2016-08-08 07:52:54	&	1148	\\
00034659003	&	57610.44377	&	2016-08-10 10:39:01	&	2016-08-10 18:59:52	&	1610	\\
00034659004	&	57611.37894	&	2016-08-11 09:05:40	&	2016-08-11 09:21:53	&	973	        \\
00034659005	&	57612.17438	&	2016-08-12 04:11:06	&	2016-08-12 04:28:54	&	1068	\\
00034659006	&	57615.42504	&	2016-08-15 10:12:03	&	2016-08-15 10:29:54	&	1071	\\
00034659007	&	57616.22967	&	2016-08-16 05:30:43	&	2016-08-16 05:46:54	&	970	        \\
00034659009	&	57618.08242	&	2016-08-18 01:58:40	&	2016-08-18 02:15:53	&	1033	\\
00034659010	&	57619.47689	&	2016-08-19 11:26:42	&	2016-08-19 11:44:53	&	1091	\\
00034659011	&	57620.47378	&	2016-08-20 11:22:14	&	2016-08-20 11:39:55	&	1061	\\
 \noalign{\smallskip}
00043517001	&	56171.76098	&	2012-09-01 18:15:48	&	2012-09-01 18:19:56	&	248	\\
00043517002	&	56173.31138	&	2012-09-03 07:28:23	&	2012-09-03 07:32:56	&	273	\\
00043517003	&	56177.38995	&	2012-09-07 09:21:31	&	2012-09-07 09:29:55	&	504	\\
00043524001	&	56172.23429	&	2012-09-02 05:37:22	&	2012-09-02 05:46:54	&	572	\\
00043524002	&	56178.26009	&	2012-09-08 06:14:31	&	2012-09-08 06:22:55	&	504	\\
  \noalign{\smallskip}
  \hline
  \end{tabular}
\tablefoot{We report the observing sequence and date (MJD of the middle of the observation), start and end times (UT), and XRT exposure time (Exp.). Note that that
    start and stop time of each observation is given in the format yyyy-mm-dd hh:mm:ss.}
  \end{center}
  \end{table}

As \5IGR\ was initially announced to be a likely additional HMXB discovered by \inte,\ this source was included among the monitored targets in our observational campaigns carried out with \swift.  XRT performed observations in the direction of the source between 2012 and 2016, summing up to a total available exposure time of 12.1\,ks (see Table~\ref{i17503:tab:swift_xrt_log}, where the target ID 34659 identifies our 10, 1\,ks pointed observations, the remainder being archival ones). We extracted first an image of the XRT field of view (FoV) for each observation and then a single image stacking all data together in the 0.3--10\,keV energy band. No X-ray source is detected within 2\,arcmin from the best reported position of \5IGR\ in the single images or in the stacked image obtained with all available data. A 3\,$\sigma$ upper limit in the 0.3--10\,keV was calculated by using {\sc sosta} within the {\sc XIMAGE} task and a circular background region (radius of 56 pixels) away from field sources, at $1.7\times10^{-3}$ cts\,s$^{-1}$; by using PIMMS (v4.11b), when assuming a photon index of 2.1 and an absorbing column of $1.23\times10^{22}$ cm$^{-2}$, we obtain an observed (unabsorbed) flux of $9.9\times10^{-14}$ ($2.30\times10^{-13}$ \,erg~cm$^2$~s$^{-1}$).
\begin{figure}
    \includegraphics[width=\columnwidth]{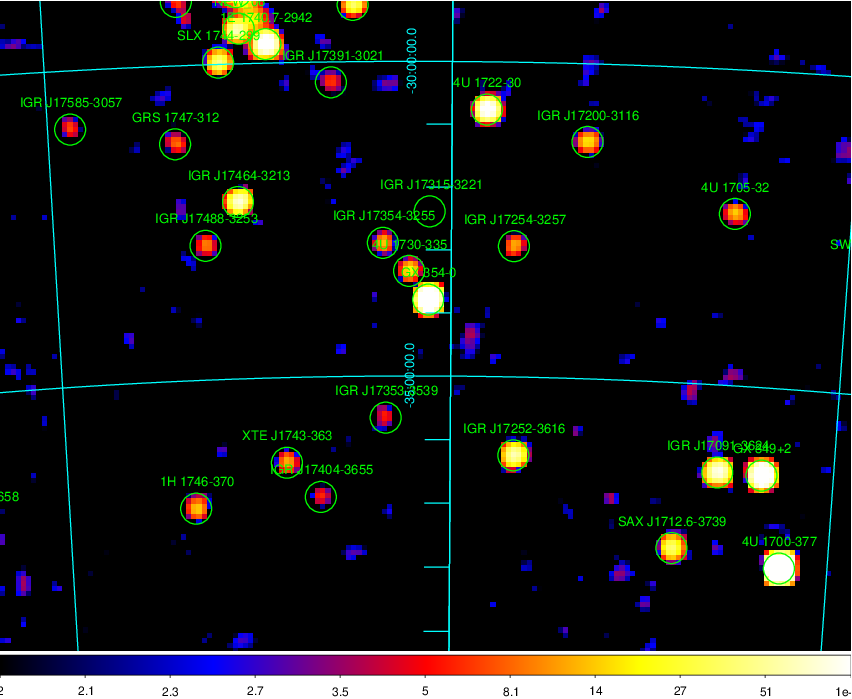}
    \caption{\inte\ IBIS/ISGRI significance map in the 20--60\,keV energy range of the region around \5IGR\ obtained from all available pointings at less than twelve degrees from the source collected in the time period 2004-2012. No significant detection of the previously reported source could be found.}
    \label{fig:int_ul}
\end{figure}

Motivated by the lack of any soft X-ray counterpart in the XRT data, we re-analyzed all \inte\ archival data available around the position of \5IGR.\ Using the latest available calibration and software version (Offline Scientific Analysis 11.2) from the multimessenger online analysis platform\footnote{\url{https://www.astro.unige.ch/mmoda/}}, we extracted a mosaic image in the 20--60~keV energy band using first the same IBIS/ISGRI data considered by \citet{krivonos12} and then from 2014 to 2021 (restricting the pointings to a maximum allowed off-axis angle of 12$^{\circ}$ to avoid large instrument systematics\footnote{See details in the IBIS/ISGRI data analysis manual at
\href{https://www.isdc.unige.ch/integral/analysis}{https://www.isdc.unige.ch/integral/analysis}.}).
Neither of the two images shows a significant detection of any source at a location compatible with the reported position of \5IGR.\
From January 2004 to December 2012, we obtained a 3~$\sigma$ upper limit on the source flux in the 20--60~keV energy band of 0.4 mCrab ($5\times10^{-12}\,\mathrm{erg\,s^{-1}\,cm^{-2}}$) in 4.2\,Ms of effective exposure on the source. From January 2013 to October 2021, we obtained
a slightly worse upper limit of 0.6~mCrab ($7\times10^{-12}\,\mathrm{erg\,s^{-1}\,cm^{-2}}$) in 3.7\,Ms,
owing to the degraded instrument response at low energy during the mission lifetime. To obtain these upper limits, we have elaborated the original mosaics using \texttt{Sextractor} \citep{Bertin1996} in its most recent implementation (version 2.25.2)\footnote{The documentation is available at the URL \url{https://sextractor.readthedocs.io/en/latest/index.html}.}. This is done to correct for the coding noise that is
particularly pronounced in the Galactic center region, due to the large number of bright sources and diffuse emission.
We apply the software with a box size of four pixels and a detection threshold for sources of three.
We show the cleaned significance map for the period 2004--2012 in Fig.~\ref{fig:int_ul} (the map in the following period is similar).

From this re-analysis of \inte\ data and the \swift/XRT non-detection, we conclude that \5IGR\ is not a real astrophysical source, but rather an analysis artifact.

\section{\swift\ data log tables}
\label{app:tables}

In the following tables, we show the logs of the \swift/XRT observations for each source included in the paper.\\
In Tables~\ref{4u1907:tab:swift_xrt_log}--\ref{i16393:tab:swift_xrt_log}, the columns represent: observing sequence, observation date (mid-observation MJD), orbital phase, start and end times of the observations (UT), exposure time in seconds, and the 0.3--10\,keV flux in units of $10^{-11}$\,erg\,cm$^{-2}$\,s$^{-1}$; missing values are due to insufficient counts to perform any spectral analysis.\\In the case of \xte\ (Table~\ref{x1855:tab:swift_xrt_log}), also the log of the BAT data are reported.\\Note that in these tables uncertainties are at 90\% confidence level.
\clearpage
\onecolumn
%
%
%
%
\renewcommand*{\arraystretch}{1.2}
\begin{longtable}{ lllll ll }
\caption{\label{4u1907:tab:swift_xrt_log}{\it Swift}/XRT observation log for 4U~J1907$+$097.}\\
\hline\hline
\small
 Sequence        & MJD                            &     Phase   & Start time  (UT)                     & End time   (UT)                      & Exposure  & Flux    \\
                        &                                   &                 & (yyyy-mm-dd hh:mm:ss)     & (yyyy-mm-dd hh:mm:ss)     &(s)               &             \\
\hline
\endfirsthead
\caption{continued.}\\
\hline
 Sequence        & MJD                            &     Phase   & Start time  (UT)                     & End time   (UT)                      & Exposure  & Flux    \\
\hline
\endhead
\hline
\endfoot
%
%
00033483001    &       57069.19720     &       0.96    &       2015-02-16 04:36:01     &       2015-02-16 04:51:54     &       953    & 	 $13.3^{+2.4}_{-2.0}$ \\
00033483003     &       57076.77906     &       0.87    &       2015-02-23 18:34:47     &       2015-02-23 18:48:54     &       847    & 	 $28.3^{+7.7}_{-6.2}$ \\
00033483004     &       57079.45028     &       0.19    &       2015-02-26 10:39:53     &       2015-02-26 10:56:54     &       1020   & 	 $14.6^{+2.2}_{-1.9}$ \\
00033483005     &       57083.24917     &       0.64    &       2015-03-02 05:49:41     &       2015-03-02 06:07:54     &       1093   & 	 $69.8^{+7.8}_{-7.0}$ \\
00033483006     &       57086.50529     &       0.03    &       2015-03-05 11:59:17     &       2015-03-05 12:15:55     &       998    & 	 $9.5^{+2.1}_{-1.7}$ \\
00033483007     &       57090.35591     &       0.49    &       2015-03-09 08:24:05     &       2015-03-09 08:40:55     &       1010   & 	 $2.8^{+0.7}_{-0.5}$ \\
00033483008     &       57093.95062     &       0.92    &       2015-03-12 22:40:50     &       2015-03-12 22:56:55     &       965    & 	 $17.2^{+4.1}_{-3.3}$ \\
00033483009     &       57097.34327     &       0.32    &       2015-03-16 08:06:41     &       2015-03-16 08:21:56     &       915    & 	 $21.3^{+3.0}_{-2.6}$ \\
00033483010     &       57100.28330     &       0.67    &       2015-03-19 06:39:57     &       2015-03-19 06:55:55     &       958    & 	 $27.2^{+3.9}_{-3.3}$ \\
00033483011     &       57104.94361     &       0.23    &       2015-03-23 22:30:39     &       2015-03-23 22:46:56     &       978    & 	 $19.0^{+2.7}_{-2.3}$ \\
00033483012     &       57107.87126     &       0.58    &       2015-03-26 20:46:19     &       2015-03-26 21:02:54     &       995    & 	 $63.9^{+7.6}_{-6.8}$ \\
00033483014     &       57114.66061     &       0.39    &       2015-04-02 15:43:43     &       2015-04-02 15:58:50     &       908    & 	 $41.9^{+5.8}_{-5.0}$ \\
00033483015     &       57118.05140     &       0.80    &       2015-04-06 01:05:06     &       2015-04-06 01:22:54     &       1068   & 	 $17.4^{+3.6}_{-3.0}$ \\
00033483016     &       57121.11656     &       0.16    &       2015-04-09 02:39:45     &       2015-04-09 02:55:56     &       970    & 	 $1.5^{+0.4}_{-0.3}$ \\
00033483017     &       57125.11082     &       0.64    &       2015-04-12 07:28:14     &       2015-04-13 21:50:54     &       1336   & 	 $35.9^{+3.8}_{-3.4}$ \\
00033483018     &       57128.81887     &       0.08    &       2015-04-16 19:30:25     &       2015-04-16 19:47:55     &       1051   & 	 $40.6^{+4.2}_{-3.8}$ \\
00033483019     &       57132.02167     &       0.46    &       2015-04-20 00:22:30     &       2015-04-20 00:39:56     &       1043   & 	 $6.8^{+1.3}_{-1.1}$ \\
00033483020     &       57135.02061     &       0.82    &       2015-04-23 00:21:26     &       2015-04-23 00:37:54     &       988    & 	 $1.1^{+0.4}_{-0.3}$ \\
00033483021     &       57139.20940     &       0.32    &       2015-04-27 04:53:09     &       2015-04-27 05:09:54     &       1005   & 	 $0.6^{+0.4}_{-0.3}$ \\
00033483022     &       57142.33898     &       0.70    &       2015-04-30 07:59:21     &       2015-04-30 08:16:54     &       1053   & 	 $63.2^{+8.9}_{-7.8}$ \\
00033483023     &       57146.33073     &       0.17    &       2015-05-04 07:47:35     &       2015-05-04 08:04:55     &       1041   & 	 $36.1^{+4.6}_{-4.1}$ \\
00033483024     &       57149.25551     &       0.52    &       2015-05-07 05:55:57     &       2015-05-07 06:19:54     &       1437   & 	 $49.1^{+5.3}_{-4.8}$ \\
00033483025     &       57153.17402     &       0.99    &       2015-05-11 04:10:32     &       2015-05-11 04:30:55     &       5        &    -- \\ 
00033483026     &       57156.70637     &       0.41    &       2015-05-14 16:49:25     &       2015-05-14 17:04:55     &       930    & 	 $0.5^{+0.3}_{-0.2}$ \\
00033483027     &       57160.75299     &       0.89    &       2015-05-18 17:56:41     &       2015-05-18 18:11:56     &       915    & 	 $14.5^{+2.6}_{-2.1}$ \\
00033483028     &       57163.65917     &       0.24    &       2015-05-21 14:59:29     &       2015-05-21 16:38:54     &       908    & 	 $13.4^{+1.8}_{-1.6}$ \\
00033483029     &       57167.87186     &       0.74    &       2015-05-25 20:50:01     &       2015-05-25 21:00:55     &       654    & 	 $22.8^{+4.6}_{-3.8}$ \\
00033483030     &       57170.07622     &       0.01    &       2015-05-28 01:44:37     &       2015-05-28 01:54:56     &       617    & 	 $26.4^{+5.1}_{-4.2}$ \\
00033483031     &       57174.06796     &       0.48    &       2015-06-01 01:29:49     &       2015-06-01 01:45:54     &       965    & 	 $12.5^{+2.4}_{-2.0}$ \\
00033483033     &       57181.70856     &       0.40    &       2015-06-08 16:47:44     &       2015-06-08 17:12:54     &       1509   & 	 $91.4^{+8.8}_{-8.0}$ \\
00033483034     &       57184.38300     &       0.72    &       2015-06-11 07:32:07     &       2015-06-11 10:50:54     &       1056   & 	 $42.1^{+4.3}_{-3.9}$ \\
00033483035     &       57188.64007     &       0.22    &       2015-06-15 15:14:28     &       2015-06-15 15:28:55     &       868    & 	 $2.9^{+0.7}_{-0.5}$ \\
00033483036     &       57191.42814     &       0.56    &       2015-06-18 10:08:07     &       2015-06-18 10:24:55     &       1008   & 	 $32.6^{+3.8}_{-3.4}$ \\
00033483037     &       57195.15266     &       0.00    &       2015-06-22 03:31:44     &       2015-06-22 03:47:54     &       970    & 	 $42.2^{+5.2}_{-4.7}$ \\
00033483038     &       57198.68399     &       0.42    &       2015-06-25 16:16:57     &       2015-06-25 16:32:55     &       958    & 	 $20.0^{+3.3}_{-2.8}$ \\
00033483039     &       57202.91001     &       0.93    &       2015-06-29 20:59:52     &       2015-06-29 22:40:57     &       955    & 	 $25.5^{+3.8}_{-3.3}$ \\
00033483040     &       57205.11780     &       0.19    &       2015-07-02 02:42:20     &       2015-07-02 02:56:55     &       875    & 	 $20.2^{+3.0}_{-2.6}$ \\
00033483041     &       57209.84092     &       0.76    &       2015-07-06 20:02:54     &       2015-07-06 20:18:55     &       960    & 	 $33.0^{+5.4}_{-4.6}$ \\
00033483042     &       57212.30136     &       0.05    &       2015-07-09 07:08:01     &       2015-07-09 07:19:53     &       712    & 	 $42.1^{+5.0}_{-4.4}$ \\
00033483043     &       57216.97833     &       0.61    &       2015-07-13 23:24:41     &       2015-07-13 23:32:53     &       491    & 	 $50.1^{+8.7}_{-7.2}$ \\
00033483044     &       57219.17442     &       0.87    &       2015-07-16 04:08:27     &       2015-07-16 04:13:53     &       326    & 	 $16.9^{+4.9}_{-3.7}$ \\
00033483045     &       57223.31399     &       0.36    &       2015-07-20 04:56:23     &       2015-07-20 10:07:54     &       762    & 	 $12.0^{+2.3}_{-2.0}$ \\
00033483046     &       57226.51238     &       0.75    &       2015-07-23 06:55:46     &       2015-07-23 17:39:52     &       933    & 	 $38.2^{+4.4}_{-3.9}$ \\
00033483047     &       57230.33604     &       0.20    &       2015-07-27 07:55:52     &       2015-07-27 08:11:55     &       963    & 	 $17.7^{+3.2}_{-2.7}$ \\
00033483048     &       57233.28949     &       0.56    &       2015-07-30 02:46:48     &       2015-07-30 11:06:55     &       905    & 	 $21.8^{+3.4}_{-2.9}$ \\
00033483049     &       57237.31020     &       0.04    &       2015-08-03 07:19:28     &       2015-08-03 07:33:53     &       865    & 	 $11.8^{+2.1}_{-1.7}$ \\
00033483050     &       57240.04352     &       0.36    &       2015-08-06 00:54:24     &       2015-08-06 01:10:55     &       990    & 	 $10.6^{+2.9}_{-2.2}$ \\
00033483051     &       57244.70114     &       0.92    &       2015-08-10 16:41:22     &       2015-08-10 16:57:54     &       993    & 	 $5.3^{+1.2}_{-0.9}$ \\
00033483052     &       57247.73589     &       0.28    &       2015-08-13 12:08:28     &       2015-08-13 23:10:53     &       888    & 	 $17.9^{+2.5}_{-2.2}$ \\
00033483053     &       57251.75651     &       0.76    &       2015-08-17 18:00:50     &       2015-08-17 18:17:53     &       1023   & 	 $32.7^{+4.6}_{-4.0}$ \\
00033483054     &       57254.88463     &       0.13    &       2015-08-20 21:05:50     &       2015-08-20 21:21:53     &       963    & 	 $24.8^{+3.0}_{-2.7}$ \\
00033483055     &       57258.20110     &       0.53    &       2015-08-24 04:41:15     &       2015-08-24 04:57:53     &       998    & 	 $56.0^{+7.6}_{-6.6}$ \\
00033483056     &       57261.38728     &       0.91    &       2015-08-27 09:08:27     &       2015-08-27 09:26:53     &       1106   & 	 $20.8^{+4.1}_{-3.3}$ \\
00033483057     &       57265.72559     &       0.43    &       2015-08-31 17:16:46     &       2015-08-31 17:32:56     &       970    & 	 $36.5^{+4.2}_{-3.8}$ \\
00033483058     &       57268.24792     &       0.73    &       2015-09-03 04:16:07     &       2015-09-03 07:37:53     &       978    & 	 $25.7^{+4.0}_{-3.4}$ \\
00033483059     &       57273.62788     &       0.37    &       2015-09-08 13:15:21     &       2015-09-08 16:52:55     &       1156   & 	 $4.1^{+0.9}_{-0.7}$ \\
00033483060     &       57275.08861     &       0.55    &       2015-09-10 01:59:16     &       2015-09-10 02:15:54     &       998    & 	--  \\ 
00033483061     &       57279.21154     &       0.04    &       2015-09-14 04:56:19     &       2015-09-14 05:12:55     &       995    & 	 $3.8^{+0.7}_{-0.6}$ \\
00033483062     &       57282.26433     &       0.40    &       2015-09-17 06:20:29     &       2015-09-17 06:20:46     &       18      &     -- \\   
00033483063     &       57286.87282     &       0.95    &       2015-09-21 20:48:49     &       2015-09-21 21:04:54     &       965    & 	 $41.8^{+5.6}_{-4.9}$ \\
00033483064     &       57289.39770     &       0.25    &       2015-09-24 09:23:28     &       2015-09-24 09:41:54     &       1106   & 	 $30.2^{+4.2}_{-3.7}$ \\
00033483065     &       57293.24991     &       0.71    &       2015-09-28 05:50:49     &       2015-09-28 06:08:55     &       1086   & 	 $33.7^{+4.2}_{-3.8}$ \\

%


  \end{longtable}
%
%
%
%
\begin{longtable}{ lllll ll }
 \caption{{\it Swift}/XRT observation log for IGR~J19140$+$0951.\label{i19140:tab:swift_xrt_log}}\\
 \hline
 \hline
  \small
 Sequence        & MJD                            &     Phase   & Start time  (UT)                     & End time   (UT)                      & Exposure  & Flux   \\
                        &                                   &                 & (yyyy-mm-dd hh:mm:ss)     & (yyyy-mm-dd hh:mm:ss)     &(s)               &           \\
 \hline
\endfirsthead
\caption{continued.}\\
\hline
Sequence        & MJD                            &     Phase   & Start time  (UT)                     & End time   (UT)                      & Exposure  & Flux   \\
\hline
\endhead
\hline
\endfoot
%
%
00030393003	&	57071.13152	&	0.65		&	2015-02-18 03:00:51	&	2015-02-18 03:17:54	&	1023& $3.1^{+0.9}_{-0.7}$ \\
00030393004	&	57074.05215	&	0.86		&	2015-02-21 01:06:15	&	2015-02-21 01:23:56	&	1061& $0.7^{+0.6}_{-0.3}$ \\
00030393005	&	57078.11378	&	0.16		&	2015-02-25 02:35:45	&	2015-02-25 02:51:55	&	970	& $0.3^{+0.2}_{-0.1}$ \\
00030393006	&	57081.17754	&	0.39		&	2015-02-28 04:07:22	&	2015-02-28 04:23:55	&	993	& $11.4^{+2.0}_{-1.7}$ \\
00030393007	&	57085.50884	&	0.71		&	2015-03-04 12:04:32	&	2015-03-04 12:20:55	&	983	& $4.9^{+1.7}_{-1.2}$ \\
00030393008	&	57088.50679	&	0.93		&	2015-03-07 12:01:37	&	2015-03-07 12:17:55	&	925	& $0.6^{+0.8}_{-0.3}$ \\
00030393009	&	57092.50049	&	0.22		&	2015-03-11 11:51:30	&	2015-03-11 12:09:53	&	1103& $4.3^{+1.2}_{-0.9}$ \\
00030393010	&	57095.29009	&	0.43		&	2015-03-14 06:49:31	&	2015-03-14 07:05:56	&	985	& $4.4^{+1.2}_{-0.9}$ \\
00030393011	&	57099.67249	&	0.75		&	2015-03-18 15:59:50	&	2015-03-18 16:16:56	&	1025& $4.7^{+1.0}_{-0.8}$ \\
00030393012	&	57102.74830	&	0.98		&	2015-03-21 17:49:24	&	2015-03-21 18:05:54	&	978	& 	-- \\  
00030393013	&	57106.13932	&	0.23		&	2015-03-25 03:12:37	&	2015-03-25 03:28:54	&	960	& $7.9^{+1.6}_{-1.4}$ \\
00030393014	&	57109.73308	&	0.50		&	2015-03-28 17:27:19	&	2015-03-28 17:43:57	&	998	& $10.0^{+2.0}_{-1.6}$ \\
00030393015	&	57113.91388	&	0.80		&	2015-04-01 21:47:02	&	2015-04-01 22:04:55	&	1073& 	-- \\  
00030393016	&	57116.84270	&	0.02		&	2015-04-04 20:05:02	&	2015-04-04 20:21:55	&	1013& $4.7^{+0.9}_{-0.8}$ \\
00030393017	&	57120.50776	&	0.29		&	2015-04-08 12:00:26	&	2015-04-08 12:21:54	&	1158& $23.0^{+3.9}_{-3.3}$ \\
00030393018	&	57123.03947	&	0.48		&	2015-04-11 00:48:43	&	2015-04-11 01:04:56	&	973	& $3.2^{+1.1}_{-0.8}$ \\
00030393019	&	57127.82063	&	0.83		&	2015-04-15 19:32:31	&	2015-04-15 19:50:54	&	1103& $9.9^{+1.5}_{-1.3}$ \\
00030393020	&	57130.02916	&	0.99		&	2015-04-18 00:39:02	&	2015-04-18 00:44:56	&	354	& $7.8^{+3.1}_{-2.1}$ \\
00030393021	&	57134.08242	&	0.29		&	2015-04-22 01:50:27	&	2015-04-22 02:06:54	&	988	& $19.7^{+5.3}_{-4.1}$ \\
00030393022	&	57137.54770	&	0.55		&	2015-04-25 13:00:27	&	2015-04-25 13:16:55	&	988	& $5.3^{+1.2}_{-0.9}$ \\
00030393023	&	57141.53064	&	0.84		&	2015-04-29 12:35:19	&	2015-04-29 12:52:55	&	1056& $6.0^{+1.3}_{-1.0}$ \\
00030393024	&	57144.12763	&	0.03		&	2015-05-02 02:55:41	&	2015-05-02 03:11:54	&	973	& 	-- \\  
00030393025	&	57148.05961	&	0.32		&	2015-05-06 01:17:45	&	2015-05-06 01:33:55	&	970	& $7.7^{+2.1}_{-1.6}$ \\
00030393026	&	57151.78003	&	0.60		&	2015-05-09 16:53:34	&	2015-05-09 20:32:54	&	1364& $7.9^{+1.4}_{-1.2}$ \\
00030393027	&	57155.04283	&	0.84		&	2015-05-13 00:53:26	&	2015-05-13 01:09:54	&	988	& $0.4^{+0.5}_{-0.2}$ \\
00030393028	&	57158.82522	&	0.12		&	2015-05-16 19:44:42	&	2015-05-16 19:51:55	&	434	&  -- \\ 
00030393029	&	57162.89551	&	0.42		&	2015-05-20 21:28:09	&	2015-05-20 21:30:54	&	165	&  -- \\ 
00030393030	&	57165.87459	&	0.64		&	2015-05-23 20:51:53	&	2015-05-23 21:06:55	&	903	& $2.9^{+1.1}_{-0.8}$	\\
00030393031	&	57169.87156	&	0.93		&	2015-05-27 20:48:10	&	2015-05-27 21:01:55	&	825	& 	-- \\  
00030393033	&	57176.66578	&	0.43		&	2015-06-03 15:48:33	&	2015-06-03 16:08:54	&	1221& $1.5^{+0.7}_{-0.4}$ \\
00030393034	&	57179.78132	&	0.66		&	2015-06-06 18:36:16	&	2015-06-06 18:53:54	&	1058& $4.1^{+1.0}_{-0.8}$ \\
00030393035	&	57183.21927	&	0.92		&	2015-06-10 04:21:34	&	2015-06-10 06:09:55	&	722	& $0.7^{+0.7}_{-0.3}$ \\
00030393036	&	57186.50262	&	0.16		&	2015-06-13 11:54:38	&	2015-06-13 12:12:54	&	1096& $0.9^{+0.6}_{-0.3}$ \\
00030393037	&	57190.17122	&	0.43		&	2015-06-17 03:59:12	&	2015-06-17 04:13:54	&	883	& 	-- \\  
00030393038	&	57193.93137	&	0.71		&	2015-06-20 21:31:26	&	2015-06-20 23:10:54	&	1108& $0.03^{+0.03}_{-0.02}$ 	 \\         
00030393039	&	57197.54561	&	0.97		&	2015-06-24 12:55:27	&	2015-06-24 13:15:54	&	1226& $17.6^{+3.2}_{-2.7}$ \\
00030393040	&	57200.63216	&	0.20		&	2015-06-27 14:24:41	&	2015-06-27 15:55:54	&	1294& $5.6^{+1.4}_{-1.1}$ \\
00030393041	&	57204.52040	&	0.49		&	2015-07-01 12:20:50	&	2015-07-01 12:37:53	&	1003& $3.0^{+1.0}_{-0.7}$ \\
00030393042	&	57207.51383	&	0.71		&	2015-07-04 12:10:56	&	2015-07-04 12:28:54	&	1078& $5.8^{+1.2}_{-1.0}$ \\
00030393043	&	57211.71016	&	0.02		&	2015-07-08 16:54:20	&	2015-07-08 17:10:55	&	995	& $1.3^{+1.2}_{-0.8}$ \\
00030393044	&	57214.07191	&	0.19		&	2015-07-11 01:08:13	&	2015-07-11 02:18:52	&	918	& $2.7^{+0.9}_{-0.7}$ \\
00030393045	&	57218.44256	&	0.52		&	2015-07-15 10:35:40	&	2015-07-15 10:38:54	&	193	&    -- \\  
00030393046	&	57221.76517	&	0.76		&	2015-07-18 18:14:48	&	2015-07-18 18:28:53	&	845	& $0.5^{+0.3}_{-0.2}$ \\
00030393047	&	57225.28945	&	0.02		&	2015-07-22 06:48:41	&	2015-07-22 07:04:54	&	973	& $2.8^{+0.7}_{-0.6}$ \\
00030393048	&	57228.40191	&	0.25		&	2015-07-25 09:31:04	&	2015-07-25 09:46:27	&	920	& $28.4^{+5.2}_{-4.3}$ \\
00030393049	&	57232.40124	&	0.55		&	2015-07-29 09:29:40	&	2015-07-29 09:45:53	&	973	& $6.0^{+1.6}_{-1.2}$ \\
00030393050	&	57235.62146	&	0.78		&	2015-08-01 10:37:54	&	2015-08-01 19:11:53	&	1038& 	-- \\ 
00030393051	&	57239.04772	&	0.04		&	2015-08-05 01:03:30	&	2015-08-05 01:13:55	&	624	& 	-- \\ 
00030393052	&	57242.03565	&	0.26		&	2015-08-08 00:42:47	&	2015-08-08 00:59:53	&	1025& $27.1^{+6.3}_{-5.1}$ \\
00030393053	&	57246.23508	&	0.57		&	2015-08-12 05:30:08	&	2015-08-12 05:46:53	&	1005& $1.8^{+0.9}_{-0.6}$ \\
00030393054	&	57249.03230	&	0.77		&	2015-08-15 00:38:08	&	2015-08-15 00:54:53	&	1005& $0.9^{+0.4}_{-0.3}$ \\
00030393055	&	57253.55333	&	0.11		&	2015-08-19 13:05:40	&	2015-08-19 13:27:54	&	1334& $1.5^{+0.5}_{-0.4}$ \\
00030393056	&	57256.47959	&	0.32		&	2015-08-22 11:21:20	&	2015-08-22 11:39:53	&	1113& $4.1^{+0.9}_{-0.7}$ \\
00030393057	&	57260.33788	&	0.61		&	2015-08-26 07:58:11	&	2015-08-26 08:14:54	&	1003& $0.8^{+0.4}_{-0.2}$ \\
00030393058	&	57263.36187	&	0.83		&	2015-08-29 07:53:17	&	2015-08-29 09:28:53	&	968	& $2.0^{+0.7}_{-0.5}$ \\
00030393059	&	57267.31970	&	0.12		&	2015-09-02 07:30:49	&	2015-09-02 07:49:55	&	1146& $1.5^{+0.9}_{-0.6}$ \\
00030393061	&	57274.10016	&	0.62		&	2015-09-09 02:08:32	&	2015-09-09 02:39:54	&	1063& $4.9^{+1.1}_{-0.9}$ \\
00030393062	&	57277.89862	&	0.90		&	2015-09-12 21:30:08	&	2015-09-12 21:37:54	&	466	& $3.3^{+1.2}_{-0.9}$ \\
00030393063	&	57281.68750	&	0.18		&	2015-09-16 16:22:04	&	2015-09-16 16:37:55	&	950	& $0.9^{+0.4}_{-0.3}$ \\
00030393064	&	57284.27302	&	0.37		&	2015-09-19 06:14:24	&	2015-09-19 06:51:53	&	1033& $5.9^{+1.8}_{-1.3}$ \\
00030393065	&	57288.07229	&	0.65		&	2015-09-23 00:03:16	&	2015-09-23 03:24:55	&	797	& $7.8^{+1.5}_{-1.2}$ \\
00030393066	&	57291.92352	&	0.94		&	2015-09-26 22:01:49	&	2015-09-26 22:17:54	&	965	& $1.6^{+0.6}_{-0.4}$ \\
00030393067	&	57295.78035	&	0.22		&	2015-09-30 18:35:30	&	2015-09-30 18:51:53	&	983	& $3.9^{+1.0}_{-0.7}$ \\
  \end{longtable}

%
%
%
%
%
%
%
\renewcommand*{\arraystretch}{1.2}
\begin{longtable}{ lllll ll }
\caption{{\it Swift}/XRT observation log for IGR~J16393$-$4643.\label{i16393:tab:swift_xrt_log}}\\
\hline
\hline
\small
 Sequence        & MJD                            &     Phase   & Start time  (UT)                     & End time   (UT)                      & Exposure  & Flux     \\
                        &                                   &                 & (yyyy-mm-dd hh:mm:ss)     & (yyyy-mm-dd hh:mm:ss)     &(s)               &            \\
 \hline
\endfirsthead
\caption{continued.}\\
\hline
 Sequence        & MJD                            &     Phase   & Start time  (UT)                     & End time   (UT)                      & Exposure  & Flux     \\
 \hline
 \endhead
 \hline
 \endfoot
00034135001	&	57408.33228	&	0.49	&	2016-01-21 07:50:05	&	2016-01-21 08:06:53	&	1008		 & $1.9^{+0.7}_{-0.5}$ \\
00034135002	&	57409.06321	&	0.66	&	2016-01-22 01:23:07	&	2016-01-22 01:38:55	&	948		 & $0.6^{+0.9}_{-0.3}$ \\ 
00034135003	&	57410.47124	&	0.99	&	2016-01-23 11:10:17	&	2016-01-23 11:26:53	&	995	& -- \\	
00034135004	&	57411.19096	&	0.16	&	2016-01-24 04:26:05	&	2016-01-24 04:43:53	&	1068		 & $2.4^{+0.8}_{-0.6}$ \\
00034135005	&	57412.31988	&	0.43	&	2016-01-25 07:32:22	&	2016-01-25 07:48:53	&	990		 & $4.6^{+1.1}_{-0.9}$ \\
00034135006	&	57413.31668	&	0.67	&	2016-01-26 07:28:08	&	2016-01-26 07:43:53	&	945		 & $2.8^{+1.0}_{-0.8}$ \\
00034135008	&	57415.38120	&	0.15	&	2016-01-28 08:59:56	&	2016-01-28 09:17:55	&	1078		 & $2.3^{+1.1}_{-0.7}$ \\
00034135009	&	57416.10815	&	0.32	&	2016-01-29 02:27:54	&	2016-01-29 02:43:34	&	940		 & $4.8^{+1.2}_{-1.0}$ \\
00034135010	&	57417.24288	&	0.59	&	2016-01-30 05:41:35	&	2016-01-30 05:57:55	&	980		 & $3.0^{+0.9}_{-0.7}$ \\
00034135011	&	57418.04097	&	0.78	&	2016-01-31 00:51:04	&	2016-01-31 01:06:54	&	950		 & $1.9^{+0.9}_{-0.6}$ \\
00034135012	&	57419.23200	&	0.06	&	2016-02-01 05:25:16	&	2016-02-01 05:42:54	&	1058		 & $2.6^{+1.1}_{-0.7}$ \\
00034135013	&	57420.83151	&	0.44	&	2016-02-02 19:48:49	&	2016-02-02 20:05:55	&	1025		 & $3.4^{+1.0}_{-0.8}$ \\
00034135014	&	57421.03322	&	0.49	&	2016-02-03 00:38:46	&	2016-02-03 00:56:54	&	1088		 & $2.9^{+0.9}_{-0.7}$ \\
00034135015	&	57422.49361	&	0.83	&	2016-02-04 03:47:43	&	2016-02-04 19:53:53	&	1785		 & $2.1^{+0.8}_{-0.6}$ \\
00034135016	&	57423.22185	&	0.00	&	2016-02-05 05:11:01	&	2016-02-05 05:27:54	&	1013		 & $7.6^{+1.9}_{-1.6}$ \\
00034135017	&	57424.94619	&	0.41	&	2016-02-06 22:35:08	&	2016-02-06 22:49:53	&	885		 & $3.9^{+2.9}_{-1.8}$ \\
00034135018	&	57425.67767	&	0.58	&	2016-02-07 16:06:46	&	2016-02-07 16:24:54	&	1088		 & $2.5^{+0.8}_{-0.6}$ \\
00034135019	&	57426.87456	&	0.86	&	2016-02-08 20:50:49	&	2016-02-08 21:07:54	&	1025		 & $0.1^{+0.3}_{-0.1}$ \\
00034135020	&	57427.80319	&	0.08	&	2016-02-09 19:09:18	&	2016-02-09 19:23:53	&	875		 & $2.2^{+1.1}_{-0.7}$ \\
00034135021	&	57428.14085	&	0.16	&	2016-02-10 03:14:44	&	2016-02-10 03:30:54	&	970		 & $2.7^{+0.9}_{-0.7}$ \\
00034135022	&	57429.14492	&	0.40	&	2016-02-11 00:16:27	&	2016-02-11 06:40:54	&	953		 & $2.9^{+0.9}_{-0.7}$ \\
00034135023	&	57430.10943	&	0.63	&	2016-02-12 00:03:16	&	2016-02-12 05:11:53	&	1339         & $2.3^{+0.6}_{-0.5}$ \\
00034135024	&	57431.37563	&	0.93	&	2016-02-13 08:07:54	&	2016-02-13 09:53:54	&	1013 & -- \\	
00034135025	&	57432.07308	&	0.09	&	2016-02-14 01:38:34	&	2016-02-14 01:51:54	&	800		 & $3.3^{+1.3}_{-0.9}$ \\
00034135026	&	57433.26786	&	0.37	&	2016-02-15 06:18:31	&	2016-02-15 06:32:54	&	863	 & -- \\	
00034135027	&	57434.36777	&	0.63	&	2016-02-16 01:35:15	&	2016-02-16 16:03:54	&	890		 & $2.4^{+0.9}_{-0.6}$ \\
00034135028	&	57435.79345	&	0.97	&	2016-02-17 18:55:15	&	2016-02-17 19:09:53	&	878		 & $3.3^{+1.2}_{-0.9}$ \\
00034135029	&	57436.78924	&	0.20	&	2016-02-18 18:49:38	&	2016-02-18 19:03:21	&	822		 & $2.8^{+1.4}_{-0.9}$ \\
00034135030	&	57437.78604	&	0.44	&	2016-02-19 18:42:51	&	2016-02-19 19:00:55	&	1083		 & $4.5^{+1.3}_{-1.0}$ \\
00034135031	&	57438.78265	&	0.67	&	2016-02-20 18:38:08	&	2016-02-20 18:55:54	&	1066		 & $1.2^{+0.5}_{-0.4}$ \\
00034135032	&	57439.77768	&	0.91	&	2016-02-21 18:30:49	&	2016-02-21 18:48:52	&	1083		 & $1.3^{+0.8}_{-0.5}$ \\
00034135033	&	57440.77698	&	0.14	&	2016-02-22 18:30:46	&	2016-02-22 18:46:56	&	970		 & $4.3^{+1.9}_{-1.2}$ \\
00034135034	&	57441.70582	&	0.36	&	2016-02-23 16:47:51	&	2016-02-23 17:04:54	&	1023		 & $2.7^{+0.9}_{-0.7}$ \\
00034135035	&	57442.09999	&	0.46	&	2016-02-24 02:15:03	&	2016-02-24 02:32:54	&	1071		 & $2.3^{+0.7}_{-0.5}$ \\
00034135036	&	57443.68719	&	0.83	&	2016-02-25 16:22:12	&	2016-02-25 16:36:52	&	880	 & -- \\	
00034135037	&	57444.41879	&	0.00	&	2016-02-26 09:55:11	&	2016-02-26 10:10:54	&	943		 & $2.4^{+1.0}_{-0.7}$ \\
00034135038	&	57445.23192	&	0.20	&	2016-02-27 05:32:02	&	2016-02-27 05:35:52	&	231	  & -- \\ 
00034135039	&	57446.35042	&	0.46	&	2016-02-28 03:34:19	&	2016-02-28 13:14:53	&	1371		 & $2.6^{+0.8}_{-0.6}$ \\
00034135040	&	57446.75943	&	0.56	&	2016-02-28 03:52:15	&	2016-02-29 08:34:53	&	1356		 & $2.9^{+0.8}_{-0.6}$ \\
00034135041	&	57448.60692	&	0.99	&	2016-03-01 14:26:00	&	2016-03-01 14:41:55	&	955		 & $2.7^{+1.3}_{-0.8}$ \\
00034135042	&	57449.35112	&	0.17	&	2016-03-02 08:19:19	&	2016-03-02 08:31:54	&	755		 & $5.9^{+1.9}_{-1.4}$ \\
00034135044	&	57451.40197	&	0.65	&	2016-03-04 09:30:46	&	2016-03-04 09:46:54	&	968		 & $2.8^{+1.0}_{-0.7}$ \\
00034135045	&	57452.60000	&	0.93	&	2016-03-05 14:15:07	&	2016-03-05 14:32:52	&	1066		 & $1.2^{+1.0}_{-0.7}$ \\
00034135046	&	57453.73040	&	0.20	&	2016-03-06 17:23:37	&	2016-03-06 17:39:55	&	978		 & $3.2^{+1.5}_{-1.1}$ \\
00034135047	&	57454.26308	&	0.33	&	2016-03-07 06:10:46	&	2016-03-07 06:26:54	&	968		 & $4.5^{+1.4}_{-1.1}$ \\
00034135048	&	57455.49565	&	0.62	&	2016-03-08 11:09:33	&	2016-03-08 12:37:54	&	953		 & $2.9^{+1.0}_{-0.8}$ \\
00034135049	&	57456.12555	&	0.77	&	2016-03-09 02:53:41	&	2016-03-09 03:07:53	&	852		 & $2.6^{+1.2}_{-0.8}$ \\
00034135050	&	57457.12260	&	0.00	&	2016-03-10 02:49:10	&	2016-03-10 03:03:53	&	883		 & $3.9^{+1.4}_{-1.0}$ \\
00034135051	&	57458.39444	&	0.30	&	2016-03-11 09:19:05	&	2016-03-11 09:36:53	&	1068		 & $1.4^{+0.5}_{-0.3}$ \\
00034135052	&	57459.25599	&	0.51	&	2016-03-12 06:00:21	&	2016-03-12 06:16:54	&	993		 & $3.0^{+1.2}_{-0.9}$ \\
00034135053	&	57460.31769	&	0.76	&	2016-03-13 07:30:02	&	2016-03-13 07:44:55	&	893		 & $2.6^{+1.1}_{-0.8}$ \\
00034135054	&	57461.44428	&	0.02	&	2016-03-14 10:31:36	&	2016-03-14 10:47:54	&	978		 & $1.9^{+0.8}_{-0.5}$ \\
00034135055	&	57462.83928	&	0.35	&	2016-03-15 20:00:13	&	2016-03-15 20:16:53	&	1000		 & $4.3^{+1.1}_{-0.9}$ \\
00034135056	&	57463.90409	&	0.60	&	2016-03-16 21:33:51	&	2016-03-16 21:49:54	&	963		 & $2.1^{+0.8}_{-0.5}$ \\
00034135057	&	57464.63881	&	0.78	&	2016-03-17 15:11:52	&	2016-03-17 15:27:53	&	960		 & $0.6^{+0.6}_{-0.3}$ \\
00034135058	&	57465.33851	&	0.94	&	2016-03-18 07:27:00	&	2016-03-18 08:47:54	&	812		 & $1.8^{+0.8}_{-1.1}$ \\
00034135059	&	57466.33907	&	0.18	&	2016-03-19 00:52:37	&	2016-03-19 15:23:53	&	1041		 & $2.5^{+1.1}_{-0.7}$ \\
00034135060	&	57467.16779	&	0.37	&	2016-03-20 03:53:19	&	2016-03-20 04:09:54	&	995		 & $0.5^{+0.8}_{-0.3}$ \\
00034135061	&	57468.01586	&	0.57	&	2016-03-21 00:22:35	&	2016-03-21 00:38:52	&	30  & -- \\ 
00034135062	&	57469.62087	&	0.95	&	2016-03-22 14:47:12	&	2016-03-22 15:00:54	&	822		 & $1.8^{+1.1}_{-0.7}$ \\
00034135063	&	57470.08529	&	0.06	&	2016-03-23 02:02:35	&	2016-03-23 02:18:53	&	28	 & -- \\	
00034135064	&	57471.88419	&	0.49	&	2016-03-24 21:05:35	&	2016-03-24 21:20:52	&	321	& -- \\	
00034135065	&	57472.87287	&	0.72	&	2016-03-25 20:56:39	&	2016-03-25 21:12:55	&	33	 & -- \\	
00034135066	&	57473.71689	&	0.92	&	2016-03-26 14:47:45	&	2016-03-26 19:36:54	&	802	 & -- \\	
00034135067	&	57474.24053	&	0.04	&	2016-03-27 00:08:48	&	2016-03-27 11:23:54	&	822		 & $2.8^{+1.0}_{-0.7}$ \\
00034135068	&	57475.72972	&	0.39	&	2016-03-28 17:22:41	&	2016-03-28 17:38:54	&	973		 & $3.1^{+1.2}_{-0.8}$ \\
00034135069	&	57476.60286	&	0.60	&	2016-03-29 12:53:21	&	2016-03-29 16:02:53	&	790		 & $3.1^{+1.3}_{-0.9}$ \\
00034135070	&	57477.00666	&	0.69	&	2016-03-30 00:03:16	&	2016-03-30 00:15:54	&	705		 & $3.7^{+1.8}_{-1.2}$ \\
00034135071	&	57478.58739	&	0.07	&	2016-03-31 13:56:47	&	2016-03-31 14:14:53	&	1086         & $2.7^{+0.9}_{-0.7}$ \\
00034135072	&	57479.78202	&	0.35	&	2016-04-01 18:37:19	&	2016-04-01 18:54:53	&	1053         & $2.1^{+0.7}_{-0.5}$ \\
00034135073	&	57480.06030	&	0.41	&	2016-04-02 01:18:47	&	2016-04-02 01:34:55	&	965		 & $2.6^{+0.9}_{-0.6}$ \\
00034135074	&	57481.70841	&	0.80	&	2016-04-03 16:52:18	&	2016-04-03 17:07:54	&	935		 & $2.2^{+1.2}_{-0.8}$ \\
00034135076	&	57483.76959	&	0.29	&	2016-04-05 18:19:29	&	2016-04-05 18:36:55	&	1046	& $2.9^{+1.0}_{-0.7}$ \\
00034135077	&	57484.97279	&	0.57	&	2016-04-06 23:12:44	&	2016-04-06 23:28:54	&	970		 & $1.9^{+0.9}_{-0.6}$ \\
00034135078	&	57485.04257	&	0.59	&	2016-04-07 00:53:43	&	2016-04-07 01:08:53	&	910		 & $15.6^{+4.9}_{-3.7}$ \\
00034135079	&	57486.07752	&	0.83	&	2016-04-08 00:59:21	&	2016-04-08 02:43:53	&	963	 & -- \\	
00034135080	&	57487.17528	&	0.09	&	2016-04-09 04:05:54	&	2016-04-09 04:18:54	&	780		 & $1.4^{+0.8}_{-0.5}$ \\
00034135081	&	57488.96274	&	0.51	&	2016-04-10 23:02:46	&	2016-04-10 23:09:55	&	429		 & $2.0^{+2.2}_{-1.0}$ \\
00034135082	&	57489.75842	&	0.70	&	2016-04-11 18:11:19	&	2016-04-11 18:12:55	&	95	 & -- \\	
00034135083	&	57490.69493	&	0.92	&	2016-04-12 13:25:28	&	2016-04-12 19:55:55	&	863		 & $1.7^{+1.0}_{-0.6}$ \\
00034135084	&	57491.48572	&	0.11	&	2016-04-13 11:31:12	&	2016-04-13 11:47:55	&	988		 & $3.6^{+1.1}_{-0.8}$ \\
00034135085	&	57492.94566	&	0.45	&	2016-04-14 22:33:36	&	2016-04-14 22:49:54	&	978		 & $4.0^{+1.2}_{-0.9}$ \\
00034135086	&	57493.01501	&	0.47	&	2016-04-15 00:13:19	&	2016-04-15 00:29:54	&	995		 & $2.8^{+0.8}_{-0.6}$ \\
00034135087	&	57494.08169	&	0.72	&	2016-04-16 01:50:21	&	2016-04-16 02:04:54	&	873		 & $3.1^{+1.3}_{-0.9}$ \\
00034135088	&	57495.39796	&	0.03	&	2016-04-17 09:26:12	&	2016-04-17 09:39:54	&	822		 & $1.2^{+0.7}_{-0.4}$ \\
00034135089	&	57496.72762	&	0.35	&	2016-04-18 17:18:38	&	2016-04-18 17:36:53	&	1096         & $3.8^{+1.0}_{-0.8}$ \\
00034135090	&	57497.92956	&	0.63	&	2016-04-19 22:10:13	&	2016-04-19 22:26:53	&	998		 & $2.7^{+1.0}_{-0.7}$ \\
00034135091	&	57498.39649	&	0.74	&	2016-04-20 09:22:59	&	2016-04-20 09:38:54	&	930		 & $1.9^{+1.0}_{-0.6}$ \\
00034135092	&	57499.13246	&	0.91	&	2016-04-21 03:05:36	&	2016-04-21 03:15:53	&	617	 & -- \\	
00034135093	&	57500.52289	&	0.24	&	2016-04-22 12:25:02	&	2016-04-22 12:40:53	&	950		 & $3.6^{+1.3}_{-0.9}$ \\
00034135094	&	57501.25418	&	0.42	&	2016-04-23 05:58:07	&	2016-04-23 06:13:55	&	948		 & $3.7^{+1.2}_{-0.9}$ \\
00034135095	&	57502.05761	&	0.60	&	2016-04-24 01:14:00	&	2016-04-24 01:31:53	&	1073		 & $4.5^{+1.1}_{-0.8}$ \\
00034135096	&	57503.38791	&	0.92	&	2016-04-25 09:09:16	&	2016-04-25 09:27:54	&	1118		 & $2.8^{+1.0}_{-0.8}$ \\
00034135097	&	57504.71843	&	0.23	&	2016-04-26 17:06:10	&	2016-04-26 17:22:53	&	1003		 & $1.4^{+0.5}_{-0.3}$ \\
00034135098	&	57505.31402	&	0.37	&	2016-04-27 07:23:12	&	2016-04-27 07:40:55	&	1046		 & $2.2^{+0.8}_{-0.6}$ \\
00034135099	&	57506.60238	&	0.68	&	2016-04-28 13:41:58	&	2016-04-28 15:12:53	&	1053		 & $2.1^{+1.1}_{-0.7}$ \\
00034135100	&	57507.57783	&	0.91	&	2016-04-29 13:43:15	&	2016-04-29 14:00:53	&	1058		 & $2.4^{+1.1}_{-0.8}$ \\
00034135101	&	57508.57364	&	0.14	&	2016-04-30 13:37:10	&	2016-04-30 13:54:53	&	1063  & -- \\	
00034135102	&	57509.56819	&	0.38	&	2016-05-01 13:29:27	&	2016-05-01 13:46:55	&	1048		 & $2.4^{+1.0}_{-0.7}$ \\
00034135103	&	57510.70142	&	0.64	&	2016-05-02 16:42:10	&	2016-05-02 16:57:53	&	943		 & $3.8^{+1.0}_{-0.8}$ \\
00034135104	&	57511.69894	&	0.88	&	2016-05-03 16:39:01	&	2016-05-03 16:53:54	&	893		 & $0.3^{+0.4}_{-0.1}$ \\
00034135105	&	57512.69870	&	0.12	&	2016-05-04 16:37:20	&	2016-05-04 16:54:53	&	1053		 & $4.4^{+1.4}_{-1.0}$ \\
00034135106	&	57513.21925	&	0.24	&	2016-05-05 05:08:31	&	2016-05-05 05:22:54	&	863		 & $1.0^{+0.7}_{-0.4}$ \\
00034135107	&	57514.28210	&	0.49	&	2016-05-06 06:38:33	&	2016-05-06 06:53:53	&	918		 & $3.2^{+1.0}_{-0.8}$ \\
00034135109	&	57516.60700	&	0.04	&	2016-05-08 14:26:16	&	2016-05-08 14:41:53	&	938	 & -- \\ 
00034135110	&	57517.74974	&	0.31	&	2016-05-09 17:52:18	&	2016-05-09 18:06:55	&	878		 & $11.9^{+4.0}_{-2.9}$ \\ 
00034135111	&	57518.66796	&	0.52	&	2016-05-10 15:54:50	&	2016-05-10 16:08:53	&	842		 & $5.1^{+1.5}_{-1.1}$ \\
00034135112	&	57519.60390	&	0.74	&	2016-05-11 14:25:18	&	2016-05-11 19:01:53	&	517		 & $1.9^{+1.4}_{-1.0}$ \\
00034135113	&	57520.66917	&	1.00	&	2016-05-12 15:55:18	&	2016-05-12 16:11:54	&	995		 & $3.8^{+1.4}_{-1.0}$ \\
00034135114	&	57521.07151	&	0.09	&	2016-05-13 01:35:03	&	2016-05-13 01:50:54	&	950		 & $1.5^{+0.7}_{-0.4}$ \\
00034135115	&	57522.71594	&	0.48	&	2016-05-14 17:09:48	&	2016-05-14 17:23:53	&	35	 & -- \\ 
00034135116	&	57523.31809	&	0.62	&	2016-05-15 07:30:11	&	2016-05-15 07:45:54	&	943		 & $3.0^{+1.0}_{-0.8}$ \\
00034135117	&	57524.31462	&	0.86	&	2016-05-16 07:24:11	&	2016-05-16 07:41:54	&	1063  & -- \\	
00034135118	&	57525.64240	&	0.17	&	2016-05-17 15:17:12	&	2016-05-17 15:32:53	&	940		 & $2.8^{+1.2}_{-0.8}$ \\
00034135119	&	57526.30889	&	0.33	&	2016-05-18 07:16:41	&	2016-05-18 07:32:54	&	973		 & $1.7^{+0.6}_{-0.4}$ \\
00034135120	&	57527.62959	&     0.64  &     2016-05-19 15:06:36	&	2016-05-19 15:22:56	&	350	  & -- \\  
00034135120	&	57527.63526	&	0.64	&	2016-05-19 15:06:40	&	2016-05-19 15:22:51	&	213	  & -- \\ 
00034135121	&	57528.08381	&	0.75	&	2016-05-20 01:07:28	&	2016-05-20 02:53:53	&	1133		 & $2.8^{+0.9}_{ -0.7}$ \\
00034135122	&	57529.27390	&	0.03	&	2016-05-21 02:44:55	&	2016-05-21 10:23:53	&	787		 & $1.6^{+1.5}_{ -0.7}$ \\
00034135123	&	57530.52718	&	0.32	&	2016-05-22 11:51:24	&	2016-05-22 13:26:53	&	878		 & $1.2^{+0.8}_{ -0.5}$ \\
00034135124	&	57531.42992	&	0.54	&	2016-05-23 10:11:14	&	2016-05-23 10:26:55	&	940		 & $2.0^{+1.0}_{ -0.6}$ \\
00034135125	&	57532.83370	&	0.87	&	2016-05-24 19:52:10	&	2016-05-24 20:08:52	&	1003	 & -- \\ 	
00034135126	&	57533.83165	&	0.10	&	2016-05-25 19:50:14	&	2016-05-25 20:04:54	&	880		 & $7.7^{+1.9}_{-1.6}$ \\
00034135127	&	57534.34778	&	0.22	&	2016-05-26 08:12:42	&	2016-05-26 08:28:53	&	970		 & $3.5^{+1.0}_{-0.8}$ \\
00034135128	&	57535.54997	&	0.51	&	2016-05-27 13:04:00	&	2016-05-27 13:19:55	&	955		 & $3.1^{+1.0}_{-0.7}$ \\
00034135129	&	57536.35117	&	0.70	&	2016-05-28 08:25:34	&	2016-05-28 08:39:54	&	860		 & $1.7^{+1.1}_{-0.7}$ \\   
00034135130	&	57537.60277	&	0.99	&	2016-05-29 14:20:03	&	2016-05-29 14:35:54	&	950		 & $1.5^{+0.7}_{-0.4}$ \\
00034135131	&	57538.34686	&	0.17	&	2016-05-30 08:11:11	&	2016-05-30 08:27:54	&	993		 & $2.8^{+0.9}_{-0.7}$ \\
00034135132	&	57539.34113	&	0.40	&	2016-05-31 08:03:32	&	2016-05-31 08:18:55	&	920		 & $2.3^{+1.2}_{-0.7}$ \\
00034135133	&	57540.72807	&	0.73	&	2016-06-01 17:19:54	&	2016-06-01 17:36:55	&	1020		 & $2.7^{+1.0}_{-0.7}$ \\
00034135134	&	57541.46951	&	0.90	&	2016-06-02 09:44:16	&	2016-06-02 12:47:54	&	567		 & $0.6^{+0.3}_{-0.2}$ \\
00034135135	&	57542.59067	&	0.17	&	2016-06-03 14:04:12	&	2016-06-03 14:16:54	&	762		 & $3.6^{+1.2}_{-0.9}$ \\
00034135136	&	57543.38816	&	0.36	&	2016-06-04 09:11:59	&	2016-06-04 09:25:54	&	835		 & $3.9^{+1.6}_{-1.1}$ \\
00034135137	&	57544.46000	&	0.61	&	2016-06-05 10:54:52	&	2016-06-05 11:09:54	&	903		 & $1.4^{+0.7}_{-0.4}$ \\
00034135138	&	57545.91938	&	0.95	&	2016-06-06 21:55:54	&	2016-06-06 22:11:54	&	960		 & $3.8^{+2.5}_{-2.7}$ \\
00034135139	&	57546.71955	&	0.14	&	2016-06-07 17:08:24	&	2016-06-07 17:23:54	&	930		 & $2.0^{+0.8}_{-0.5}$ \\
00034135140	&	57547.74795	&	0.39	&	2016-06-08 17:04:11	&	2016-06-08 18:49:54	&	847		 & $6.1^{+1.6}_{-1.3}$ \\
00034135141	&	57548.90854	&	0.66	&	2016-06-09 21:39:42	&	2016-06-09 21:56:53	&	1031	& $3.8^{+1.1}_{-0.8}$ \\
00034135142	&	57549.76590	&	0.86	&	2016-06-10 18:14:53	&	2016-06-10 18:30:54	&	960	 & -- \\ 
00034135143	&	57550.76243	&	0.10	&	2016-06-11 18:09:52	&	2016-06-11 18:25:54	&	963		 & $2.0^{+0.8}_{-0.6}$ \\
00034135144	&	57551.69627	&	0.32	&	2016-06-12 16:34:22	&	2016-06-12 16:50:53	&	990		 & $7.0^{+1.6}_{-1.3}$ \\
00034135145	&	57552.69249	&	0.55	&	2016-06-13 16:28:28	&	2016-06-13 16:45:54	&	1046	& $6.1^{+1.5}_{-1.2}$ \\
00034135146	&	57553.62090	&	0.77	&	2016-06-14 14:45:17	&	2016-06-14 15:02:55	&	1058	& $1.0^{+0.8}_{-0.4}$ \\
00034135147	&	57554.55088	&	0.99	&	2016-06-15 13:04:37	&	2016-06-15 13:21:55	&	1038	& $4.3^{+1.1}_{-0.9}$ \\
00034135148	&	57555.41071	&	0.19	&	2016-06-16 08:13:55	&	2016-06-16 11:28:54	&	572		 & $2.8^{+1.5}_{-1.0}$ \\
00034135150	&	57557.21669	&	0.62	&	2016-06-18 05:04:09	&	2016-06-18 05:19:54	&	945		 & $2.2^{+0.8}_{-0.6}$ \\
\end{longtable}
\FloatBarrier
%
%
%
%
%
%
%
 \begin{table*}
  \begin{center}
 \caption{{\it Swift} observation log of XTE~J1855$-$026}
 \label{x1855:tab:swift_xrt_log}
  \small
\begin{tabular}{ lllllll  }
 \hline
 \hline
 \noalign{\smallskip}
 Sequence/Instr./mode              & MJD                             & Start time  (UT)                      & End time   (UT)                         & Exposure    & Time since \\
                      &                                 & (yyyy-mm-dd hh:mm:ss)    & (yyyy-mm-dd hh:mm:ss)         &(s)                &  trigger (s)        \\
   \noalign{\smallskip}
 \hline
 \noalign{\smallskip}

00503434000/BAT/evt	 &	   55822.48886     &       2011-09-18 10:03:35    &       2011-09-18T13:24:19	    &       1257	 & $-$239--11805 \\
 00503434000/XRT/WT    &        55822.49860     &       2011-09-18 10:32:44     &       2011-09-18 13:23:13     &      629	& 1510--11740 \\ 
00503434000/XRT/PC     &        55822.49932     &       2011-09-18 10:34:42     &       2011-09-18 13:23:20     &       1264 & 1629--11746 \\ 
  \noalign{\smallskip}
  \hline
  \end{tabular}
\tablefoot{We show: observing sequence/instrument/mode, date (MJD middle), start and end times (UT),  exposure time and time since the trigger.}
  \end{center}
  \end{table*}

%
%
%
%
%
 \begin{table*}[th]
\tabcolsep 4pt
 \begin{center}
 \caption{{\it Swift}/XRT observation log for IGR~J17503$-$2636.}
 \label{i17503:tab:swift_xrt_log}
 \small
\begin{tabular}{ lllll llll }
 \hline
 \hline
 \noalign{\smallskip}
Sequence        & MJD                      & Start time  (UT)                     & End time   (UT)                      & Exposure  & $N_{\rm H}$       & $\Gamma$ & $F_{\rm 0.3-10~keV}$ &Cstat/d.o.f. \\
  \noalign{\smallskip}
 \hline
 \noalign{\smallskip}
%
00010980001	&	58576.71546	&	2019-04-03 17:10:15	&	2019-04-03 22:17:53	&	5178	& $8.3^{+1.8}_{-1.6}$ & $2.8^{+ 0.9}_{-0.8}$  & $0.9^{+0.1}_{-0.1}$   &   120.2/142 \\
00010980002	&	58583.49265	&	2019-04-10 11:49:25	&	2019-04-10 15:26:52	&	3372	& $5.7^{+1.5}_{-1.2}$ & $2.6^{+ 1.0}_{-0.9}$  & $1.0^{+0.2}_{-0.1}$    &    83.9/116\\
00010980003	&	58590.53303	&	2019-04-17 12:47:33	&	2019-04-17 20:56:53	&	4944	& $2.0^{+1.4}_{-0.8}$ & $-0.4^{+ 0.9}_{-0.7}$  & $0.7^{+0.2}_{-0.1}$ &     95.2/79\\
00010980005	&	58611.77889	&	2019-05-08 18:41:35	&	2019-05-08 22:20:52	&	4012	& $5.3^{+0.8}_{-0.7}$ & $1.7^{+ 0.5}_{-0.5}$  & $3.4^{+0.3}_{-0.3}$ &    229.8/278\\
00010980006	&	58618.10236	&	2019-05-15 02:27:23	&	2019-05-15 15:10:53	&	4897	& $8.4^{+1.2}_{-1.1}$ & $1.3^{+ 0.5}_{-0.5}$  & $3.3^{+0.3}_{-0.2}$ &    216.9/279\\
00010980007	&	58643.79713	&	2019-06-09 19:07:51	&	2019-06-10 00:06:53	&	4734	& $4.3^{+0.7}_{-0.6}$ & $1.5^{+ 0.5}_{-0.4}$  & $2.5^{+0.2}_{-0.2}$ &    219.1/269\\
 \noalign{\smallskip}
%
00010807001     &       58343.82179     &       2018-08-13 19:43:22     &       2018-08-13 20:13:11     &       990     & $2.0^{+0.7}_{-0.5}$ & $0.1^{+ 0.5}_{-0.5}$  & $14.7^{+1.9}_{-1.6}$&       118.2/156 \\
%
00088805001     &       58353.19412     &       2018-08-23 04:39:32     &       2018-08-23 06:38:52     &       1858    &  $8.2^{+2.7}_{-2.2}$ & $1.1^{+ 1.2}_{-1.0}$  & $1.9^{+0.4}_{-0.3}$ &     70.0/82 \\
 %
%
%
00048022007	&	56224.21009	&	2012-10-24 05:02:31	&	2012-10-24 05:30:54	&	1702	      & $0.5^{+0.4}_{-0.2}$ & $1.9^{+ 1.0}_{-0.8}$  & $0.5^{+0.2}_{-0.1}$ &   44.4/47 \\
\noalign{\smallskip}
  \hline
  \end{tabular}
\tablefoot{We show: observing sequence, date (MJD of the middle of the observation), start and end times (UT), and XRT exposure time in seconds.
    We also report the absorption column density $N_{\rm H}$ in units of  10$^{23}$\,cm$^{-2}$, the powerlaw photon index $\Gamma$,
    and the flux in the 0.3--10\,keV energy band (not corrected for absorption in units of 10$^{-11}$\,erg\,cm$^{-2}$\,s$^{-1}$) as determined from the spectral fit.}
  \end{center}
  \end{table*}

\end{appendix}
\end{document}